\documentclass[prb,aps,nofootinbib,amssymb,twocolumn,superscriptaddress,10pt,floatfix]{revtex4-2}

\usepackage[T1]{fontenc}
\usepackage{graphicx}
\usepackage{xcolor}
\usepackage{booktabs}
\usepackage{dcolumn}
\usepackage{comment}
\usepackage{amsmath,amssymb,bbold,bm}
\usepackage{hyperref}
\usepackage{amsfonts}
\usepackage{float,latexsym}
\usepackage{multirow}
\usepackage{color}
\usepackage[normalem]{ulem}
\usepackage{cleveref}
\usepackage{mhchem}
\usepackage{physics}
\usepackage{tabularx}
\usepackage{array}
\usepackage[caption=false]{subfig}
\usepackage{siunitx}
\usepackage{gensymb}
\usepackage{pifont}
\usepackage{placeins}
\usepackage{mathtools}
\usepackage{longtable}
\usepackage{multirow}
\usepackage{nicematrix}
\usepackage{tikz}
\usepackage{xstring}
\usepackage{makecell}
\usepackage{xr}
\usepackage{ifthen}
\usetikzlibrary{decorations.markings}
\usetikzlibrary{positioning}
\usetikzlibrary{arrows.meta}
\usepackage{tikz-feynman}
\usetikzlibrary{calc}
\tikzfeynmanset{warn luatex=false}
\let\vec\mathbf

\usepackage{chemformula}

\makeatletter
\def\maketag@@@#1{\hbox{\m@th\normalfont\normalsize#1}}
\makeatother

\crefname{appendix}{Appendix}{Appendices}
\crefname{equation}{Eq.}{Eqs.}
\crefname{figure}{Fig.}{Figs.}
\crefname{table}{Table}{Tables}
\crefname{section}{Section}{Sections}
\crefname{enumi}{Point}{Points}
\creflabelformat{appendix}{[#2#1#3]}
\crefmultiformat{figure}{Figs.~#2#1\xdef\mycreffirstarg{#1}#3}%
{ and~#2#1#3}{, #2#1#3}{ and~#2#1#3}
\makeatletter     
\renewcommand\onecolumngrid{% <<<<<<
	\do@columngrid{one}{\@ne}%
	\def\set@footnotewidth{\onecolumngrid}% <<<<<<<<<<<<<<<<
	\def\footnoterule{\kern-6pt\hrule width 1.5in\kern6pt}%
}

\makeatletter
\AtBeginDocument{\let\LS@rot\@undefined}
\makeatother 
\renewcommand{\arraystretch}{1.2}

\newcommand{\cre}[2]{\hat{#1}^\dagger_{#2}}
\newcommand{\des}[2]{\hat{#1}_{#2}}

\newcommand{\ie}{{\it i.e.}}
\newcommand{\eg}{{\it e.g.}}

\newcommand{\be}[0]{\begin{equation}}
	\newcommand{\ee}[0]{\end{equation}}

\def\ba#1\ea{\begin{align}#1\end{align}}

\newcommand{\bmat}[0]{\begin{bmatrix}}
	\newcommand{\emat}[0]{\end{bmatrix}}

\newcommand{\Uone}{$\mathrm{U}(1)$ }

\newcommand{\vk}{\vec{k}}

\newcommand{\vQ}{\vec{Q}}
\newcommand{\vq}{\vec{q}}
\newcommand{\vG}{\vec{G}}
\newcommand{\vR}{\vec{R}}

\def\kk{\mathbf{k}}

\def\rr{\mathbf{r}}

\newcommand{\titlePaper}{Engineering topological flat bands in $\Gamma$-valley moir\'e with Ising-type SOC: twisted 1T–\ch{ZrS2} and 1T–\ch{SnSe2}}

\newcommand{\paperAuthors}{%
	\author{Hanqi Pi}
	\thanks{These authors contributed equally to this work.}
	\affiliation{Donostia International Physics Center (DIPC), Paseo Manuel de Lardizábal. 20018, San Sebastián, Spain}
	\author{Yves H. Kwan}
	\thanks{These authors contributed equally to this work.}
	\affiliation{Department of Physics, University of Texas at Dallas, Richardson, Texas 75080, USA}
	\affiliation{Princeton Center for Theoretical Science, Princeton University, Princeton, NJ 08544}
	\author{Haoyu Hu}
	\affiliation{Donostia International Physics Center (DIPC), Paseo Manuel de Lardizábal. 20018, San Sebastián, Spain}
	\affiliation{Department of Physics, Princeton University, Princeton, NJ 08544, USA}
	\affiliation{Department of Physics, University of Science and Technology of China, Hefei, Anhui 230026, China}
	\author{Yi Jiang}
	\affiliation{Donostia International Physics Center (DIPC), Paseo Manuel de Lardizábal. 20018, San Sebastián, Spain}
	\author{Dumitru C\u{a}lug\u{a}ru}
	\affiliation{Rudolf Peierls Centre for Theoretical Physics, University of Oxford, Oxford OX1 3PU, United Kingdom}
	\author{Jie Shan}
	\affiliation{School of Applied and Engineering Physics, Cornell University, Ithaca, NY 14850, USA}
	\affiliation{Department of Physics, Cornell University, Ithaca, NY 14850, USA}
	\affiliation{Kavli Institute at Cornell for Nanoscale Science, Ithaca, NY 14850, USA}
	\affiliation{Max Planck Institute for the Structure and Dynamics of Matter, Hamburg, Germany}
	\author{Kin Fai Mak}
	\affiliation{School of Applied and Engineering Physics, Cornell University, Ithaca, NY 14850, USA}
	\affiliation{Department of Physics, Cornell University, Ithaca, NY 14850, USA}
	\affiliation{Kavli Institute at Cornell for Nanoscale Science, Ithaca, NY 14850, USA}
	\affiliation{Max Planck Institute for the Structure and Dynamics of Matter, Hamburg, Germany}
	\author{Miguel M.~Ugeda}
	\affiliation{Donostia International Physics Center (DIPC), Paseo Manuel de Lardizábal. 20018, San Sebastián, Spain}
	\affiliation{IKERBASQUE, Basque Foundation for Science, Bilbao, Spain}
	\author{Dmitri K.~Efetov}
	\affiliation{Faculty of Physics, Ludwig-Maximilians-University Munich, Munich 80799, Germany}
	\affiliation{Munich Center for Quantum Science and Technology (MCQST), Ludwig-Maximilians-University Munich, Munich 80799, Germany}
	\author{Maia G.~Vergniory}
	\affiliation{Donostia International Physics Center (DIPC), Paseo Manuel de Lardizábal. 20018, San Sebastián, Spain}
	\affiliation{Département de Physique et Institut Quantique, Université de Sherbrooke, Sherbrooke, J1K 2R1 Québec, Canada}
	\author{B.~Andrei Bernevig}
	\email{bernevig@princeton.edu}
	\affiliation{Department of Physics, Princeton University, Princeton, NJ 08544, USA}
	\affiliation{Donostia International Physics Center (DIPC), Paseo Manuel de Lardizábal. 20018, San Sebastián, Spain}
	\affiliation{IKERBASQUE, Basque Foundation for Science, Bilbao, Spain}
}

\newcommand{\tAA}{\text{AA}}
\newcommand{\tAB}{\text{AB}}

\newcommand{\vg}{\vec{g}}
\newcommand{\tMoire}{\text{moir\'e}}
\newcommand{\tSl}{\text{sl}}

\crefname{appendix}{Appendix}{Appendices}
\crefname{equation}{Eq.}{Eqs.}
\crefname{figure}{Fig.}{Figs.}
\crefname{table}{Table}{Tables}
\crefname{section}{Section}{Sections}
\creflabelformat{appendix}{[#2#1#3]}

\makeatletter
\AtBeginDocument{\let\LS@rot\@undefined}
\makeatother

\makeatletter
\renewcommand\onecolumngrid{% <<<<<<
\do@columngrid{one}{\@ne}%
\def\set@footnotewidth{\onecolumngrid}% <<<<<<<<<<<<<<<<
\def\footnoterule{\kern-6pt\hrule width 1.5in\kern6pt}%
}

\allowdisplaybreaks

\begin{document}
\title{\titlePaper}
\paperAuthors
\let\oldaddcontentsline\addcontentsline
% !TEX root = ./main.tex

\begin{abstract}
Twisted moir\'e superlattices hosting topological flat bands provide a platform to explore the interplay between topology and correlations. Here we investigate topological band structures in $\Gamma$-valley moir\'e systems based on 1T–\ch{ZrS2} and 1T–\ch{SnSe2}. Using large-scale \emph{ab initio} calculations and continuum modelling, we demonstrate that both materials exhibit an approximate spin-\Uone symmetry and host isolated topological moir\'e valence bands, including quantum spin Hall and high spin Chern states. By constructing a hierarchy of $\Gamma$-valley moir\'e continuum models, we show that isolated moir\'e bands carry a trivial $C_{3}$ symmetry indicator when the low-energy physics is described by a single effective orbital and a single layer-hybridized branch (either bonding or antibonding). Topological bands therefore arise from the inter-branch and/or inter-orbital coupling. Moreover, we determine the interaction-driven phase diagrams using Hartree–Fock and exact diagonalization, finding various phases tunable by twist angle, interaction strength, and displacement field. We identify specific conditions under which fractional Chern insulators are favored. Together with previous work showing that the moir\'e conduction bands of 1T–\ch{ZrS2} and 1T–\ch{SnSe2} realize $M$-valley twisting and host quasi-one-dimensional physics~\cite{cualuguaru2024new}, our results establish these systems as ideal platforms for strongly correlated moir\'e physics and provide a systematic framework for understanding topological band structures in $\Gamma$-valley moir\'e materials.

\end{abstract}

\maketitle

\section{Introduction}
Moir\'e superlattices have become a highly tunable platform for exploring correlated and topological quantum matter, with bandwidth, correlations, and topology controlled by twist angle, displacement field, and carrier density~\cite{DAI16,AND21,andrei2020graphene,KOS19,BAL20,CHE19}. Experiments have realized a wide variety of correlated phases, including unconventional superconductivity in graphene moir\'e systems~\cite{CAO18a, CAO18,YAN19,SHA19,LU19,SER20,PO18a,CAO20a,CHE19a}, as well as Mott insulators, generalized Wigner crystals~\cite{regan2020mott,li2021imaging,huang2021correlated,XU20a,TAN20}, and superconductivity~\cite{XIA24a,GUO24,xu2025signatures} in TMD moir\'e heterostructures. The nontrivial band topology of moir\'e flat bands further intertwines with correlations~\cite{WU19b,dongCompositeFermiLiquid2023,XU18,HER24c} to produce phenomena such as the fractional Chern insulators (FCIs) at zero magnetic field~\cite{regnault2011fractional,sun2011nearly,neupert2011fractional,WAN24a,CAI23,PAR23,ZEN23,XIE21d,xie2025tunable} and the fractional quantum spin Hall insulator~\cite{kangEvidenceFractionalQuantum2024a} observed in twisted \ch{MoTe2}, although more work is needed to determine the origin of experimental observation. Exploring and understanding new moir\'e platforms that naturally host isolated topological flat bands is a key step toward robustly engineering these exotic correlated and topological quantum phases.
Research has extended moir\'e physics beyond the familiar $K$-valley to platforms whose low-energy states reside at other momenta and in distinct symmetry environments~\cite{JIA24b,WAN22,KEN20,CLA22a,klebl2022moire}. These include $\Gamma$-valley dominated regimes in TMD homobilayers and twisted double bilayers~\cite{ANG21,pei2022observation,foutty2023tunable,campbell2024interplay,maRelativisticMottTransition2025a,qi2025chern}, as well as $M$-valley systems that realize quasi-one-dimensional or highly anisotropic triangular/rectangular moir\'e lattices~\cite{cualuguaru2024new,lei2025moire,bao2025anisotropic}. Although these non-$K$-valley platforms broaden the accessible correlation physics, the single-particle moir\'e bands are often topologically trivial~\cite{ANG21,qi2025chern}.
This motivates the search for new moir\'e materials that intrinsically host topological minibands. In the weak-coupling regime, several works have proposed systematic theoretical frameworks based on symmetry-based continuum modelling and topological quantum chemistry to predict and diagnose topological moir\'e flat bands~\cite{crepel2025efficient,lhachemi2025efficient,liu2025symmetry,yang2025engineering}.

\begin{figure*}[!t]
    \centering
    \includegraphics[width=\textwidth]{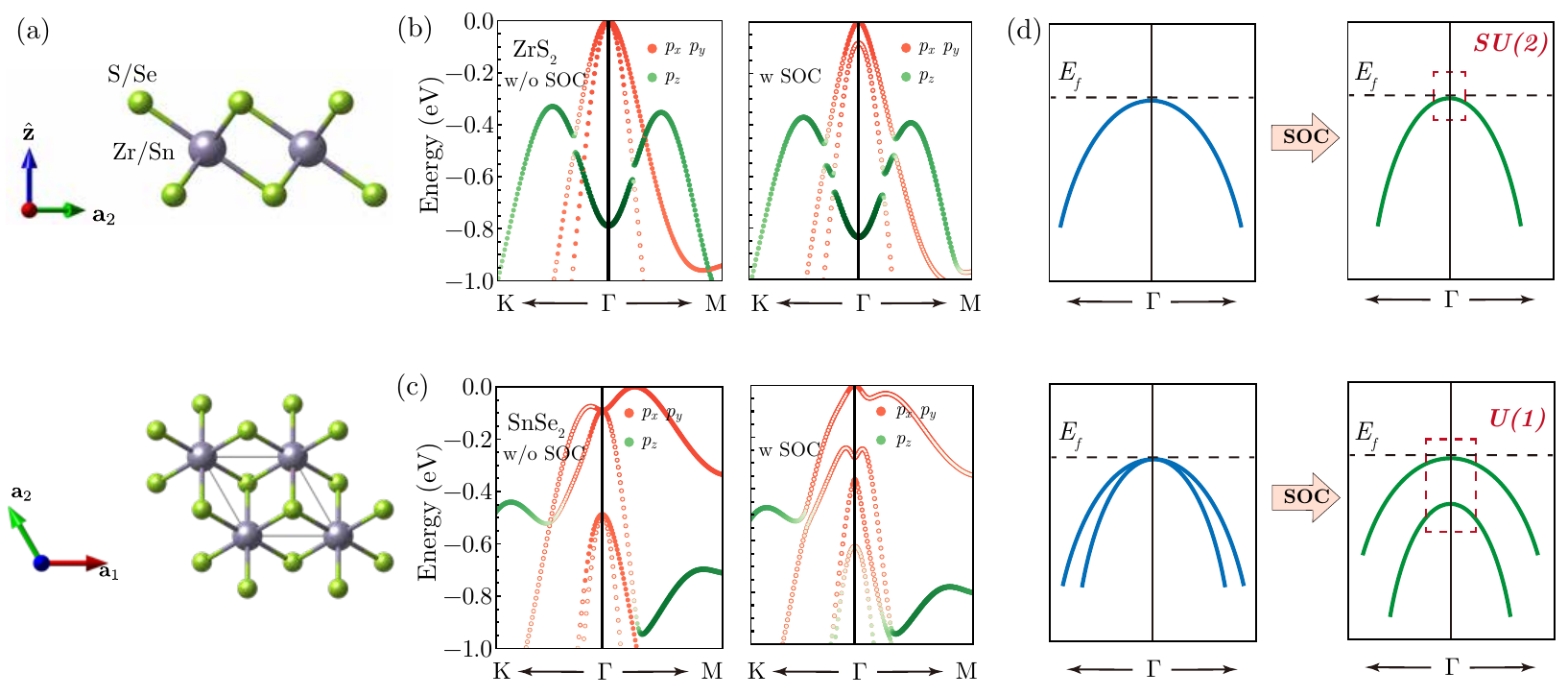}
     \subfloat{\label{fig:mono-topo:a}}
     \subfloat{\label{fig:mono-topo:b}}
     \subfloat{\label{fig:mono-topo:c}}
     \subfloat{\label{fig:mono-topo:d}}
    \caption{Crystal structure and orbital-resolved band structures of monolayer 1T-\ch{SnSe2} and 1T-\ch{ZrS2}. (a) Top and side views of the 1T-TMD crystal structure. (b, c) Band structures of monolayer (b) \ch{ZrS2} and (c) \ch{SnSe2} calculated without (left subpanels) and with (right subpanels) spin-orbit coupling (SOC). The red and blue dots represent the projection weights of the chalcogen in-plane ($p_x,p_y$) and out-of-plane $p_z$ orbitals, respectively. (d) Schematic illustrating the influence of the IRREP dimension on the $\Gamma$ states. Top: The state contributed by $p_z$ orbital transforms as a 1D IRREP (left) without SOC, which preserves an approximate spin-$\mathrm{SU}(2)$ symmetry near $\Gamma$ and evolves into a single degenerate Kramers pair upon including SOC (right). Bottom: The state contributed by $(p_x,p_y)$ orbital transforms as a 2D IRREP (left) without SOC, which preserves an Ising-type SOC near $\Gamma$ and splits into two energetically separated Kramers pairs upon including SOC (right).}
    \label{fig:fig-mono-topo}
\end{figure*}

In this work, we use twisted bilayer 1T-\ch{ZrS2} and 1T-\ch{SnSe2} as concrete material platforms and study the origin of topological moir\'e bands in $\Gamma$-valley moir\'e system. Both systems host topological moir\'e valence bands, including quantum spin Hall (QSH) and high-spin Chern states, as established by large-scale density functional theory (DFT) calculations and continuum modelling. The QSH state in AB-stacked twisted 1T-\ch{ZrS2} has been reported in Ref.~\cite{CLA22a}, where the topology was attributed to strong spin-orbit coupling (SOC) and an emergent kagome lattice. Here we study both AA- and AB-stacked twisted 1T-\ch{ZrS2} and 1T-\ch{SnSe2} and develop a symmetry-based microscopic framework to understand isolated topological moir\'e bands in systems with Ising-type SOC. Based on the symmetry and orbital analysis of the monolayer, we identify the origin of an approximate spin-\Uone symmetry in the low-energy states of twisted bilayer \ch{ZrS2} and \ch{SnSe2}. Because $\Gamma$-valley states in the two layers remain momentum-aligned under twisting, interlayer tunneling is generally strong and naturally forms layer-hybridized bonding and antibonding branches. Hereafter, ``branch'' refers to these bonding/antibonding layer-hybridized states. By analyzing a hierarchy of symmetry-constrained continuum models, we show that isolated topological bands in spin-\Uone-symmetric moir\'e systems require inter-branch and/or inter-orbital coupling. Finally, using Hartree–Fock and exact diagonalization (ED), we observe spin-polarized Chern states and QSH states, and discuss the conditions required to realize FCIs in these platforms. Our framework provides a symmetry- and material-based route to systematically analyze and design topological moir\'e bands based on 1T-TMDs, which can be straightforwardly generalized to other moir\'e and superlattice platforms.

\section{Topological moir\'e bands in twisted bilayer \ch{ZrS2} and \ch{SnSe2}}\label{sec:twist ZrS2 and SnSe2}

\begin{figure*}[!t]
    \centering
    \includegraphics[width=\textwidth]{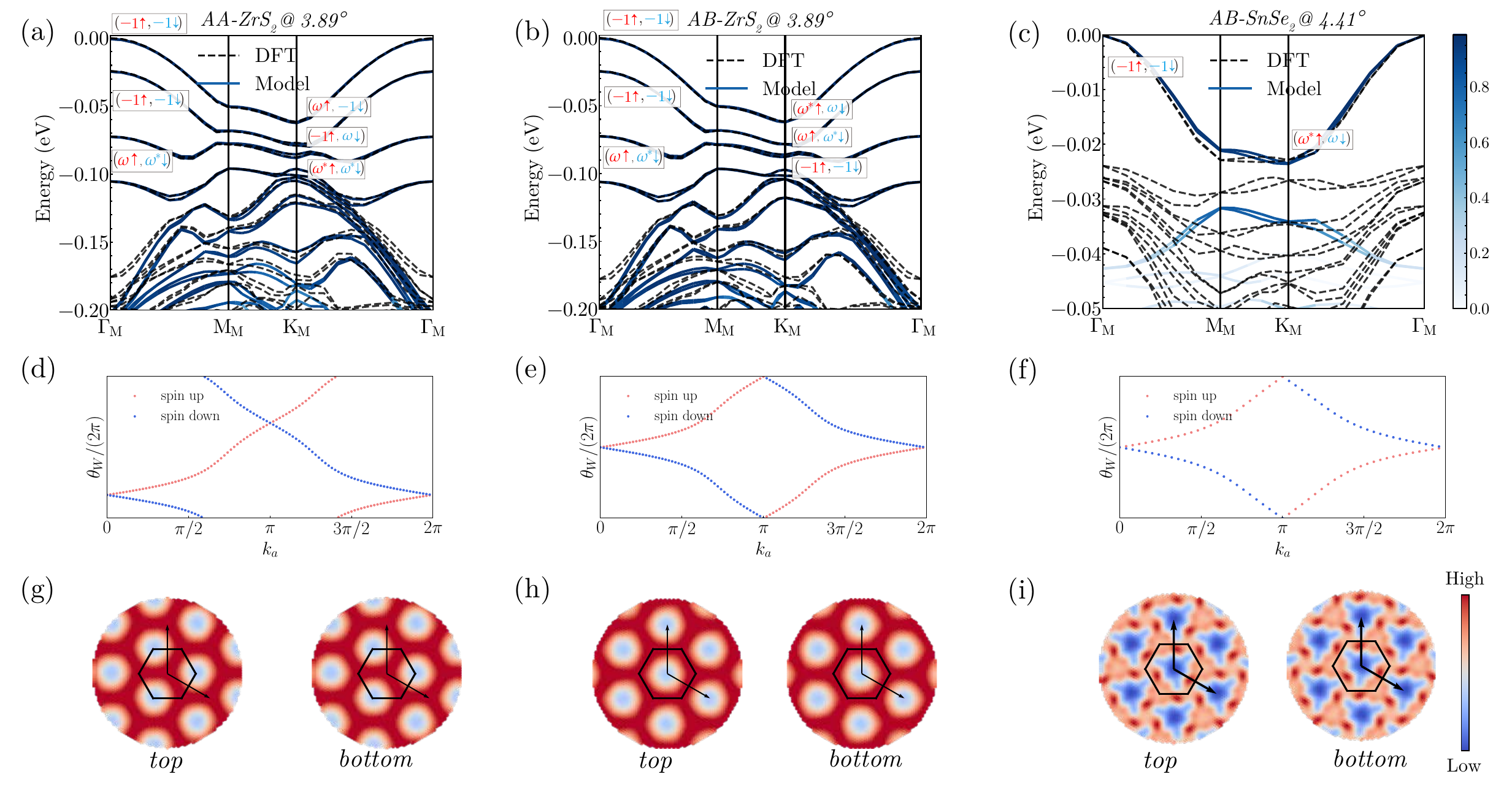}
     \subfloat{\label{fig:model-dft:a}}
     \subfloat{\label{fig:model-dft:b}}
     \subfloat{\label{fig:model-dft:c}}
     \subfloat{\label{fig:model-dft:d}}
     \subfloat{\label{fig:model-dft:e}}
     \subfloat{\label{fig:model-dft:f}}
     \subfloat{\label{fig:model-dft:g}}
     \subfloat{\label{fig:model-dft:h}}
     \subfloat{\label{fig:model-dft:i}}
    \caption{Electronic and topological properties of twisted bilayer \ch{ZrS2} and \ch{SnSe2}. (a–c) Band structures for (a) AA-stacked \ch{ZrS2} at \SI{3.89}{\degree}, (b) AB-stacked \ch{ZrS2} at \SI{3.89}{\degree}, and (c) AB-stacked \ch{SnSe2} at \SI{4.41}{\degree}. Black dashed lines and blue solid lines represent the results from DFT calculations and the first harmonic continuum model with parameters in \cref{app:tab:continuum_parameters for first harmonic AA ZrS2,app:tab:continuum_parameters for first harmonic AB ZrS2,app:tab:continuum_parameters for second harmonic AB SnSe2}, respectively. The color intensity indicates the wavefunction overlap between the DFT and continuum model states. The $C_{3z}$ eigenvalue of $\Gamma_M$ and $K_M$ are listed near the momentum. (d–f) The spin-projected Wilson loop for the topmost isolated bands in (a–c). (g–i) Local density of states (LDOS) distributions of the topmost isolated bands for (g) AA-stacked \ch{ZrS2}, (h) AB-stacked \ch{ZrS2}, and (i) AB-stacked \ch{SnSe2}. The left and right subpanels correspond to the LDOS of the top and bottom layers, respectively.
    }
    \label{fig:fig-model-dft}
\end{figure*}

We first examine the monolayer band structures of 1T–\ch{ZrS2} and 1T–\ch{SnSe2}. Both materials crystallize in a hexagonal lattice as shown in \cref{fig:mono-topo:a} and belong to the space group (SG) $P\overline{3}m1$ (No. 164). In this work, we focus on the valence bands, and the analysis of the conduction bands can be found in Ref.~\cite{cualuguaru2024new}. As shown in \cref{fig:mono-topo:b,fig:mono-topo:c}, the monolayer valence band maxima (VBM) of both compounds lie at $\Gamma$ (in \ch{SnSe2}, SOC shifts the VBM from the $\Gamma$–$M$ line to $\Gamma$), and are dominated by chalcogen $(p_x,p_y)$ orbitals. Crucially, the topmost state at $\Gamma$ transforms according to a 2D irreducible representation (IRREP) without SOC, and the inclusion of SOC splits it into two Kramers pairs. Although both compounds have the same crystalline symmetry and chalcogen $(p_x,p_y)$-dominated $\Gamma$-valley valence states, the two Kramers pairs are assigned to different IRREPs. In \ch{ZrS2} the split spinful $\Gamma$ states transform as $\overline{\Gamma}_6\overline{\Gamma}_7$ and $\overline{\Gamma}_9$ with odd parity under inversion, while in \ch{SnSe2} they transform as $\overline{\Gamma}_4\overline{\Gamma}_5$ and $\overline{\Gamma}_8$ with even parity. The IRREP notation follows the convention of \textit{Bilbao Crystallographic Server} (BCS)~\cite{aroyo2011crystallography, aroyo2006bilbao1, aroyo2006bilbao2}. 
The $\overline{\Gamma}_6\overline{\Gamma}_7$ ($\overline{\Gamma}_4\overline{\Gamma}_5$) pair corresponds to $p$-like effective orbital with total angular momentum $|J_z|=\frac{3}{2}$ and $\overline{\Gamma}_9$ ($\overline{\Gamma}_8$) exhibits $s$-like character with $|J_z|=\frac{1}{2}$. The different IRREPs in the two compounds can be understood from the fact that the $\Gamma$-valley states arise from the hybridization between $(p_x,p_y)$ orbitals on the two chalcogen sublayers through the intermediate metal sublayer, which forms combinations with even- and odd-parity since the two sublayers are related by $\mathcal{I}$. Because the chalcogen orbitals hybridize with different metal-orbitals, \textit{i.e.}, \ch{Zr}-$d$ in \ch{ZrS2} and \ch{Sn}-$s$ in \ch{SnSe2}, the relative ordering of these two combinations can be reversed, leading to different IRREP assignments for the topmost Kramers pairs.

The orbital component at $\Gamma$ plays a central role in determining the effective spin symmetry of the low-energy moir\'e bands.
Depending on the orbital character, the resulting moir\'e low-energy states can have either approximate spin-\Uone or spin-SU(2) symmetry.
To see this, we project the onsite atomic SOC, $
    H_{\mathrm{SOC}}=\lambda \vec{L}\cdot\vec{S}
    =\frac{\lambda}{2}\left(2L_zS_z+L_+S_-+L_-S_+\right)$, 
onto the low-energy $(p_x,p_y)$ manifold. Introducing the circular basis $ \ket{p_\pm}=\frac{\ket{p_x}\pm i\ket{p_y}}{\sqrt2}$ for the topmost $\Gamma$-state without SOC, the ladder operators $L_\pm$ connect the low-energy $p_\pm$ sector to the higher-lying $p_z$ sector and therefore have no matrix elements within the projected subspace. Thus, within the low-energy $(p_x,p_y)$ manifold, the SOC is diagonal in $S_z$ and is therefore Ising-like, which splits the doubly degenerate states without SOC into two spinful Kramers doublets as shown in bottom panel of \cref{fig:mono-topo:d}.
By contrast, if the low-energy $\Gamma$-valley manifold is dominated by an $m_\ell=0$ orbital such as $d_{z^2}$ in \ch{MoTe2}, the projected first-order SOC vanishes and spin anisotropy arises only through virtual mixing with remote bands. This is the origin of the approximate spin-$\mathrm{SU}(2)$ symmetry of the $\Gamma$-valley moir\'e bands in bilayer and multilayer twisted \ch{MoTe2}~\cite{qi2025chern,JIA24}. In the exact spin-$\mathrm{SU}(2)$ limit with time-reversal symmetry, the $Z_2$ index is trivial.

Because the monolayer structure lacks the $C_{2z}$ symmetry, the untwisted bilayer structure allows for two distinct configurations. The first, referred to as AA stacking, is constructed by placing one monolayer directly on top of the other and preserves the symmetry of the monolayer \ie, $P\overline{3}m1$ (No. 164). The second, referred to as AB stacking, is obtained by applying $C_{2z}$ to the bottom layer and belongs to SG $P\overline{6}m2$ (No. 187). Upon twisting, the inversion symmetry $\mathcal{I}$ (or in-plane mirror $\mathcal M_z$) present in the untwisted AA (AB) configuration is broken, while the $C_{3z}$ and in-plane $C_{2}$ about [010] ([210]) are preserved. As shown in \cref{fig:model-dft:a,fig:model-dft:b,fig:model-dft:c}, the large-scale DFT calculations at twist angle $3.89^\circ$ show that AB–\ch{SnSe2} hosts a single isolated set of bands composed of the top two spinful valence bands. By contrast, AA-\ch{SnSe2} does not exhibit an isolated valence-band manifold for $\theta\ge 3.89^\circ$ (see \cref{app:fig:SnSe2-moire-band}), and we therefore do not further discuss it. Twisted AA– and AB–\ch{ZrS2} instead organize the top six spinful bands into three isolated sets. In all cases, the isolated band sets exhibit approximately double degeneracy along high-symmetry lines. By inspecting the DFT wavefunctions, we find that spin-\Uone remains an excellent approximate symmetry for these low-energy isolated bands, with symmetry breaking below 2\% according to the measure defined in \cref{app:eqn:definetion of U1 breaking}. This approximate symmetry is inherited from the monolayer $\Gamma$-valley states discussed above. Therefore, we can diagnose the topology from the spin Chern number $C_s$ defined as $C_s=\ln \left(-\xi_{\Gamma_M} \xi_{K_M}\xi_{K_M^{\prime}}\right) /\left(i \frac{2 \pi}{3}\right) \bmod 3$ with $C_{3z}$ eigenvalues $\xi$ at $C_{3z}$-invariant momenta~\cite{fang2012bulk}. As summarized in \cref{fig:model-dft:a,fig:model-dft:b,fig:model-dft:c}, all isolated bands in these materials carry a nonzero spin Chern number, realizing either a quantum spin Hall state $\lvert C_s\rvert=1$ or a high spin Chern state $\lvert C_s\rvert=2$ as proved later by Wilson loop calculations. The band structures at the other angles can be found in \cref{app:sec:twist bilayer calculation}.

To gain a deeper understanding, we construct continuum models~\cite{ZHA24,cualuguaru2024new} for twisted AA/AB-stacked \ch{ZrS2} and AB-stacked \ch{SnSe2} (details in \cref{app:sec:extracted continuum model}). The general continuum model reads
$\mathcal{H}=\sum_{m,n}\int d^2 r \left[h(\vec{r})\right]_{mn}\cre{\psi}{m,\vec{r}}\des{\psi}{n,\vec{r}}$,
where $\cre{\psi}{m,\vec{r}}$ creates an electron at $\vec{r}$, and the composite index $m\equiv(l,\alpha,s)$ denotes the layer $l$, orbital $\alpha$, and spin $s$. Based on the monolayer IRREP analysis, the basis for \ch{ZrS2} includes both $s$-like molecular orbitals with $J_z=\pm\frac{1}{2}$ and $p$-like molecular orbitals with $J_z=\pm\frac{3}{2}$, whereas for \ch{SnSe2}, we keep only the $p$-like molecular orbital since the $s$-like state lies about \SI{270}{meV} below it due to the large SOC splitting as represented in \cref{fig:mono-topo:c}. As shown in \cref{fig:model-dft:a,fig:model-dft:b,fig:model-dft:c}, the continuum model dispersions agree excellently with DFT and achieve high wavefunction overlap for isolated band sets along high-symmetry lines ($>99\%$ in \ch{ZrS2} and $>93\%$ for \ch{SnSe2}), and the overlap is defined in \cref{app:eqn: definetion for overlap}. Due to the presence of approximate spin-\Uone symmetry, we calculated the spin-projected Wilson loops in \cref{fig:model-dft:d,fig:model-dft:e,fig:model-dft:f}, which confirm that the topmost isolated band set in each system realizes a QSH state. Furthermore, in AA/AB-\ch{ZrS2}, the subsequent two isolated sets carry a high spin Chern number $\lvert C_s\rvert=2$ as shown in \cref{app:fig:ZrS2-first-harmonic-model-wcc}. To gain analytic insight, we simplify the continuum models while retaining low-energy accuracy.
The intralayer part of the simplified Hamiltonian reads,
{\small
\begin{equation}
\begin{aligned}
[h(\mathbf r)]_{ls;ls}^{\mathrm{SnSe}_2}
=&
-\frac{\hbar^2\nabla^2}{2m_1^*}+2V_1\sum_{j=1}^3\cos(\vec{g}_j\cdot\vec{r})\\
&-2lV_2\sum_{j=1}^3\sin(\vec{g}_j\cdot\vec{r})+2V_3\sum_{j=1}^3\cos{(\vec{g}_{2j}\cdot\vec{r})},\\
[h(\mathbf r)]_{ls;ls}^{\mathrm{ZrS}_2}
=&
-\frac{\hbar^2\nabla^2}{2m_1^*}\sigma_0-\frac{\hbar^2}{2m_2^*}\left(\partial_+^{\,2}\,\sigma_s^{-}+\partial_-^{\,2}\,\sigma_s^{+}\right)+V_1\sigma_z\\
&+2lV_2
\sum_{j=1}^3
\Big[e^{i\phi_j}\sigma_s^-+e^{-i\phi_j}\sigma_s^+\Big]
\cos\!\big(\mathbf g_j\cdot\mathbf r+\phi_1\big),
\end{aligned}
\end{equation}
}where $\partial_{\pm}=\partial_x\pm i\partial_y$, $\sigma_s^{\pm}=\tfrac12(\sigma_x\pm is\sigma_y)$, $\phi_j=\tfrac{2\pi j}{3}-\tfrac{\pi}{6}$. Here $l=\pm1$ denotes the top/bottom layer, $s=\pm1$ denotes spin, and $\sigma_i$ are Pauli matrices acting on the orbital subspace spanned by $s$-like and $p$-like molecular orbitals. For \ch{SnSe2}, only the $p$-like molecular orbital is retained, so the orbital index is omitted.
The interlayer coupling terms are given by,
{\small
\begin{equation}
\begin{aligned}
[h(\mathbf r)]_{ls;-ls}^{\mathrm{SnSe}_2}
=&-\frac{\hbar^2\nabla^2}{2m_2^*}+w_1+2w_2\sum_{j=1}^3\cos(\vec{g}_j\cdot\vec{r})\\
&+2w_3\sum_{j=1}^3\cos(\vec{g}_{2j}\cdot\vec{r})-2ilsw_4\sum_{j=1}^3\sin(\vec{g}_{2j}\cdot\vec{r}),\\
[h(\mathbf r)]_{ls;-ls}^{\mathrm{ZrS}_2}
=
&w_1\sigma_0+2w_2\sigma_0\sum_{j=1}^3\cos(\vec{g}_{j}\cdot\vec{r}+\phi_2).\\
\end{aligned}
\end{equation}
}Parameter values are listed in \cref{app:tab:ZrS2_simplified_model_para_values,app:tab:SnSe2_simplified_model_para_values}. These reduced models make the emergent approximate symmetries transparent. Twisted AB \ch{SnSe2} has an additional inversion $\mathcal I$ in the simplified continuum model, which is a zero-twist symmetry~\cite{cualuguaru2024new} and originates from $\mathcal{M}_z$ of the untwisted AB-stacked bilayer as proved in \cref{app:sec:zero-twist symmetry}. For twisted AB-\ch{ZrS2}, the zero-twist symmetry remains inversion-like in its action on moir\'e coordinates, but its representation in the orbital subspace is no longer that of a pure inversion. Because the retained $s$-like and $p$-like orbitals have the same inversion parity, a pure inversion would act trivially, whereas they are distinguished through different $J_z$. The action on orbital subspace of zero-twist symmetry is therefore equivalent, up to an overall phase, to that of $C_{2z}$, so the simplified model of twisted AB-\ch{ZrS2} exhibits emergent $C_{2z}$ instead of $\mathcal I$. Besides, the simplified Hamiltonians for twisted AA- and AB-stacked \ch{ZrS2} are related by the fractional translation $T_{\bm{\tau}_0}=\{E|\bm{\tau}_0\}$ with $\bm{\tau}_0=\tfrac{2}{3}\mathbf a_{M,1}+\tfrac{1}{3}\mathbf a_{M,2}$ where $a_{M,1}$ and $a_{M,2}$ are moir\'e lattice vectors. This relation is clearly reflected in the LDOS patterns shown in \cref{fig:model-dft:g,fig:model-dft:h} and leads to a corresponding phase shift of the \(C_{3z}\) eigenvalues at \(K_M/K'_M\), as discussed in \cref{app:sec: topology origin in ZrS2}. The origin of $T_{\bm{\tau}_0}=\{E|\bm{\tau}_0\}$ can be understood from the interlayer-distance patterns of relaxed twisted structure at \SI{3.89}{\degree}. Strictly speaking, the AA and AB corrugation fields cannot be exactly related by a moir\'e translation, since the two stackings have different symmetries. However, if we quantify the similarity between the two corrugation patterns by the Pearson correlation coefficient defined in \cref{app:eqn:pearson correlation}, which is computed between the relaxed interlayer-distance field in AB-case $d_{\mathrm{AB}}(\mathbf r)$ and AA-case $d_{\mathrm{AA}}(\mathbf r+\bm{\tau}_0)$, we obtain $\rho=0.93$. This value shows that the two corrugation fields are close to being $T_{\bm{\tau}_0}$-related in the first-harmonic simplified model, and in turn motivates the emergent approximate symmetry $\tilde C_{2z}=\{C_{2z}|\bm{\tau}_0\}$ in twisted AA \ch{ZrS2}. 
These emergent symmetries, $\mathcal I$ and $C_{2z}/\tilde C_{2z}$, in addition to the spin-\Uone symmetry, account for the near-double degeneracy observed in the low-energy moir\'e bands of twisted AB-stacked \ch{SnSe2} and AA/AB-stacked \ch{ZrS2} in \cref{fig:model-dft:a,fig:model-dft:b,fig:model-dft:c}.

\section{Engineering topological flat band in $\Gamma$-valley}\label{sec:topo_mechanism}
\begin{figure*}[!t]
    \centering
    \includegraphics[width=\textwidth]{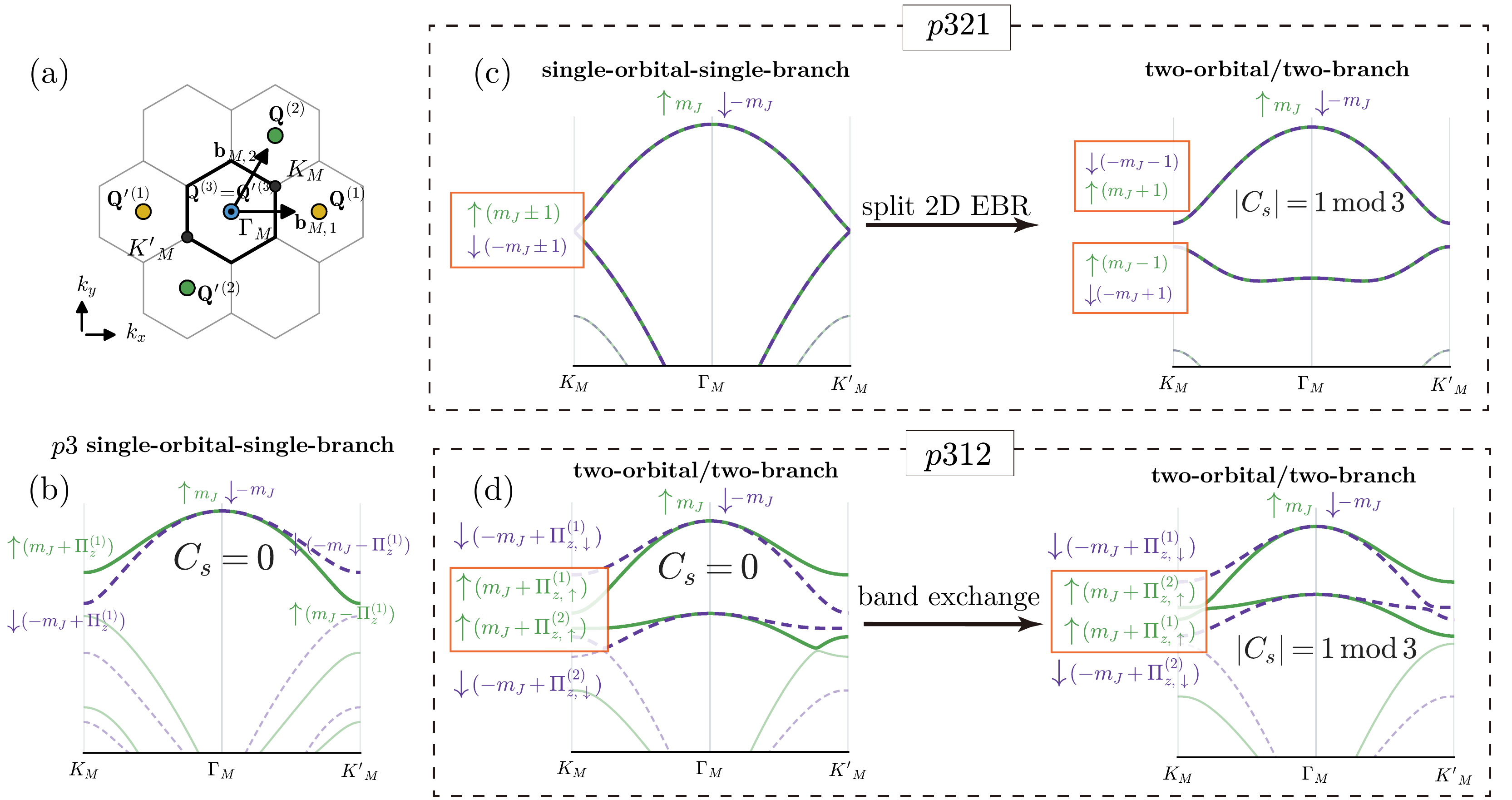}
     \subfloat{\label{fig:SBSO-SB-SO:a}}
     \subfloat{\label{fig:SBSO-SB-SO:b}}
     \subfloat{\label{fig:SBSO-SB-SO:c}}
     \subfloat{\label{fig:SBSO-SB-SO:d}}
    \caption{Triple-$\vQ$ description of isolated moir\'e bands in spin-\Uone-symmetric $\Gamma$-valley systems. Green solid and purple dashed curves denote spin-up and spin-down bands, respectively, and the $C_{3z}$ eigenvalues annotated next to top bands are color-matched to their spin sector. $m_J$ is the angular momentum of the spinful orbital, and $\Pi_z\in\{0,\pm1\}$ is the $\vQ$-lattice angular momentum. (a) Moir\'e Brillouin zone with the three nearest $\vQ$-sites $\vQ^{(j)}$ around $K_M$ and $\vQ'^{(j)}=-\vQ^{(j)}$ around $K'_M$. (b) Schematic band structure of layer group $p3$ in the \emph{single-orbital-single-branch} limit, and the constraint \cref{eqn:SBSO constraint} pins the topmost isolated Kramers pair to $C_s=0\mod 3$. (c) Schematic band structure of layer group $p321$. Left: in the \emph{single-orbital-single-branch} limit, the upper manifold is a fourfold Dirac-like multiplet at $K_M/K'_M$ protected by $C_{2y}$ together  with the constraint in \cref{eqn:SBSO constraint}, realizing a connected 2D EBR. Right: once an additional orbital or layer-hybridized branch is included, the constraint in \cref{eqn:SBSO constraint} is violated and the multiplet splits into two spin-degenerate doublets. The isolated upper doublet carries $|C_s|=1\mod 3$. (d) Schematic band structure of layer group $p312$ in both \emph{single-orbital-two-branch} and \emph{two-orbital-single-branch} cases. Left (trivial): $\Pi_{z,\uparrow}^{\mathrm{top}}=\Pi_{z,\downarrow}^{\mathrm{top}}$ gives $C_\uparrow=0\mod 3$. Right (topological): tuning the relative branch splitting drives a band inversion between blocks with different $\Pi_z$ in one spin sector, so $\Pi_{z,\uparrow}^{\mathrm{top}}\neq\Pi_{z,\downarrow}^{\mathrm{top}}$ and $|C_s|=1\mod 3$}
    \label{fig:fig-SBSO-SB-SO}
\end{figure*}

Building on the results for twisted bilayer \ch{SnSe2} and \ch{ZrS2}, we now investigate the microscopic origin of their band topology and derive the condition for isolated topological bands in spin-\Uone-symmetric $\Gamma$-valley moir\'e systems. Closely related symmetry-based analyses were developed in Ref.~\cite{liu2025symmetry,crepel2025efficient}, where the topology of the low-energy moir\'e bands is inferred from the monolayer symmetry group, the monolayer IRREP at $\Gamma$, and the moir\'e symmetry group.  For the moir\'e systems with $P31m$ or $P3m1$ symmetry group that is constructed by monolayer system with $C_{3z}$ symmetry, an isolated topological insulating phase is symmetry-enforced by the band representation when the relevant monolayer $\Gamma$ state transforms as $\bar{\Gamma}_4\bar{\Gamma}_5$. Nonetheless, due to the absence of symmetry indicator in the moir\'e symmetry group, the band representation analysis is not able to distinguish fragile from stable topology. In our case, the approximate spin-\Uone symmetry inherited from the monolayer allows us to distinguish the stable topology from the fragile topology through the $C_{3z}$ symmetry indicator~\cite{fang2012bulk}. Besides, the spin-\Uone symmetry forbids the isolated moir\'e bands to have stable topology in the \textit{single-orbital-single-branch} limit as shown below. 

We begin with the simplest \emph{single-orbital-single-branch} limit with $C_3$ symmetry, in which the low-energy Hilbert space is described by a single effective orbital with total angular momentum $J_z=\pm m_J$ and a single interlayer-hybridized branch (either layer bonding or antibonding). 
At the moir\'e $\Gamma_M$ point, we assume the moir\'e potential is not strong enough to induce level crossings. Thus the topmost Kramers pair is adiabatically connected to the topmost Kramers pair of the untwisted multilayer at $\Gamma$, and its $C_{3z}$ eigenvalues are $\xi_{\Gamma_M,s}=e^{- i\frac{2\pi}{3}\sigma m_J}$ with $\sigma=+1$ for $s=\uparrow$ and $\sigma=-1$ for $s=\downarrow$.
To determine $C_s$, we next analyze the moir\'e $K_M$ and $K'_M$ points. In the continuum description, a basis state at fixed moir\'e crystal momentum $\vk$ is denoted by $\ket{\vk,\vQ,s}$, where $\vQ$ is a moir\'e reciprocal-lattice vector and the physical momentum is $\vk-\vQ$. In the \emph{single-orbital-single-branch} limit, the six spinful low-energy states at $K_M$ are described by the three basis states, $\ket{K_M,\vQ^{(j)},s}$, where $j=1,2,3$, $\vQ^{(1)}=\mathbf b_{M,1}$, $\vQ^{(2)}=\mathbf b_{M,2}$, and $\vQ^{(3)}=\mathbf 0$. Similarly, the corresponding manifold at $\vk=K'_M$ is spanned by  $\ket{K'_M,\vQ'^{(j)},s}$ with  $\vQ'^{(j)}=-\vQ^{(j)}$, as shown in \cref{fig:SBSO-SB-SO:a}. When the moir\'e potential is turned off, the three basis states at $K_M$ and $K'_M$ are related by $C_{3z}$ and have the same energy within a fixed spin sector, which therefore form a threefold-degenerate state at $K_M$ and $K'_M$. Turning on the moir\'e potential generically splits each such degenerate state into three nondegenerate eigenstates carrying $\vQ$-lattice angular momentum $\Pi_z\in\{0,\pm1\}$ and $C_{3z}$ eigenvalue $\xi_{K_M/K'_M,s}=e^{-i\frac{2\pi}{3}(\sigma m_J+\Pi_z)}$. 

Now we use the leading-harmonic approximation, in which the moir\'e potential $T_{\vQ,\vQ'}$ is treated as a long-wavelength periodic modulation and depends only on the momentum transfer $\vQ-\vQ'$~\cite{BIS11,ANG21}.  In the \emph{single-orbital-single-branch} limit, the hopping among the three $\vQ$-sites near $K_M$ is described by a single complex number $t\equiv T_{\vQ^{(1)},\vQ^{(2)}}$, while the corresponding hopping near $K'_M$ is $t^*$, \textit{i.e.}, $T_{\vQ^{(i)},\vQ^{(i+1)}}
    =
    T_{\vQ'^{(i+1)},\vQ'^{(i)}}
    =
\big(T_{\vQ'^{(i)},\vQ'^{(i+1)}}\big)^*$ with $i+1$ understood modulo $3$. A detailed proof can be found in \cref{app:subsec:triple model}. Consequently, the Hamiltonians within the same spin sector at $K_M$ and $K'_M$ have the same splitting pattern, but the corresponding states carry opposite $\vQ$-lattice angular momentum. More explicitly, if the isolated band in spin sector $s$ belongs to the $j$-th split state, its $C_{3z}$ eigenvalues at $K_M$ and $K'_M$ are
$\xi^{(j)}_{K_M,s} = e^{-i\frac{2\pi}{3}(\sigma m_J+\Pi_{z,j})}$ and  $\xi^{(j)}_{K'_M,s} = e^{-i\frac{2\pi}{3}(\sigma m_J-\Pi_{z,j})}$.  
Hence 
\begin{equation} \label{eqn:SBSO constraint}
\xi^{(j)}_{K_M,s}\xi^{(j)}_{K'_M,s} = e^{-i\frac{4\pi}{3}\sigma m_J}, 
\end{equation} 
which is independent of the lattice angular momentum $\Pi_{z,j}$. For a spin-resolved band, we obtain $e^{-i\frac{2\pi}{3}C_s} = -\,e^{-i2\pi \sigma m_J}$. Since $m_J$ is always a half-integer for a spinful orbital, one has $e^{-i2\pi \sigma m_J}=-1$, and therefore $e^{-i\frac{2\pi}{3}C_s}=1$, \textit{i.e.}, $C_s=0 \mod 3$. Thus, in the \emph{single-orbital-single-branch} limit, any isolated moir\'e Kramers pair has a trivial $C_3$ symmetry indicator, as shown in \cref{fig:SBSO-SB-SO:b}

The trivial isolated band derived above relies on the \emph{single-orbital-single-branch} limit. Once an additional orbital and/or branch is included, the moir\'e hopping is no longer a scalar but a matrix in the orbital/branch subspace,
    $T_{\vQ_1 m_1,\vQ_2 m_2}$,
where $m=(\lambda,\alpha)$ is a composite index, and \(\lambda\) and \(\alpha\) denote the branch and orbital indices, respectively. In the leading-harmonic approximation, for fixed spin sector \(s\),
    $T_{\vQ^{(i)}m_1,\vQ^{(i+1)}m_2}
    =
    T_{\vQ'^{(i+1)}m_1,\vQ'^{(i)}m_2}
    =
    \big(T_{\vQ'^{(i)}m_2,\vQ'^{(i+1)}m_1}\big)^*$.
Thus the relation between the \(K_M\) and \(K'_M\) triples is now a conjugate transpose in the orbital/branch subspace, rather than a scalar complex conjugation. The constraint \cref{eqn:SBSO constraint} is therefore lost once more than one orbital or more than one interlayer-hybridized branch participates in the low-energy sector. The full derivation is given in \cref{app:subsec:triple model}.

We now consider the two moir\'e layer groups associated with the two stackings, both of which possess an additional in-plane \(C_2\) symmetry besides \(C_{3z}\). For the layer group \(p321\) corresponding to twisted AB stacking, it has \(C_{2y}\) with \(x\) and \(y\) axes defined in \cref{fig:mono-topo:a}. As a result, the bands along this line are spin degenerate. Moreover, in the \emph{single-orbital-single-branch} limit, the exact \(C_{2y}\) symmetry together with the constraint in \cref{eqn:SBSO constraint} forces the topmost six spinful states at \(K_M\) to decompose into one spin-degenerate doublet with
$\xi_{K_M,s}=e^{-i\frac{2\pi}{3}\sigma m_J}$,
and one fourfold Dirac-like multiplet with
$\xi_{K_M,s}
=
e^{-i\frac{2\pi}{3}(\sigma m_J-1)}
\oplus
e^{-i\frac{2\pi}{3}(\sigma m_J+1)}$,
as shown in the left panel of \cref{fig:SBSO-SB-SO:b}. A detailed analysis is given in \cref{app:sec:p3211' moire system}. When the moir\'e potential is maximized on the honeycomb sites, this fourfold Dirac-like multiplet connects to the topmost Kramers pair at \(\Gamma_M\), as shown in the left panel of \cref{fig:SBSO-SB-SO:c}. Because the constraint in \cref{eqn:SBSO constraint} holds only in the \emph{single-orbital-single-branch} limit, once a second orbital or an additional interlayer-hybridized branch is included, the constraint in \cref{eqn:SBSO constraint} is lost, and the fourfold multiplet at \(K_M\) and \(K'_M\) splits into two spin-degenerate doublets. Under the exact \(C_{2y}\) symmetry, the upper isolated doublet obtained from this splitting necessarily carries a nonzero spin Chern number, and therefore realizes a stable topological band, as shown in the right panel of \cref{fig:SBSO-SB-SO:c}.

For layer group \(p312\) corresponding to twisted AA-stacking with additional $C_{2x}$ symmetry, the spin splitting along $\Gamma_M-K_M$ exists in the general case unless fine-tuned. Similar to the case with only $C_{3z}$, isolated moir\'e band in \emph{single-orbital-single-branch} limit is also trivial constrained by \cref{eqn:SBSO constraint}. We need to include multiple orbitals or multiple layer-hybridized branches to break the constraint and obtain isolated band with stable topology. We illustrate the route to realize the stable topology in the \emph{single-orbital-two-branch} case, while the \emph{two-orbital-single-branch} case is analogous and given in \cref{app:sec:p3121' moire system}.
Assuming that the orbital carries $J_z=\pm m_z$ and the topmost Kramers-pair band is isolated from lower bands, the corresponding \(C_{3z}\) eigenvalues at $K_M$ in spin sector \(s\) have $\xi_{K_M,s}
    =
    e^{-i\frac{2\pi}{3}(\sigma m_J+\Pi_{z,s}^{\text{top}})}$, and 
    $\xi_{\Gamma_M,s}
    =
    e^{-i\frac{2\pi}{3}\sigma m_J}$.
Because \(C_{2x}\) and \(\mathcal T\) relate \(K_M\) and \(K'_M\) while flipping spin, the isolated states satisfy $\xi_{\Gamma_M,\bar s}\xi_{\Gamma_M,s}=1$, and $\xi_{K'_M,s}\xi_{K_M,\bar s}=1$.
Using \((\xi_{\Gamma_M,s})^3=-1\) for a spinful \(C_{3z}\) eigenvalue, we have $-\,\xi_{\Gamma_M,s}
    =
    \frac{\xi_{\Gamma_M,\bar s}}{\xi_{\Gamma_M,s}}$.
Substituting this relation into the indicator formula gives
\begin{equation}\label{eqn:p312-maintext-ratio}
    e^{-i\frac{2\pi}{3}C_s}
    =
    \frac{\xi_{K_M,s}\,\xi_{\Gamma_M,\bar s}}
    {\xi_{K_M,\bar s}\,\xi_{\Gamma_M,s}}.
\end{equation}
For the spin-up sector, this becomes
\begin{equation}
\begin{aligned}
    e^{-i\frac{2\pi}{3}C_\uparrow}
    &=
    \frac{
    e^{-i\frac{2\pi}{3}(m_J+\Pi_{z,\uparrow}^{\text{top}})}
    e^{+i\frac{2\pi}{3}m_J}
    }{
    e^{-i\frac{2\pi}{3}(-m_J+\Pi_{z,\downarrow}^{\text{top}})}
    e^{-i\frac{2\pi}{3}m_J}
    }\\
    &=
    e^{-i\frac{2\pi}{3}(\Pi_{z,\uparrow}^{\text{top}}-\Pi_{z,\downarrow}^{\text{top}})}.
\end{aligned}
\end{equation}
Therefore, we arrive at $C_\uparrow
    \equiv
    (\Pi_{z,\uparrow}^{\text{top}}-\Pi_{z,\downarrow}^{\text{top}})
    \mod 3$.
Hence the isolated band pair is topological precisely when the highest \(K_M\) block in the two spin sectors carries different \(\Pi_z\). A topological transition occurs when, upon varying model parameters, the highest \(K_M\) block in one spin sector exchanges with another block carrying a different \(\Pi_z\), as shown in \cref{fig:SBSO-SB-SO:d}. 
In the first-harmonic continuum models, the energies of these blocks are controlled by the competition between splitting resulted from interlayer hybridization and moir\'e potential. Consequently, pressure or changes of twist angle might modify the relative strength between moir\'e potential and interlayer coupling. Such tuning can therefore drive the required band inversion and change \(C_s\) by \(\pm1\). 

The topological bands found in twisted bilayer \ch{SnSe2} and \ch{ZrS2} can be understood within this framework. In the cases where the interlayer distance is minimized on the honeycomb sites, the \emph{single-orbital-single-branch} limit produces a Dirac-like manifold at the top of the valence spectrum. Inter-orbital or inter-branch couplings then gap and split this connected manifold, producing the isolated topological band pairs observed in the continuum model and DFT calculations.
Because twisted AA-stacked \ch{ZrS2} at \SI{3.89}{\degree} is approximately related to the twisted AB-stacked \ch{ZrS2} by $T_{\boldsymbol \tau_0}$, the moir\'e potential maxima also form a honeycomb lattice, except that they occupy the $1a$ Wyckoff position and one of the honeycomb sites, as shown by the LDOS distribution in \cref{fig:model-dft:g}. Consequently, the topological bands are again engineered by the splitting induced by inter-orbital coupling.

\section{Hartree-Fock and Exact-Diagonalization Results}\label{sec:interactions}

\begin{figure*}[!t]
    \centering
    \includegraphics[width=\textwidth]{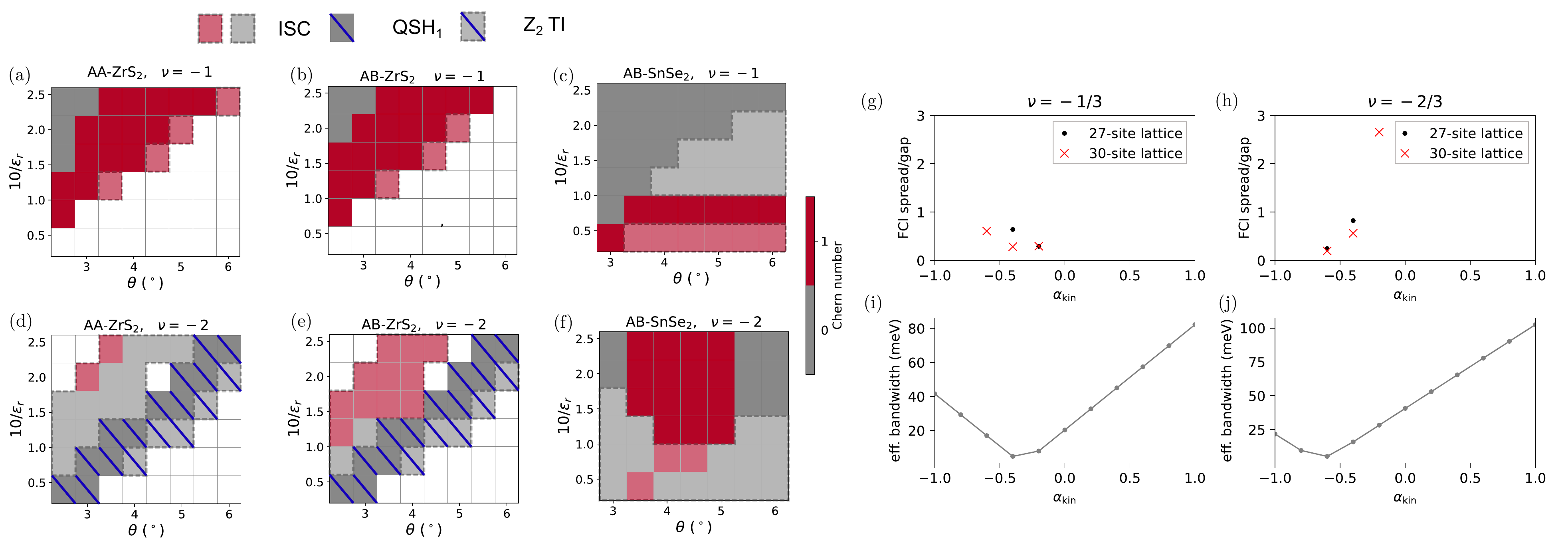}
     \subfloat{\label{fig:HF-phase-diagram:a}}
     \subfloat{\label{fig:HF-phase-diagram:b}}
     \subfloat{\label{fig:HF-phase-diagram:c}}
     \subfloat{\label{fig:HF-phase-diagram:d}}
     \subfloat{\label{fig:HF-phase-diagram:e}}
     \subfloat{\label{fig:HF-phase-diagram:f}}
     \subfloat{\label{fig:HF-phase-diagram:g}}
     \subfloat{\label{fig:HF-phase-diagram:h}}
     \subfloat{\label{fig:HF-phase-diagram:i}}
     \subfloat{\label{fig:HF-phase-diagram:j}}
    \caption{Hartree–Fock (HF) phase diagrams at integer fillings and exact-diagonalization (ED) results at fractional fillings. 
    Panels (a–f) show HF phase diagrams. White regions denote gapless phases. Red and grey regions denote gapped phases with Chern number $C=1$ and $C=0$, respectively. Light shaded regions indicate phases with inter-spin coherence (ISC). In regions with finite ISC marked by dashed outlines, ``$Z_2$ TI'' denotes a topological insulator with broken spin-\Uone symmetry, in contrast to the QSH$_1$ phase which preserves spin-\Uone. Panels (a, d) correspond to AA-stacked \ch{ZrS2}, (b, e) to AB-stacked \ch{ZrS2}, and (c, f) to AB-stacked \ch{SnSe2}. The top row (a–c) shows $\nu=-1$, and the bottom row (d–f) shows $\nu=-2$. All HF calculations use a $12\times 12$ moir\'e Brillouin-zone mesh, screening length $d_{\mathrm{sc}}=\SI{20}{nm}$, and projection onto the top four valence moir\'e bands per spin. Panels (g–j) show calculations at fractional fillings for AA-stacked \ch{ZrS2} at $\theta=3.89^\circ$ with $d_{\mathrm{sc}}=$\SI{20}{nm} and $\epsilon_r=5$. Panels (g,h) display the FCI spread/gap ratio from ED as a function of the kinetic-energy scaling factor $\alpha_{\mathrm{kin}}=-1.0,-0.8,\ldots,0.8,1.0$ for (g) $\nu=-1/3$ and (h) $\nu=-2/3$. Values of $\alpha_{\mathrm{kin}}$ without markers indicate the absence of clear FCI signatures. The ED is restricted to the highest noninteracting valence band within a single spin sector. (i,j) show the effective HF bandwidth [see \cref{eq:E_eff}] for (i) $\nu=-1/3$ and (j) $\nu=-2/3$.
    }
    \label{fig:HF-phase-diagram}
\end{figure*}

We now analyze the interacting phases of AA- and AB-stacked twisted bilayer ZrS$_2$ and AB-stacked twisted bilayer SnSe$_2$ using self-consistent Hartree-Fock and ED.
Experimentally, accessing these hole fillings in the valence moir\'e bands requires carrier densities of order \SIrange{e13}{e14}{\per\centi\meter\squared}, which is within reach of ionic gating via the electric double layer it forms at the surface (see \cref{app:sec:ionic-gating}).
We start from the spin-\Uone symmetric continuum model and include density-density interactions described by a dual-gate screened Coulomb potential
$V(\mathbf q)=\tfrac{e^2}{2\epsilon_0\epsilon_r|\mathbf q|}\tanh(|\mathbf q|d_{\mathrm{sc}})$, 
with dielectric constant $\epsilon_r$ setting the effective interaction strength. In HF calculations, we project to the top four moir\'e bands per spin, and freeze all lower remote bands to be filled. We enforce moir\'e translation symmetry with the possible exception of inter-spin coherence (ISC) at wavevector $\bm{q}$ corresponding to spontaneous breaking of \Uone symmetry. We constrain $\bm{q}$ to the $C_{3z}$-symmetric momenta $\Gamma_M,K_M,K'_M$. More details of the implementation are given in \cref{app:sec:HF and ED}.

\cref{fig:HF-phase-diagram} summarizes HF phase diagrams as a function of interaction strength $10/\epsilon_r$ and twist angle $\theta$ at integer fillings $\nu$. 
For both stackings of twisted bilayer \ch{ZrS2}, a spin-polarized Chern insulator with $C=1$ at $\nu=-1$ appears only beyond a critical interaction strength as shown in \cref{fig:HF-phase-diagram:a,fig:HF-phase-diagram:b}. For weaker interactions, the system remains metallic. The critical interaction strength grows with $\theta$, consistent with the increasing noninteracting bandwidth of the top valence band. In AB-stacked SnSe$_2$, the first moir\'e valence band is narrower than in twisted ZrS$_2$, and therefore the insulating $C=1$ phase at $\nu=-1$ emerges at interaction strengths as weak as $10/\epsilon_r=0.4$, as shown in \cref{fig:HF-phase-diagram:c}. Furthermore, at stronger interaction strengths, both \ch{SnSe2} and \ch{ZrS2} undergo a transition from the $C=1$ insulator to a $C=0$ insulator at $\nu=-1$.
Due to the indirect energy overlap between the first and second single-particle moir\'e valence band within each spin sector, a finite interaction strength is required to open a gap in both AA- and AB-stacked ZrS$_2$ at $\nu=-2$, as shown in \cref{fig:HF-phase-diagram:d,fig:HF-phase-diagram:e}. Nonetheless, the critical interaction strength for obtaining insulating phases is smaller than at $\nu=-1$ for the same twist angle. For example, in AB-stacked \ch{ZrS2} at \SI{5}{\degree}, a gap opens only for $10/\varepsilon_r>2.0$ at $\nu=-1$, whereas it already opens for $10/\varepsilon_r>1.2$ at $\nu=-2$. Close to the metal-insulator boundary in both AA- and AB-stacked ZrS$_2$, the gapped phase is a time-reversal-symmetric topological phase: either a quantum spin Hall state with spin Chern number $C=1$ (QSH$_1$) or a $Z_2$ topological insulator with weak ISC. At stronger interactions, the system develops sizable ISC and the resulting insulator has $C=0$ for AA stacking or $C=1$ for AB stacking. In \ch{SnSe2} at $\nu=-2$, we do not observe  a stable QSH$_1$ or $Z_2$ topological insulator phase for the interaction strengths $10/\epsilon_r \ge 0.4$ explored here due to strong ISC and band inversion. Nonetheless, in an intermediate twist angle region $\SI{3.2}{\degree}<\theta<\SI{5.2}{\degree}$, we observe a $C=1$ Chern insulator with sufficient interaction strength as shown in \cref{fig:HF-phase-diagram:f}. We also study the effect of a displacement field $D$ on twisted bilayer \ch{ZrS2} at $\theta=3.89^\circ$ as shown in \cref{app:fig:D_interaction_HF_phase_diagram}. The displacement field tunes a topological transition between $C=1$ and $C=0$ spin-polarized insulators at $\nu=-1$, and it suppresses the QSH$_1$/$Z_2$ topological phase in favor of a time-reversal-symmetric gapless state at $\nu=-2$. More results can be found in \cref{app:sec:HF and ED}.

To assess the potential for fractional Chern insulators (FCIs)~\cite{neupert2011fractional,sheng2011fqhe,regnault2011fractional,sun2011nearly,tang2011highT,yang2012arbitrary}, we complement HF with exact diagonalization (ED) calculations. We focus on AA-stacked ZrS$_2$ at $\theta=3.89^\circ$ and consider $\nu=-1/3$ and $\nu=-2/3$ restricted to the topmost valence band within one spin sector. Similar results hold for AB-\ch{ZrS2} and AB-\ch{SnSe2} with twist angles ranging from \SI{2}{\degree} to \SI{6}{\degree}. In the HF calculation for AA-stacked ZrS$_2$ at $\theta=4^\circ$, we observe that the bandwidth of the unoccupied HF band nearly doubles for the $C=1$ state at $\nu=-1$ [\cref{app:fig:HF_bandstruct_ZrS2_AA_nu-1_theta4.00_epsr6.25}], driven by the Fock self-energy difference between $\Gamma_M$ and $K_M$. To offset the interaction-induced broadening and explore flatter effective bands, we introduce an artificial scaling factor $\alpha_{\mathrm{kin}}$ on the non-interacting dispersion, $E(\mathbf k) \rightarrow \alpha_{\mathrm{kin}} E(\mathbf k)$, with $\alpha_{\mathrm{kin}}=1$ corresponding to the physical band, $\alpha_{\mathrm{kin}}=0$ to a perfectly flat band, and $\alpha_{\mathrm{kin}}<0$ to an inverted dispersion. We diagnose FCIs by identifying the putative topological ground-state manifold, which consists of the three lowest states with momenta consistent with the FCI momenta~\cite{regnault2011fractional,bernevig2012emergent}. The spread of the lowest states should be smaller than the gap between the three lowest states and the lowest state not in this manifold. As shown in \cref{fig:HF-phase-diagram:g,fig:HF-phase-diagram:h}, robust FCI signatures at $\nu=-1/3$ and $\nu=-2/3$ appear only within a window of negative $\alpha_{\mathrm{kin}}$. To rationalize this, we estimate an ``effective'' band dispersion,
\begin{equation}\label{eq:E_eff}
    E_{\mathrm{eff}}(\mathbf k)
    = \alpha_{\mathrm{kin}} E(\mathbf k)
    + |\nu|\,E^{\mathrm{HF,int}}(\mathbf k),
\end{equation}
where $E^{\mathrm{HF,int}}(\mathbf k)$ is the interaction-induced HF contribution for a fully hole-occupied valence band [see \cref{app:eqn:HF potential of one-body density}], rescaled by $|\nu|$ to mimic fractional fillings. The effective bandwidth extracted from $E_{\mathrm{eff}}(\mathbf k)$ exhibits a minimum as a function of $\alpha_{\mathrm{kin}}$ that closely coincides with the FCI region identified by ED as shown in \cref{fig:HF-phase-diagram:i,fig:HF-phase-diagram:j}. This indicates that, for the physical dispersion ($\alpha_{\mathrm{kin}}=1$), interaction-induced band broadening drives the system away from the optimal flat-band limit, and FCIs are only stabilized when the bare dispersion is inverted to partially compensate the broadening. The appearance of a $C=1$ state at $\nu=-1$ is consistent with Ref.~\cite{CLA22a}. However, we do not find an FCI at $\nu=-1/3$ without artificially inverting the band dispersion. This difference is likely related to the ordering of the second and third sets of single-particle bands, which depends on the choice of van der Waals correction. A detailed comparison is provided in \cref{app:sec:choice of vdW}.

\section{Discussion}\label{sec:conclusion}
By combining large-scale DFT calculations with faithful continuum models, we establish twisted bilayer \ch{ZrS2} and \ch{SnSe2} as concrete $\Gamma$-valley platforms that host isolated topological moir\'e bands at small twist angles. Our analysis identifies the key ingredients for such bands in spin-\Uone-symmetric $\Gamma$-valley moir\'e systems. First, the relevant monolayer state near the band edge should transform as a two-dimensional IRREP without SOC, corresponding to effective orbitals with $m_l\neq 0$. This reduces the approximate spin symmetry of the low-energy moir\'e bands from spin-$\mathrm{SU}(2)$ to spin-\Uone. Second, the moir\'e problem must go beyond the \emph{single-orbital-single-branch} limit where isolated topological bands are symmetry-constrained to be trivial. Inter-orbital or inter-branch couplings generate isolated topological bands either by splitting Dirac-like bands in layer group $p321$ or by driving band inversions in layer group $p312$, as confirmed by the simplified models for AB-stacked \ch{SnSe2} and AA/AB-stacked \ch{ZrS2}.
Analysis of the simplified models for these materials validates this framework. Moreover, the competition between bandwidth and interaction yields a rich phase diagram. In both \ch{ZrS2} and \ch{SnSe2} systems, Hartree–Fock calculations find spin-polarized Chern insulators at $\nu=-1$, and symmetry-preserving QSH$_1$/$Z_2$ phases at $\nu=-2$, with the ability to tune Chern numbers and close gaps via a displacement field. Exact diagonalization suggests that interaction-induced Fock broadening currently disfavors FCIs. As a route to stabilize FCIs, future work may explore enhancing dielectric screening asymmetrically or leveraging substrate-induced moir\'e potentials to quench the kinetic energy further, thereby realizing the full potential of these $\Gamma$-valley platforms.

% \begin{acknowledgments}
\section*{Acknowledgments}
The authors are grateful to Dante M. Kennes, Lede Xian, and Jiabin Yu for insightful discussions.
\paragraph*{Funding:} 
We thank the technical support provided by Donostia International Physics Center Supercomputing Center. 
The simulations presented in this article were partially performed on computational resources managed and supported by Princeton Research Computing, a consortium of groups including the Princeton Institute for Computational Science and Engineering (PICSciE) and the Office of Information Technology's High Performance Computing Center and Visualization Laboratory at Princeton University. 
M.G.V and H.P. were supported by the Ministry for Digital Transformation and of Civil Service of the Spanish Government through the QUANTUM ENIA project call - Quantum Spain project, and by the European Union through the Recovery, Transformation and Resilience Plan - NextGenerationEU within the framework of the Digital Spain 2026 Agenda. 
M.G.V. thanks support to the Deutsche Forschungsgemeinschaft (DFG, German Research Foundation) GA 3314/1-1 – FOR 5249 (QUAST), the Spanish Ministerio de Ciencia e Innovacion (PID2022-142008NB-I00) and the Canada Excellence Research Chairs Program for Topological Quantum Matter.
H.P. and Y.J. were supported by the European Research Council (ERC) under the European Union’s Horizon 2020 research and innovation program (Grant Agreement No. 101020833), as well as by the IKUR Strategy under the collaboration agreement between Ikerbasque Foundation and DIPC on behalf of the Department of Education of the Basque Government. 
B.A.B. was supported by the Gordon and Betty Moore Foundation through Grant No. GBMF8685 towards the Princeton theory program, the Gordon and Betty Moore Foundation’s EPiQS Initiative (Grant No. GBMF11070), the Global Collaborative Network Grant at Princeton University, the Simons Investigator Grant No. 404513, the NSF-MERSEC (Grant No. MERSEC DMR 2011750), the Simons Collaboration on New Frontiers in Superconductivity (Grant No. SFI-MPS-NFS-00006741-01), Princeton Catalysis Initiative (PCI), the Schmidt Foundation at the Princeton University and the National Science Foundation through the AI Research Institutes program Award No. DMR-2433348. 

% \paragraph*{Author contributions:} 
% % B.A.B. conceived of the study. Y.J., U.P., H.P., D.C., and H.H developed the code and performed the high-throughput calculations. 
% % J.X., R.A.M., P.H., V.H., E.M., C.F., and L.M.S. synthesized the samples.
% % A.O., S.S., J.Z., D.Y., Z.G., J.S., K.F.M., and D.K.E. exfoliated and characterized the monolayer samples. 
% % N.R. built the Topological 2D materials Database with input data from U.P. and Y.J.. 
% % Q.X., H.P., M.C., D.M.K., A.R.,. and L.X. performed calculations and provided DFT results for moderate twist angles.
% % Y.J., G.S., H.P., and B.A.B. wrote the original draft and supplementary materials. 
% % All authors contributed to the review and editing of the final draft. 

% \paragraph*{Competing interests:}
% The authors declare that they have no competing interests.

% \paragraph*{Data and materials availability:} 
% % All data are available in the supplementary materials, through our public website \webtwoDTQC. 
% Additional data, along with any code required for reproducing the figures, are available from the authors upon reasonable request.

% \end{acknowledgments}

\renewcommand{\addcontentsline}[3]{}
%\bibliographystyle{apsrev4-2}
%\bibliography{Twistronics,tbg}

\let\addcontentsline\oldaddcontentsline

\renewcommand{\thetable}{S\arabic{table}}
\renewcommand{\thefigure}{S\arabic{figure}}
\renewcommand{\theequation}{S\arabic{section}.\arabic{equation}}

\onecolumngrid
\pagebreak
\thispagestyle{empty}

\clearpage
\begin{center}
	\textbf{\large Supplementary Information for ''\titlePaper{}``}\\[.2cm]
\end{center}

\appendix
\renewcommand{\thesection}{\Roman{section}}
\tableofcontents
\let\oldaddcontentsline\addcontentsline

\newpage
\section{\textit{Ab initio} calculations}
\subsection{Untwist monolayer/bilayer results}\label{app:sec:untwist result}
\begin{figure}[htbp]
    \centering
    \includegraphics[width=\textwidth]{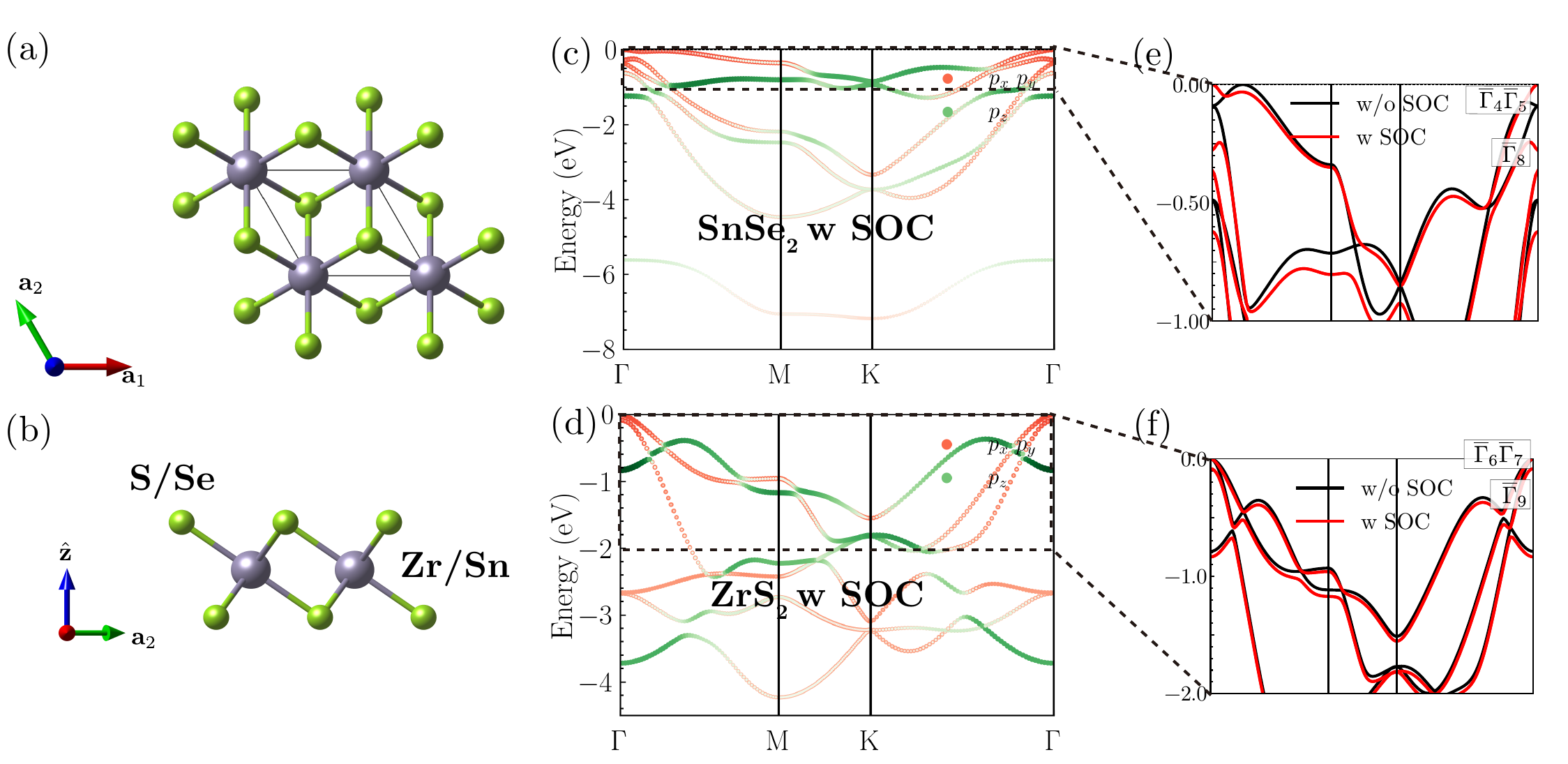}%
    \subfloat{\label{app:fig:monolayer-band:a}}%
    \subfloat{\label{app:fig:monolayer-band:b}}%
    \subfloat{\label{app:fig:monolayer-band:c}}%
    \subfloat{\label{app:fig:monolayer-band:d}}%
    \subfloat{\label{app:fig:monolayer-band:e}}%
    \subfloat{\label{app:fig:monolayer-band:f}}%
    \caption{The monolayer crystal structure of \ch{SnSe2} and \ch{ZrS2} and corresponding band structure. (a) and (b) show monolayer crystal structure of \ch{SnSe2} and \ch{ZrS2} from the top view and side view, respectively. The monolayer band structure with SOC of (c) \ch{SnSe2} and (d) \ch{ZrS2} with $p$-orbital projection of Se and S atoms, respectively. The red and green colors indicate the orbital weight of $p_{x,y}$ and $p_z$, respectively. The enlarge monolayer band structure of (e) \ch{SnSe2} and (f) \ch{ZrS2} without (black lines) and with (red lines) spin-orbit coupling. We label the double irreducible representation (IRREP) of the two topmost bands at the $\Gamma$ point. 
    }
    \label{app:fig:monolayer-bands}
\end{figure}

In this section, we present \textit{ab initio} results for the monolayer and untwisted bilayer structure of \ch{SnSe2} and \ch{ZrS2}. As shown in \cref{app:fig:monolayer-band:a,app:fig:monolayer-band:b}, the monolayer crystal structures of both materials have Shubnikov Space Group (SSG) 164.86 $P\bar{3}m11'$ symmetry, characterized by symmetry generators $C_{3z}$, $C_{2}$ with rotated axes along $[100]$ direction, inversion symmetry $\mathcal{I}$ and time-reversal symmetry $\mathcal{T}$. The hexagonal lattice vectors are given by
\begin{equation}
\vec{a}_1=a \left(1, 0 \right), \quad 
\vec{a}_2=a \left( -\frac{1}{2}, \frac{\sqrt{3}}{2} \right),
\label{app:eqn:hexagonal_cell_basis}
\end{equation}
where $a$ is the lattice constant, and the atoms are located at
\begin{equation}
	\label{app:eqn:atom_positions}
	\mathrm{Sn/Zr}:\ \left(0,0,0\right),\quad \mathrm{Se/S}:\ \left(\frac{1}{3}, \frac{2}{3}, z\right),\,\left(\frac{2}{3}, \frac{1}{3}, -z\right).  
\end{equation}
For \ch{SnSe2} and \ch{ZrS2}, the lattice constant and out-of-plane displacement of Se or S atoms are $a=\SI{3.811}{\angstrom}$, $z=\SI{1.528}{\angstrom}$~\cite{ WU19c} and $a=\SI{3.661}{\angstrom}$,  $z=\SI{1.441}{\angstrom}$~\cite{AL-77}, respectively. The Sn or Zr atom is located at the $1a$ Wyckoff position featuring $\bar{3}m$ site symmetry, while the Se or S atoms are positioned at $2d$ Wyckoff position and have $3m$ symmetry. Within a unit cell, the two $2d$ Wyckoff positions are mapped to one another by inversion.

We focus on the valence bands since the conduction bands have been fully studied in~\cite{cualuguaru2024new}. The valence band structures of monolayer \ch{SnSe2} and \ch{ZrS2} with SOC is shown in \cref{app:fig:monolayer-band:c,app:fig:monolayer-band:d}. Due to the presence of $\mathcal{PT}$ symmetry, every band in both \ch{ZrS2} and \ch{SnSe2} is doubly degenerate across the whole Brillouin Zone (BZ). In both materials, the $\Gamma$ valley is dominated by the $(p_{x},p_y)$ orbitals of chalcogenide atoms. A comparison of the low-energy bands without and with SOC is shown in \cref{app:fig:monolayer-band:e,app:fig:monolayer-band:f}. Without SOC included, the valence band maxima (VBM) of \ch{ZrS2} are located at $\Gamma$ while the VBM of \ch{SnSe2} is located on $\Gamma-M$ path that is near $\Gamma$. As shown by the red lines, SOC lifts the twofold degeneracy (without spin degree of freedom) of the topmost state at $\Gamma$ point in both \ch{SnSe2} and \ch{ZrS2}.  In \ch{SnSe2}, SOC further shifts the VBM from a high symmetry line to the $\Gamma$ point. Because Se atom is heavier than the S atom, we can find the SOC splitting, defined as the separation between the two spinful Kramers doublets obtained from the spinless twofold-degenerate state (or fourfold with spin) at $\Gamma$, in \ch{SnSe2}(\SI{278}{meV}) is much stronger than that in \ch{ZrS2}(\SI{83}{meV}). The first and second topmost double-degenerated bands at $\Gamma$ in \ch{ZrS2} transform as the $\overline{\Gamma}_6\overline{\Gamma}_7$ and $\overline{\Gamma}_9$ double irreducible representations (IRREPs) in the spinful case, while the first and second topmost double-degenerated bands at $\Gamma$ in \ch{SnSe2} transform as the $\overline{\Gamma}_4\overline{\Gamma}_5$ and $\overline{\Gamma}_8$ double IRREPs. Although both monolayers share the same lattice symmetry and have $\Gamma$-valley valence states dominated by chalcogen $(p_x,p_y)$ orbitals, the corresponding $\Gamma$-point IRREP is not fixed by orbital character alone. Because the two chalcogen atoms in one monolayer unit cell are related by inversion, the $(p_x,p_y)$ manifold forms even- and odd-parity layer combinations, and the ordering of these two doublets depends on their hybridization with nearby metal-derived states. Here the two compounds differ qualitatively: in \ch{ZrS2} the low-energy conduction bands is mainly Zr-$4d$, while in \ch{SnSe2} it is mainly Sn/Se-$(s,p)$ derived. This difference changes how the even- and odd-parity chalcogen $(p_x,p_y)$ combinations are renormalized, and hence leads to different parity ordering at $\Gamma$.

\begin{figure}[!t]
    \centering
    \includegraphics[width=\textwidth]{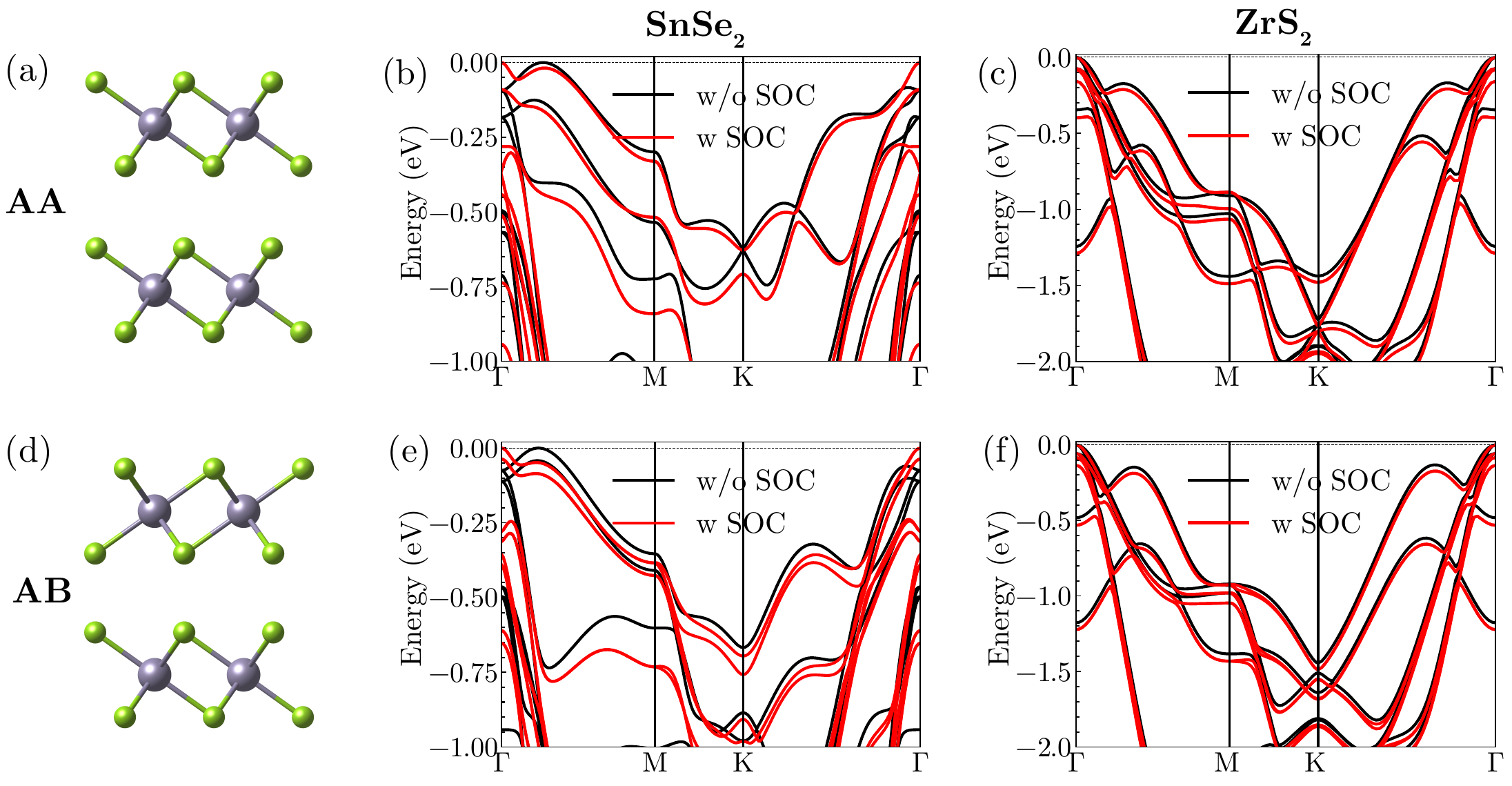}%
    \subfloat{\label{app:fig:untwist-bilayer:a}}%
    \subfloat{\label{app:fig:untwist-bilayer:b}}%
    \subfloat{\label{app:fig:untwist-bilayer:c}}%
    \subfloat{\label{app:fig:untwist-bilayer:d}}%
    \subfloat{\label{app:fig:untwist-bilayer:e}}%
    \subfloat{\label{app:fig:untwist-bilayer:f}}%
    \caption{The crystal and band structure for untwisted AA/AB-stacking bilayer SnSe$_2$ and ZrS$_2$. (a) and (d) show the AA- and AB-stacked bilayer crystal structures, respectively. (b) and (e) show the band structure for AA- and AB-stacked untwisted bilayer \ch{SnSe2}, respectively. (c) and (f) show the band structure for AA- and AB-stacked untwisted bilayer \ch{ZrS2}, respectively. The black and red colors denote the band structures without and with spin-orbital coupling included.
    }
    \label{app:fig:untwist-bilayer}
\end{figure}
By stacking one monolayer directly on top of another, we obtain the AA-stacked bilayer as shown in \cref{app:fig:untwist-bilayer:a}. The symmetry group of the AA-stacked bilayer is the same as the monolayer case, \ie, SSG 164.86 $P\bar{3}m11'$. The band structure for AA-stacked \ch{SnSe2} and \ch{ZrS2} are shown in \cref{app:fig:untwist-bilayer:b,app:fig:untwist-bilayer:c}. 
Because $\mathcal{PT}$ symmetry is present, all bands are doubly degenerate throughout the Brillouin zone. 
Comparing the monolayer band structures without SOC in \cref{app:fig:monolayer-band:e,app:fig:monolayer-band:f} with the bilayer band structures without SOC in \cref{app:fig:untwist-bilayer:b,app:fig:untwist-bilayer:c}, we find that interlayer coupling splits the topmost monolayer $\Gamma$ state into the top two interlayer bonding and antibonding states at bilayer $\Gamma$. The corresponding interlayer splitting, defined as the bonding--antibonding splitting of the topmost $\Gamma$ state, is \SI{90}{meV} and \SI{74}{meV} in \ch{SnSe2} and \ch{ZrS2}, respectively. As in the monolayer, including SOC shifts the VBM of \ch{SnSe2} from a high-symmetry line to $\Gamma$ and opens gaps at high-symmetry points/lines. Notably, in \ch{ZrS2} the SOC and interlayer energy scales are comparable, making the ordering of topmost 8 spinful states at $\Gamma$ sensitive to the choice of van der Waals (vdW) corrections (see detailed discussion in \cref{app:sec:choice of vdW}).

Because the monolayer structure lacks $C_{2z}$ symmetry, we can apply $C_{2z}$ to the bottom layer of AA-stacked bilayer and obtain the AB-stacked bilayer as seen in \cref{app:fig:untwist-bilayer:d}. The AB-stacked bilayer has $P\bar{6}m21'$ (SSG 187.210) symmetry which is generated by $C_{3z}$, vertical mirror $M_{100}$, horizontal mirror $M_z$ and $\mathcal{T}$. The head-to-head configuration of AB-stacking results in larger interlayer distances (\SI{7.016}{\angstrom} for \ch{SnSe2} and \SI{6.362}{\angstrom} for \ch{ZrS2}) compared to the AA-stacking configuration (\SI{6.166}{\angstrom} for \ch{SnSe2} and \SI{5.866}{\angstrom} for \ch{ZrS2}). Here, the interlayer distance is defined as the distance between Sn/Zr atomic sub-layers. Consequently, the interlayer splitting in AB-stacking (\SI{36}{meV} for \ch{SnSe2} and \SI{57}{meV} for \ch{ZrS2}) is smaller than that in AA-stacking. Moreover, double degeneracy in AB-stacking is only protected along the $\Gamma-M$ line by $M_{010}M_z$.

\subsection{Twisted AA/AB-stacked bilayer}\label{app:sec:twist bilayer calculation}
\begin{figure}[!t]
    \centering
    \includegraphics[width=0.5\textwidth]{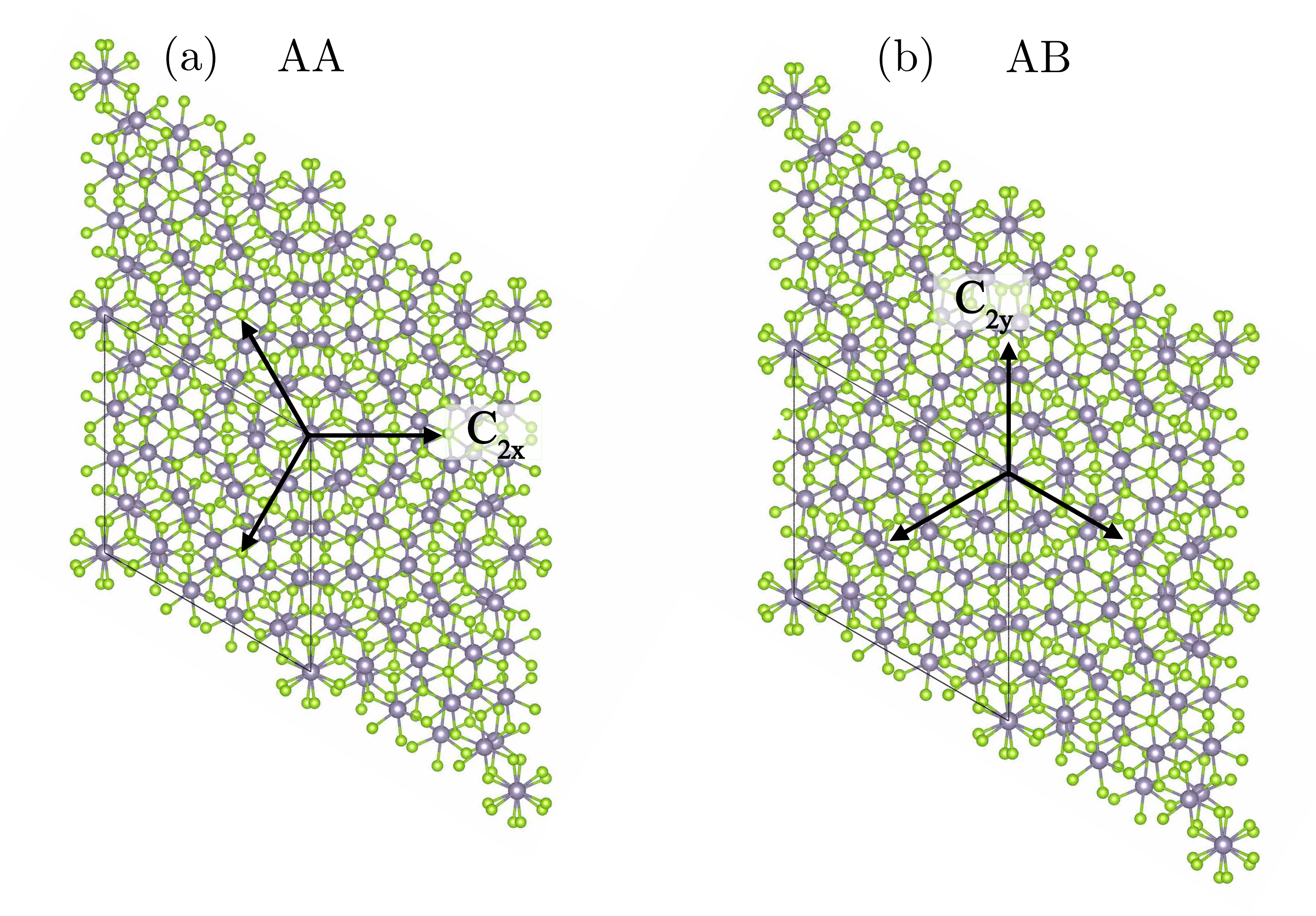}%
    \subfloat{\label{app:fig-twist-crystal:a}}%
    \subfloat{\label{app:fig-twist-crystal:b}}%
    \caption{Crystal structure for twisted (a) AA and (b) AB-stacked \ch{ZrS2} and \ch{SnSe2}. The arrows are the in-plane $C_2$ axis related by $C_{3z}$
    }
    \label{app:fig-twist-crystal}
\end{figure}

\begin{figure}[!t]
    \centering
    \includegraphics[width=\textwidth]{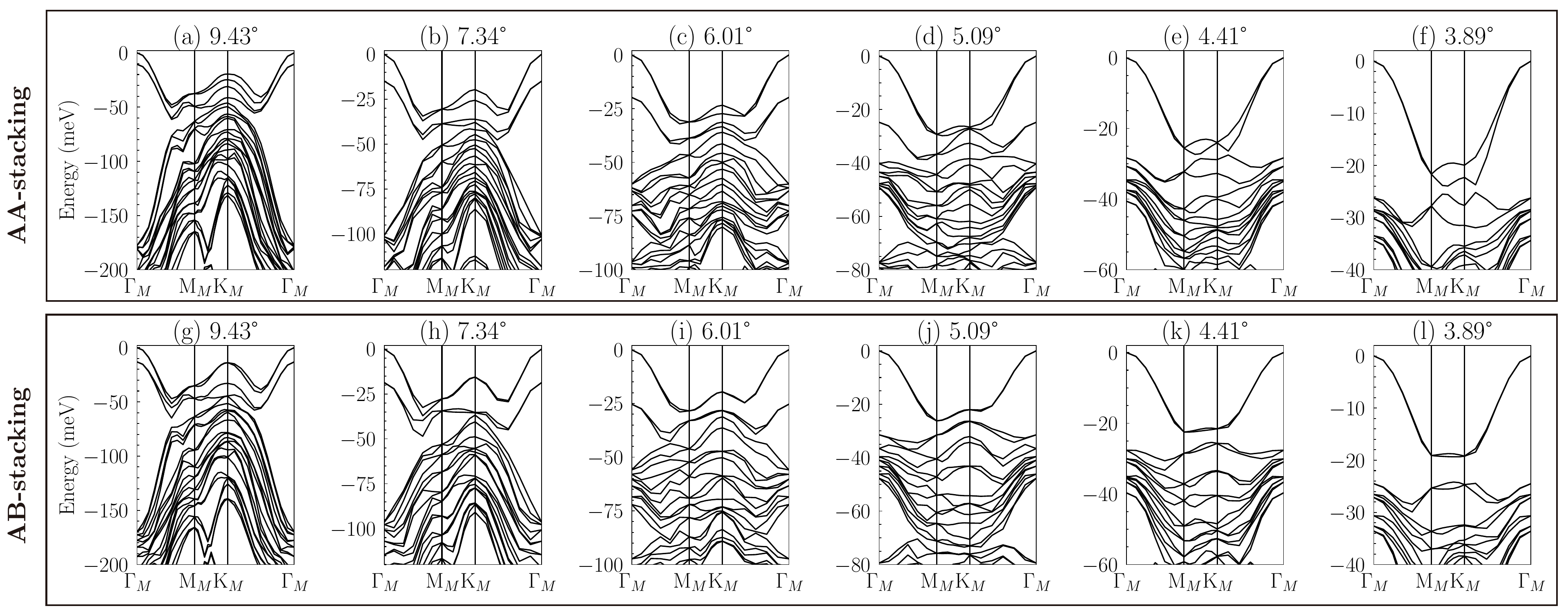}%
    \subfloat{\label{app:fig:SnSe2-moire-band:a}}%
    \subfloat{\label{app:fig:SnSe2-moire-band:b}}%
    \caption{Valence band structure for \textit{ab initio} calculated twisted bilayer \ch{SnSe2} from \SI{9.43}{\degree} to \SI{3.89}{\degree}. (a-f) and (g-l) show the AA-stacked and AB-stacked results, respectively.
    }
    \label{app:fig:SnSe2-moire-band}
\end{figure}

\begin{figure}[!t]
    \centering
    \includegraphics[width=\textwidth]{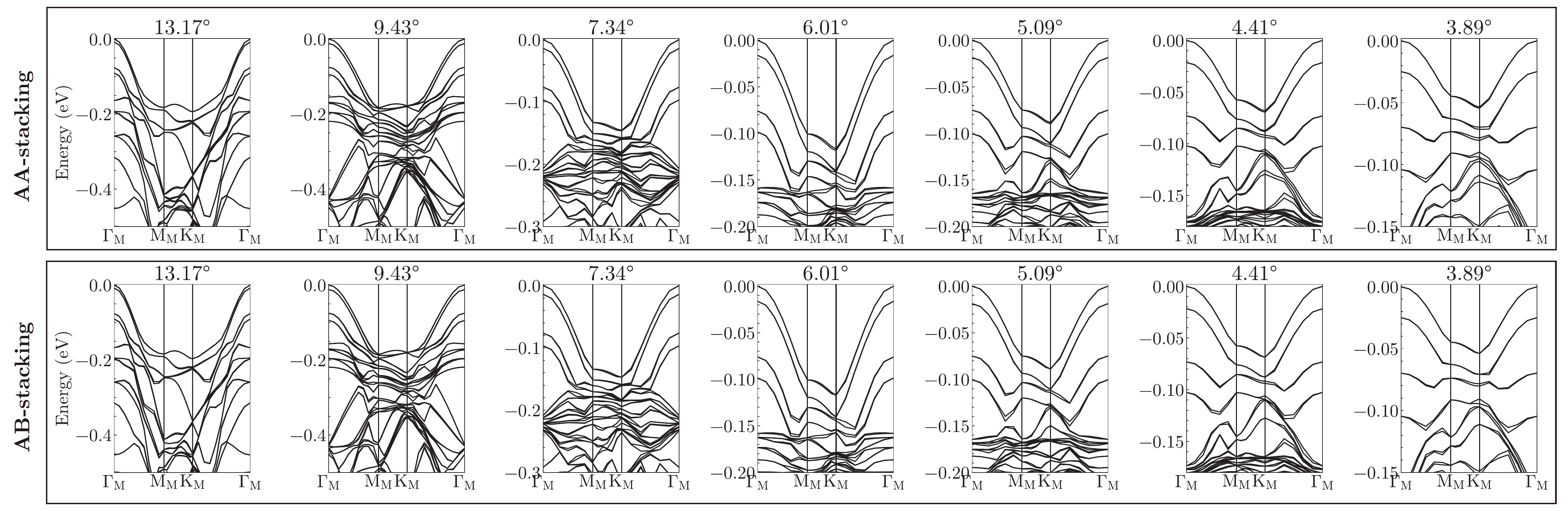}%
    \subfloat{\label{app:fig:ZrS2-moire-band:a}}%
    \subfloat{\label{app:fig:ZrS2-moire-band:b}}%
    \caption{Valence band structure for \textit{ab initio} calculated twisted bilayer \ch{ZrS2} from \SI{13.17}{\degree} to \SI{3.89}{\degree}. (a-f) and (g-l) show the AA-stacked and AB-stacked results, respectively.
    }
    \label{app:fig:ZrS2-moire-band}
\end{figure}

By rotating the top layer relative to the bottom one, we obtain the AA- and AB-stacked twisted bilayer structures at a series of angles. We take the convention that $\hat{y}$ direction aligns with the $[010]$ direction as shown in \cref{app:fig-twist-crystal}, following Ref.~\cite{cualuguaru2024new}. The twisted AA-stacked bilayer \ch{SnSe2} and \ch{ZrS2} break $\mathcal I$ symmetry in the untwist configuration and belong to SSG $P3121’$ (No. 149.22) which is characterized by symmetry generators $C_{3z}$, $C_{2x}$, $\mathcal{T}$, and translational symmetry $T_{\vec{R}}$, as shown in \cref{app:fig-twist-crystal:a}. The twisted AB-stacked bilayer \ch{SnSe2} and \ch{ZrS2} break $\mathcal M_z$ symmetry in the untwist configuration and belong to SSG $P3211’$ (No. 150.26) with symmetry generators $C_{3z}$, $C_{2y}$, $\mathcal{T}$, and translational symmetry $T_{\vec{R}}$, as shown in \cref{app:fig-twist-crystal:b}. We first study the moir\'e band structures using large-scale DFT calculation (the M-valley conduction bands are detailed in Ref.~\cite{cualuguaru2024new}). The valence bands for twisted \ch{SnSe2} (\cref{app:fig:SnSe2-moire-band}) and \ch{ZrS2} (\cref{app:fig:ZrS2-moire-band}) are shown for AA (top rows) and AB (bottom rows) from \SI{9.43}{\degree} to \SI{3.89}{\degree}. At \SI{3.89}{\degree}, the top two spinful bands in AB-\ch{SnSe2} form a single isolated set. We don't discuss the top isolated band in AA-\ch{SnSe2} since the direct gap between the second and third spinful band is too small.  In AA/AB-\ch{ZrS2} at \SI{3.89}{\degree}, the top six spinful bands organize into three isolated sets of two spinful bands each. Moreover, the bands are nearly doubly degenerate along high-symmetry lines. As shown in \cref{app:sec:extracted continuum model}, this degeneracy arises from zero-twist inversion $\mathcal I$ in twisted AB-stacked structure and from the $\tilde{C}_{2z}$ symmetry in twisted AA-\ch{ZrS2}.

Moreover, by inspecting the DFT wavefunction along high-symmetry lines, we find that the isolated bands in both \ch{SnSe2} and \ch{ZrS2} remain nearly eigenstates of $s_z$, which is the Pauli matrix acting in spin space and trivially on all other degrees of freedom. We define spin-\Uone symmetry as $[H(\mathbf{k}),s_z]=0$ and quantify its breaking by 
\begin{equation}\label{app:eqn:definetion of U1 breaking}
\delta_{s}^{(n)}(\mathbf k)=\frac{1-\big|\langle u_{n\mathbf k}|s_z|u_{n\mathbf k}\rangle\big|}{2}.
\end{equation}
Here $|u_{n\mathbf{k}}\rangle$ denotes the eigenstate of the $n$-th band at momentum $\mathbf{k}$, and
We find $\delta_s^{(n)}(\mathbf{k})\le 2\%$ for the isolated bands of both \ch{SnSe2} and \ch{ZrS2}. We will verify this again using the continuum model in \cref{app:sec:extracted continuum model}.

Therefore we could define a spin Chern number $C_s$ of each spin sector for each isolated band set,
\begin{equation}\label{app:eqn:spin chern number formula}
C_s=\ln \left(-\xi_{\Gamma_M} \xi_{K_M}\xi_{K_M^{\prime}}\right) /\left(i \frac{2 \pi}{3}\right) \bmod 3,
\end{equation}
where $\xi_{\vec{k}}$ is the $C_{3z}$ eigenvalue at $\vec{k}\in \{\Gamma_M,K_M, K'_M\}$.  Evaluating these eigenvalues for \ch{SnSe2} and \ch{ZrS2}, we summarize the resulting spin Chern numbers in \cref{app:tab:spin_chern_number_AB_ZrS2,app:tab:spin_chern_number_AA_ZrS2}, where all the isolated bands in \ch{ZrS2} and AB-stacked \ch{SnSe2} carry a nontrivial spin Chern number.

\begin{table}[]
\begin{tabular}{|c|cc|cc|cc|cc|}
\hline
\multirow{2}{*}{\ch{SnSe2}@\SI{3.89}{\degree}} &
  \multicolumn{2}{c|}{$\Gamma_M$} &
  \multicolumn{2}{c|}{$K_M$} &
  \multicolumn{2}{c|}{$K'_M$} &
  \multicolumn{2}{c|}{Spin Chern number} \\ \cline{2-9} 
        & \multicolumn{1}{c|}{spin-$\uparrow$} & spin-$\downarrow$ & \multicolumn{1}{c|}{spin-$\uparrow$}       & spin-$\downarrow$ & \multicolumn{1}{c|}{spin-$\uparrow$} & spin-$\downarrow$       & \multicolumn{1}{c|}{spin-$\uparrow$} & spin-$\downarrow$ \\ \hline
AB & \multicolumn{1}{c|}{-1} & -1   & \multicolumn{1}{c|}{$\omega^*$} & $\omega$  & \multicolumn{1}{c|}{$\omega^*$} & $\omega$ & \multicolumn{1}{c|}{-1 mod 3} & 1 mod 3\\ \hline
\end{tabular}
\label{app:tab:SnSe2-C3eigenvalue}
\caption{Table of $C_3$ eigenvalues for the top set of bands of AB-stacked twisted \ch{SnSe2} at \SI{3.89}{\degree}. 
}
\label{app:tab:spin_chern_number_AB_ZrS2}
\end{table}

\begin{table}[h!]
\begin{tabular}{|c|cc|cc|cc|cc|}
\hline
\multirow{2}{*}{AA-\ch{ZrS2}@\SI{3.89}{\degree}} &
  \multicolumn{2}{c|}{$\Gamma_M$} &
  \multicolumn{2}{c|}{$K_M$} &
  \multicolumn{2}{c|}{$K'_M$} &
  \multicolumn{2}{c|}{Spin Chern number} \\ \cline{2-9} 
        & \multicolumn{1}{c|}{spin-$\uparrow$} & spin-$\downarrow$ & \multicolumn{1}{c|}{spin-$\uparrow$}       & spin-$\downarrow$ & \multicolumn{1}{c|}{spin-$\uparrow$} & spin-$\downarrow$       & \multicolumn{1}{c|}{spin-$\uparrow$} & spin-$\downarrow$ \\ \hline
1st set & \multicolumn{1}{c|}{-1} & -1   & \multicolumn{1}{c|}{$\omega$} & -1   & \multicolumn{1}{c|}{-1} & $\omega^*$ & \multicolumn{1}{c|}{-1 mod 3}& 1 mod 3  \\ \hline
2nd set &
  \multicolumn{1}{c|}{-1} &
  -1 &
  \multicolumn{1}{c|}{-1} &
  $\omega$ &
  \multicolumn{1}{c|}{$\omega^*$} &
 -1 &
  \multicolumn{1}{c|}{-2 mod 3} &
  2 mod 3\\ \hline
3rd set &
  \multicolumn{1}{c|}{$\omega$} &
  $\omega^*$ &
  \multicolumn{1}{c|}{$\omega^*$} &
  $\omega^*$ &
  \multicolumn{1}{c|}{$\omega$} &
  $\omega$ &
  \multicolumn{1}{c|}{2 mod 3} &
  -2 mod 3 \\ \hline
\end{tabular}
\caption{Table of $C_3$ eigenvalues for the topmost three set of bands of AA-stacking twisted \ch{ZrS2} at \SI{3.89}{\degree}.}
\label{app:tab:spin_chern_number_AA_ZrS2}
\end{table}

\begin{table}[]
\begin{tabular}{|c|cc|cc|cc|cc|}
\hline
\multirow{2}{*}{AB-\ch{ZrS2}@\SI{3.89}{\degree}} &
  \multicolumn{2}{c|}{$\Gamma_M$} &
  \multicolumn{2}{c|}{$K_M$} &
  \multicolumn{2}{c|}{$K'_M$} &
  \multicolumn{2}{c|}{Spin Chern number} \\ \cline{2-9} 
        & \multicolumn{1}{c|}{spin-$\uparrow$} & spin-$\downarrow$ & \multicolumn{1}{c|}{spin-$\uparrow$}       & spin-$\downarrow$ & \multicolumn{1}{c|}{spin-$\uparrow$} & spin-$\downarrow$       & \multicolumn{1}{c|}{spin-$\uparrow$} & spin-$\downarrow$ \\ \hline

1st set & \multicolumn{1}{c|}{-1} & -1   & \multicolumn{1}{c|}{$\omega^*$} & $\omega$  & \multicolumn{1}{c|}{$\omega^*$} & $\omega$ & \multicolumn{1}{c|}{-1 mod 3} & 1 mod 3   \\ \hline
2nd set &
  \multicolumn{1}{c|}{-1} &
  -1 &
  \multicolumn{1}{c|}{$\omega$} &
  $\omega^*$ &
  \multicolumn{1}{c|}{$\omega$} &
  $\omega^*$ &
  \multicolumn{1}{c|}{-2 mod 3} &
  2 mod 3\\ \hline
3rd set &
  \multicolumn{1}{c|}{$\omega$} &
  $\omega^*$ &
  \multicolumn{1}{c|}{-1} &
  -1 &
  \multicolumn{1}{c|}{-1} &
  -1 &
  \multicolumn{1}{c|}{2 mod 3} &
  -2 mod 3\\ \hline
\end{tabular}
\caption{Table of $C_3$ eigenvalues for the topmost three set of bands of AB-stacking twisted \ch{ZrS2} at \SI{3.89}{\degree}.}
\label{app:tab:spin_chern_number_AB_ZrS2}
\end{table}

\subsection{Influence of vdW correction}\label{app:sec:choice of vdW}
\begin{figure}[!t]
    \centering
    \includegraphics[width=\textwidth]{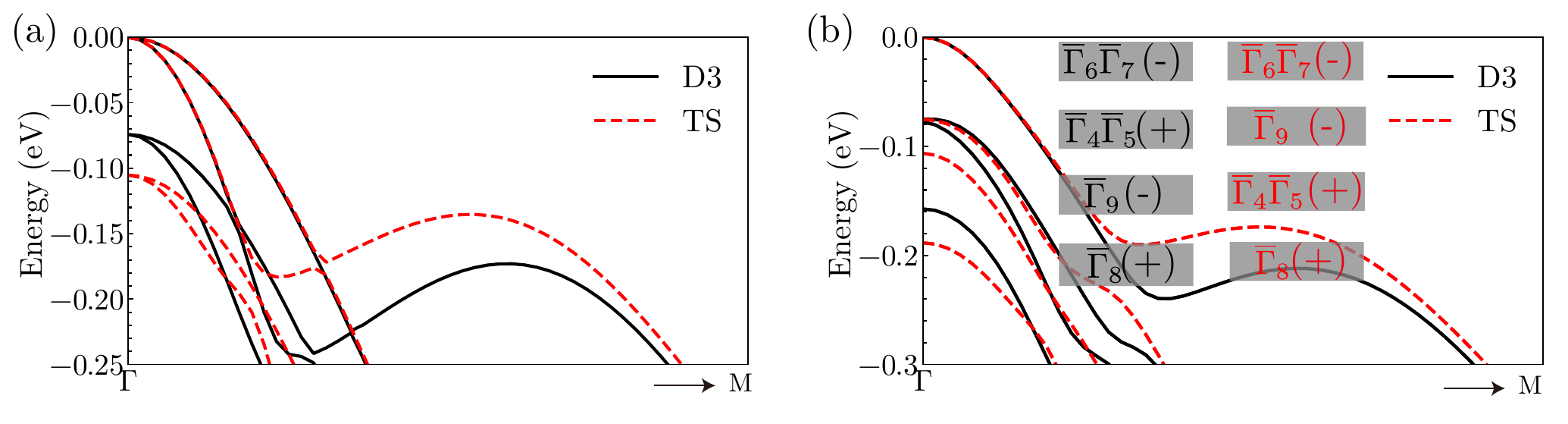}%
    \subfloat{\label{app:fig:vdw-band-compare:a}}%
    \subfloat{\label{app:fig:vdw-band-compare:b}}%
    \caption{Comparison of band structures for untwisted AA-stacked \ch{ZrS2} bilayer with D3-vdW and TS-vdW corrections, (a) without and (b) with SOC. In (b), the irreducible representations (IRREPs) of the top four bands are labeled, with $\pm$ denoting their corresponding parity. We observe that the middle two bands are switched under the two vdW corrections.  
    }
    \label{app:fig:vdw-band-compare}
\end{figure}

In \cref{app:sec:untwist result}, we mentioned that the energy scale of interlayer splitting (\SI{74}{meV} in AA-stacking and \SI{57}{meV} in AB-stacking) and SOC energy splitting (\SI{83}{meV}) is comparable in \ch{ZrS2}. Because the interlayer coupling strength is strongly related to the interlayer distance, the band structures of both untwisted and twisted bilayer \ch{ZrS2} are sensitive to the choice of vdW. To illustrate this, we relaxed untwisted AA–\ch{ZrS2} using Grimme DFT-D3 (D3-vdW) and Tkatchenko–Scheffler (TS-vdW) corrections. Bilayer structure relaxed by D3-vdW has a larger interlayer distance \SI{5.866}{\angstrom} than the one relaxed by TS-vdW \SI{5.683}{\angstrom}. Consequently, TS-relaxed structure shows a larger splitting (\SI{105}{meV}) than the D3-relaxed one (\SI{74}{meV}) due to the smaller interlayer distance as shown in  \cref{app:fig:vdw-band-compare:a}. Moreover, the interlayer splitting is slightly larger than the SOC splitting, whereas in the TS-relaxed structure, the SOC splitting is slightly larger than the interlayer splitting. Consequently, the inclusion of SOC leads to a different ordering of layer hybridized antibonding and bonding states at $\Gamma$. Specifically, in the TS-relaxed structure, the top two states at the $\Gamma$ point are both layer-hybridized antibonding states ($\bar{\Gamma}_6\bar{\Gamma}_7$ and $\bar{\Gamma}_9$) as shown in \cref{app:fig:vdw-band-compare:b}. These two states arise from the interlayer hybridization and subsequent splitting of the corresponding monolayer $\bar{\Gamma}_6\bar{\Gamma}_7$ and $\bar{\Gamma}_9$ states (\cref{app:fig:monolayer-band:f}). In contrast, in the D3-relaxed structure, the top two states are antibonding and bonding states ($\bar{\Gamma}_6\bar{\Gamma}_7$ and $\bar{\Gamma}_4\bar{\Gamma}_5$, respectively as shown in \cref{app:fig:vdw-band-compare:b}) that originates solely from the interlayer-hybridization-induced splitting of the monolayer $\bar{\Gamma}_6\bar{\Gamma}_7$ state. The band order swap is corroborated by the symmetry representations at $\Gamma$, as shown in \cref{app:fig:vdw-band-compare:b}. The different ordering is inherited by the twisted bilayer, as shown in \cref{app:fig:ZrS2-moire-band} (relaxed by D3-vdW) and in Ref.~\cite{CLA22a} (relaxed by TS-vdW).

\subsection{Ionic Gating of Two-Dimensional Semiconductors}\label{app:sec:ionic-gating}
For many two-dimensional semiconductors to be experimentally useful, it is crucial to gate them efficiently from their intrinsic semiconducting gap into the conduction or valence band, so that measurable electron or hole transport can be achieved. Ionic gating is particularly well suited for this task because it forms an electric double layer at the surface, acting like an ultra-thin capacitor with a very large capacitance (typically $\sim$\SIrange{1}{50}{\micro\farad\per\centi\meter\squared}), far exceeding that of conventional oxide gates. This allows carrier density changes on the order of \SIrange{e13}{e14}{\per\centi\meter\squared}, sufficient to move the Fermi level by $\sim$\SIrange{0.4}{0.8}{\electronvolt} and drive the system from the gap deep into either the electron or hole bands. Such densities are high enough not only to enable ambipolar transport in materials like transition metal dichalcogenides, but in some cases to induce metallic or even superconducting behavior. Depending on the application, ionic gating can be implemented using liquid ionic electrolytes, ion gels, polymer electrolytes, or solid-state ionic conductors, offering a versatile platform for strong and reversible electrostatic doping of 2D crystals.

Ionic gating has been demonstrated to induce the required carrier densities in closely related 2D materials, including monolayer and few-layer transition metal dichalcogenides~\cite{ye2012superconducting,zhang2012ambipolar,saito2015metallic} and twisted bilayer graphene-based moir\'e systems. At low temperatures, the ionic liquid is frozen and the induced charge is fixed, making ionic gating compatible with the transport and spectroscopic probes needed to resolve the Chern-insulator and QSH phases predicted in \cref{sec:interactions}. Combined with a conventional back gate, ionic gating also allows independent tuning of the carrier density and the out-of-plane displacement field, enabling the displacement-field-driven topological transitions discussed in \cref{app:fig:D_interaction_HF_phase_diagram}.

\section{Continuum model}
To elucidate the topology and related physics in twisted bilayer \ch{ZrS2} and \ch{SnSe2}, we construct continuum models following Refs.~\cite{cualuguaru2024new,ZHA24}. We briefly summarize the method here, and full details can be found in those references. We first perform Wannier and valley projection of the DFT Hamiltonian to retain only the target orbitals and valleys, yielding a reduced DFT Hamiltonian. We then parameterize a symmetry-constrained continuum model and determine the coefficients of the symmetry-allowed terms by linear least-squares fitting to the reduced Hamiltonian, as in Refs.~\cite{cualuguaru2024new,ZHA24}. The obtained model reproduces the low-energy band structure and wavefunctions with numerical accuracy. Nonetheless, this model can still contain many terms that hinder analytical understanding. Therefore, we further simplify the model by combining the nonlinear fitting with the step-wise regression procedure proposed in Ref.~\cite{cualuguaru2024new}. Finally, we can obtain a model with far fewer parameters that still captures the low-energy physics quantitatively.

\subsection{Reduced DFT Hamiltonian obtained from projection method}\label{app:sec:projection method}
In this section, we summarize the DFT calculation and the projection setup used to obtain the reduced DFT Hamiltonian, following the notation of Ref.\cite{cualuguaru2024new}. The full Kohn-Sham Hamiltonian is obtained from the OpenMX~\cite{OZA03,OZA04} package, which is expressed in the non-orthonormal pseudo-atomic orbital (PAO) basis. We denote the PAO basis as $\ket{\varphi_{il\alpha s} \left( \vR_M + \vec{\tau}_{il\alpha} \right)}$, where $\vR_M$ denotes moir\'e cell, $l$ labels the layer, $i$ indexes the monolayer unit cell within the moir\'e supercell, $\alpha$ labels the orbital, and $s$ is the spin. For \ch{ZrS2}, we use Zr-$s2p2d2$ and S-$s2p2d2$ PAO basis set. For \ch{SnSe2}, we use Sn-$s2p2d2f1$ and Se-$s2p2d2$ PAO basis set.  

The Wannier projection procedure is inspired by the construction of Wannier functions~\cite{LOW50,CLO64, MAR97b, MAR12}. In this step, our goal is to construct a set of orthonormal Wannier bases that faithfully spans the isolated low-energy subspace. To this end, we first analyze the orbital component of the isolated low-energy valence bands. As shown in~\cref{app:fig:monolayer-band:c,app:fig:monolayer-band:d}, the top 12 valence bands are separated from the rest and dominated by $p$-orbitals of chalcogen atoms in both \ch{ZrS2} and \ch{SnSe2}. Accordingly, for each monolayer unit cell we choose three $p$-orbitals on the two S/Se sites as the projected trial states $\ket{\phi_{il\alpha s}\left(\vR_M + \vec{\tau}_{il\alpha} \right)}$, which gives $1\leq \alpha \leq 6$. We then project these trial states onto the isolated low-energy subspace $\mathcal A$, spanned by the Kohn-Sham eigenstates $\{\ket{\psi_n}\}_{n\in\mathcal A}$, through the projector $\hat P_{\mathcal A}=\sum_{n\in\mathcal A}\ket{\psi_n}\bra{\psi_n}$. 
The projected trial states are thus defined as $\ket{\bar{\phi}_{i l \alpha s}(\vR_M+\vec{\tau}_{i l \alpha})}=\hat P_{\mathcal A}\ket{\phi_{i l \alpha s}(\vR_M+\vec{\tau}_{i l \alpha})}$. Since these projected states are generally not orthonormal, we introduce their overlap matrix $\mathcal S$, whose matrix elements are $\mathcal S_{i l \alpha s; j l' \beta s'}=\bra{\bar{\phi}_{i l \alpha s}(\vR_M+\vec{\tau}_{i l \alpha})}\ket{\bar{\phi}_{j l' \beta s'}(\vR_M+\vec{\tau}_{j l' \beta})}$. The refined trial basis is then obtained by L\"owdin orthonormalization\cite{cualuguaru2024new,LOW50,CLO64, MAR97b, MAR12} as $\ket{\tilde{\phi}_{i l \alpha s}(\vR_M+\vec{\tau}_{i l \alpha})}=\sum_{j,l',\beta,s'} \ket{\bar{\phi}_{j l' \beta s'}(\vR_M+\vec{\tau}_{j l' \beta})} \left(\mathcal S^{-1/2}\right)_{j l' \beta s';i l \alpha s}$, 
where $\mathcal S^{-1/2}$ is the inverse square root of the overlap matrix. By construction, the states $\ket{\tilde{\phi}_{i l \alpha s}(\vR_M+\vec{\tau}_{i l \alpha})}$ are orthonormal and exactly span the same isolated low-energy subspace $\mathcal A$, while remaining adiabatically connected to the original trial orbitals.

\begin{figure}[!t]
    \centering
    \includegraphics[width=0.8\textwidth]{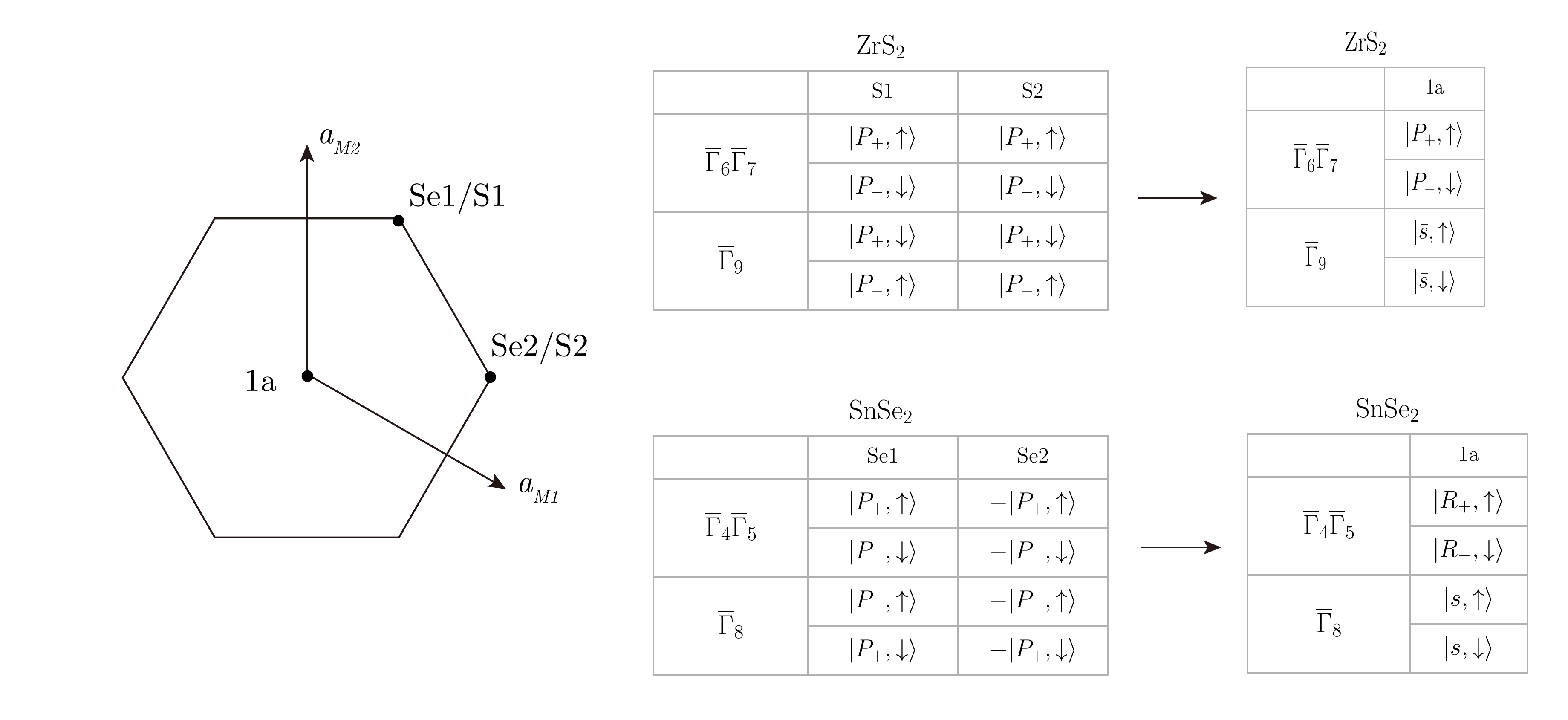}%
    \subfloat{\label{app:fig:orbital basis:a}}%
    \subfloat{\label{app:fig:orbital basis:b}}%
    \subfloat{\label{app:fig:orbital basis:c}}%
    \caption{The construction of the molecular basis corresponding to the topmost two monolayer states at $\Gamma$ in \ch{ZrS2} and \ch{SnSe2}.
    }
    \label{app:fig:orbital basis}%
\end{figure}

After the Wannier projection, we obtain a reduced Hamiltonian in the orthonormal refined trial basis $\ket{\tilde{\phi}_{i,l,\alpha,s}\left(\vR_M + \vec{\tau}_{il\alpha} \right)}$. Compared with the original DFT Hamiltonian, the Wannier-projected Hamiltonian is reduced to dimension $N_{\text{layer}}\times N_a\times N_s\times N_{\text{trial}}=24N_a$ with $N_a$ the number of monolayer unit cells per moir\'e cell, which greatly simplifies the subsequent steps. The Wannier-projected Hamiltonian is further reduced by a valley projection, analogous to the truncated atomic plane-wave (TAPW) method~\cite{MIA23} but with additional L\"{o}wdin orthogonalization procedure. We then construct trial states to further project Wannier-projected Hamiltonian into $\Gamma$-valley, and the trial states are taken as atomic plane-wave combinations of the refined projected basis in the Wannier projection procedure,
\begin{equation}\label{app:eqn:trial states of valley projection}
\begin{aligned}
    \ket{\phi_{ l\alpha s} \left( \vk, \vQ_l\right)} =& \frac{1}{\sqrt{N N_a}} \sum_{\vR_M} \sum_{i} e^{i \left( \vk + \vG_M \right) \cdot \left( \vR_M + {\tau}_{i l \alpha }\right)} \ket{\tilde{\phi}_{i l \alpha s} \left( \vR_M + \vec{\tau}_{il\alpha} \right)}\\
    =&\frac{1}{\sqrt{N N_a}} \sum_{\vR_M} \sum_{i} e^{i \left( \vk + \vec{\Gamma}_l -\vQ_l \right) \cdot \left( \vR_M + {\tau}_{i l \alpha }\right)} \ket{\tilde{\phi}_{i l \alpha s} \left( \vR_M + \vec{\tau}_{il\alpha} \right)},
\end{aligned}
\end{equation}
where $N$ is the number of moir\'e unit cell, $\vk$ lies in the first moir\'e Brillouin zone, and \(\vG_M=n_1\vec b_{M,1}+n_2\vec b_{M,2}\) with \(n_1,n_2\in\mathbb Z\) denotes a moir\'e reciprocal lattice vector. Since the low-energy moir\'e physics is related to states near a monolayer $\Gamma$-valley, we write $\mathbf G_M=\vec{\Gamma}_l-\mathbf Q_{l}$ ($\vec{\Gamma}_l$ is the $\Gamma$ momenta for layer $l$) and retain only a subset of $\mathbf Q_{l}$ vectors nearest to $\vec{\Gamma}_l$. 
Moreover, because the low-energy moir\'e bands are dominated by the lowest \(\Gamma\)-valley states of the monolayer configurations, the Hilbert space spanned by the chalcogen \(p\)-orbitals can be further reduced. Specifically, in monolayer \ch{ZrS2}, the two highest doubly degenerate spinful states at \(\Gamma\) are separated from the lower valence bands by \SI{0.73}{eV}. Their corresponding double-group IRREPs are \(\overline{\Gamma}_6\overline{\Gamma}_7\) and \(\overline{\Gamma}_9\), with out-of-plane angular momentum \(J_z=\pm \frac{3}{2}\) and \(J_z=\pm \frac{1}{2}\), respectively, as shown in \cref{app:fig:monolayer-band:f}. Since these states are dominated by the S-\((p_x,p_y)\) orbitals [\cref{app:fig:monolayer-band:d}], their orbital basis can be written approximately as
\begin{equation}
\begin{aligned}
\overline{\Gamma}_6\overline{\Gamma}_7
&:\left\{
\frac{1}{\sqrt{2}}\left(\ket{p_{+},\uparrow,\mathrm{S1}}+\ket{p_{+},\uparrow,\mathrm{S2}}\right),\ 
\frac{1}{\sqrt{2}}\left(\ket{p_{-},\downarrow,\mathrm{S1}}+\ket{p_{-},\downarrow,\mathrm{S2}}\right)
\right\},\\
\overline{\Gamma}_9
&:\left\{
\frac{1}{\sqrt{2}}\left(\ket{p_{+},\downarrow,\mathrm{S1}}+\ket{p_{+},\downarrow,\mathrm{S2}}\right),\ 
\frac{1}{\sqrt{2}}\left(\ket{p_{-},\uparrow,\mathrm{S1}}+\ket{p_{-},\uparrow,\mathrm{S2}}\right)
\right\},
\end{aligned}
\end{equation}
where \(p_{\pm}=(p_x\pm i p_y)/\sqrt{2}\), \(\uparrow\) and \(\downarrow\) denote spin up and spin down, and \(\mathrm{S1}\) and \(\mathrm{S2}\) label the two chalcogen atoms within a monolayer unit cell with fractional coordinates $\left(\frac{1}{3}, \frac{2}{3}\right)$ and $\left(\frac{2}{3}, \frac{1}{3}\right)$, respectively. Symmetry wise, the basis is equivalent to placing a spinful molecular orbital on $1a$ $(0,0)$ Wyckoff position with $J_z=\pm\frac{3}{2}$ and $J_z=\pm\frac{1}{2}$ for \(\overline{\Gamma}_6\overline{\Gamma}_7\) and \(\overline{\Gamma}_9\), respectively. For brevity, we denote the spinful molecular basis of the \(\overline{\Gamma}_6\overline{\Gamma}_7\) manifold as \(\ket{P_{\pm}}\), and the one of the \(\overline{\Gamma}_9\) manifold as \(\ket{\bar{s}_\pm}\).  Here, \(\bar{s}\) indicates a molecular orbital with odd parity under inversion, distinguishing it from an even-parity atomic \(s\) orbital.
Similarly, for \ch{SnSe2}, the highest two doubly degenerate spinful states transform according to $\overline{\Gamma}_4\overline{\Gamma}_5$ and $\overline{\Gamma}_8$, respectively, as shown in \cref{app:fig:monolayer-band:e}.  Note that these two IRREPs have even parity under inversion, which is different from the \ch{ZrS2} case. Accordingly, the orbital basis for these two states are
\begin{equation}
\begin{aligned}
\overline{\Gamma}_4\overline{\Gamma}_5
&:\left\{
\frac{1}{\sqrt{2}}\left(\ket{p_{+},\uparrow,\mathrm{Se1}}-\ket{p_{+},\uparrow,\mathrm{Se2}}\right),\ 
\frac{1}{\sqrt{2}}\left(\ket{p_{-},\downarrow,\mathrm{Se1}}-\ket{p_{-},\downarrow,\mathrm{Se2}}\right)
\right\},\\
\overline{\Gamma}_8
&:\left\{
\frac{1}{\sqrt{2}}\left(\ket{p_{+},\downarrow,\mathrm{Se1}}-\ket{p_{+},\downarrow,\mathrm{Se2}}\right),\ 
\frac{1}{\sqrt{2}}\left(\ket{p_{-},\uparrow,\mathrm{Se1}}-\ket{p_{-},\uparrow,\mathrm{Se2}}\right)
\right\},
\end{aligned}
\end{equation}
\(\mathrm{Se1}\) and \(\mathrm{Se2}\) label the two chalcogen atoms within a monolayer unit cell at fractional coordinates $\left(\frac{1}{3}, \frac{2}{3}\right)$ and $\left(\frac{2}{3}, \frac{1}{3}\right)$, respectively. Symmetry wise, the basis is equivalent to placing a spinful molecular orbital on $1a$ $(0,0)$ Wyckoff position with $J_z=\pm\frac{3}{2}$ and $J_z=\pm\frac{1}{2}$ for \(\overline{\Gamma}_4\overline{\Gamma}_5\) and \(\overline{\Gamma}_8\), respectively, as shown in \cref{app:fig:orbital basis}. For brevity, we denote the molecular basis of the $\overline{\Gamma}_4\overline{\Gamma}_5$ manifold by $\ket{R_\pm}$, and the one of the $\overline{\Gamma}_8$ manifold by $\ket{s_\pm}$. Here, $R$ denotes a molecular orbital with even parity under inversion, distinguishing it from the odd-parity molecular orbital $P$ introduced in the \ch{ZrS2} case. In practice, we retain only the highest doubly degenerate $\overline{\Gamma}_4\overline{\Gamma}_5$ band at $\Gamma$ as the continuum-model basis, since it is separated from the lower bands by \SI{0.27}{eV}, as shown in \cref{app:fig:monolayer-band:e}.

After the valley projection and the secondary Wannier projection, we obtain a projected Kohn-Sham Hamiltonian in the plane-wave basis that faithfully reproduces the low-energy \(\Gamma\)-valley moir\'e Hamiltonian, while greatly reducing the Hilbert-space dimension,
\begin{equation}\label{app:eqn:dft hamiltonian in plane wave basis}
\left[h^{\mathrm{DFT}}_{\vQ,\vQ'}(\vk)\right]_{l_1\alpha_1 s_1;\,l_2\alpha_2 s_2}
=
\bra{\phi_{l_1\alpha_1 s_1}(\vk,\vQ)}
\hat H_{\mathrm{KS}}
\ket{\phi_{l_2\alpha_2 s_2}(\vk,\vQ')}.
\end{equation}
Here, we suppress the layer index on \(\vQ\), since the two layers share the same \(\vQ\)-mesh in the \(\Gamma\)-valley moir\'e system. The index \(\alpha\) labels the retained \(\Gamma\)-valley orbital channels: \(\alpha\in\{P,\bar{s}\}\) for \ch{ZrS2}, and \(\alpha=R\) for \ch{SnSe2}.

\subsection{General continuum model constrained by symmetry}\label{app:sec:general continuum model}
In this section, we derive an explicit continuum Hamiltonian that produces the reduced DFT Hamiltonian $\left[h^{\mathrm{DFT}}_{\vQ,\vQ'}(\vk)\right]_{l_1\alpha_1 s_1;\,l_2\alpha_2 s_2}$ on a chosen set of momenta and can subsequently be evaluated at arbitrary $\mathbf{k}$ without repeating the projection procedures. To this end, we first write down a general real-space continuum Hamiltonian in terms of the continuum field operators $\cre{c}{l,\alpha, s}(\rr)$ following Ref.~\cite{ZHA24},
\begin{equation}
\begin{aligned}
	\mathcal{H} =& \frac{1}{2} \sum_{n_x, n_y\in\mathbb{N}} \sum_{\substack{l_1, \alpha_1, s_1 \\ l_2,\alpha_2,s_2}} \int \dd[2]{r} t^{n_x,n_y}_{l_1, \alpha_1,s_1;  l_2,\alpha_2, s_2} \left( \vec{r} \right) \left[ i^{n_x + n_y} \left( \partial^{n_x}_{x} \partial^{n_y}_{y} \cre{c}{l_1,\alpha_1,s_1} \left( \vec{r} \right) \right) \des{c}{l_2,\alpha_2,s_2} \left( \vec{r} \right) \right. \nonumber \\
	&\left. + (-i)^{n_x + n_y} \cre{c}{l_1,\alpha_1,s_1} \left( \vec{r} \right) \partial^{n_x}_{x} \partial^{n_y}_{y} \des{c}{l_2,\alpha_2,s_2} \left( \vec{r} \right) 	\right], \label{app:eqn:general_moire_potential}
\end{aligned}
\end{equation}
Here, $l$, $\alpha$ and $s$ label the layer, orbital and spin, respectively. As in the reduced DFT Hamiltonian, \(\alpha\in\{P,\bar{s}\}\) for \ch{ZrS2}, while \(\alpha\in\{R\}\) for \ch{SnSe2}. The coefficient \(t^{n_x,n_y}_{l_1,\alpha_1,s_1;\,l_2,\alpha_2,s_2}(\vec{r})\) denotes the generalized moir\'e potential and satisfies the Hermiticity condition
\begin{equation}
	t^{n_x,n_y}_{l_1, \alpha_1,s_1;  l_2,\alpha_2, s_2} \left( \vec{r} \right) = 
	\left(t^{n_x,n_y}_{l_2,\alpha_2, s_2;l_1, \alpha_1,s_1} \left( \vec{r} \right) \right)^*.
\end{equation}
Within this general gradient-expansion framework, the usual distinction between the intralayer kinetic term and the moir\'e potential is no longer explicit. Instead, both are incorporated into the coefficient functions $t^{n_x,n_y}_{l_1, \alpha_1,s_1;  l_2,\alpha_2, s_2} \left( \vec{r} \right)$.
Moir\'e translational symmetry requires
\begin{equation}
     t^{n_x,n_y}_{l_1,\alpha_1, s_1;  l_2,\alpha_2, s_2} \left( \vec{r} + \vR_{M} \right) = t^{n_x,n_y}_{l_1,\alpha_1, s_1;  l_2,\alpha_2, s_2} \left( \vec{r} \right) ,
\end{equation}
where \(\vR_M=m_1\vec a_{M_1}+m_2\vec a_{M_2}\) with \(m_1,m_2\in\mathbb Z\) denotes a moir\'e lattice vector. Consequently,
\begin{equation}
\label{app:eqn:potential_fourier_transform}
t^{n_x,n_y}_{l_1,\alpha_1,s_1;\,l_2,\alpha_2,s_2}(\vec{r})
=
\sum_{\vG_M}
t^{n_x,n_y}_{l_1,\alpha_1,s_1;\,l_2,\alpha_2,s_2;\vG_M}
e^{-i\vG_M\cdot\vec r},
\end{equation}
where \(\vG_M=n_1\vec b_{M,1}+n_2\vec b_{M,2}\) with \(n_1,n_2\in\mathbb Z\) denotes a moir\'e reciprocal lattice vector.
It is convenient to recast the continuum Hamiltonian in momentum space. For the \(\Gamma\)-valley problem, the two layers share the same valley center, so we suppress \(\Gamma_l\) in the following discussion. We define
\begin{equation}
\label{app:eqn:fourier_trans_field_operator}
\cre{c}{\vk,\vQ,l,\alpha,s}=
\frac{1}{\sqrt{\Omega}}
\int d^2r \cre{c}{l,\alpha,s}(\vec{r})e^{i(\vk-\vQ)\cdot \vec{r}},
\end{equation}
with inverse transform
\begin{equation}
\label{app:eqn:inv_fourier_trans_field_operator}
\cre{c}{l,\alpha,s}(\vec{r})=\frac{1}{\sqrt{\Omega}}
\sum_{\vk\in \mathrm{MBZ}}\sum_{\vQ}
\cre{c}{\vk,\vQ,l,\alpha,s}
e^{-i(\vk-\vQ)\cdot \vec{r}},
\end{equation}
where \(\Omega\) is the total area of the moir\'e system, \(\vec k\) lies in the first moir\'e Brillouin zone, and \(\vQ\) runs over the chosen plane-wave mesh. 
Substituting \cref{app:eqn:inv_fourier_trans_field_operator} and \cref{app:eqn:potential_fourier_transform} into \cref{app:eqn:general_moire_potential}, we can obtain the continuum model in the momentum space,
\begin{equation}
\begin{aligned}
\mathcal{H} 
=&\sum_{ \substack{ n_x, n_y\in\mathbb{Z}\\\vk,\vG_M}} \sum_{\substack{  \vQ_i, \vQ_j }} \sum_{\substack{s_1, l_1,\alpha_1 \\ s_2, l_2,\alpha_2}}\frac{\left( k_x - Q_{i,x} \right)^{n_x} \left( k_y - Q_{i,y} \right)^{n_y} + \left( k_x - Q_{j,x} \right)^{n_x} \left( k_y - Q_{j,y} \right)^{n_y}}{2} \\
&\times t^{n_x,n_y}_{l_1,\alpha_1, s_1;  l_2, \alpha_2,s_2; \vG_M}\delta_{\vQ_i,\vQ_j+\vG_M}\cre{c}{\vk,\vQ_i,l_1,\alpha_1,s_1} \des{c}{\vk,\vQ_j,l_2,\alpha_2,s_2}\\
=& \sum_{ \vk}\sum_{\substack{  \vQ_i, \vQ_j}} \sum_{\substack{s_1, l_1,\alpha_1 \\ s_2, l_2,\alpha_2}}\left[ h_{\vQ_i, \vQ_j} (\vk)\right]_{l_1\alpha_1 s_1 ; l_2 \alpha_2 s_2 } \cre{c}{\vk,\vQ_i,l_1,\alpha_1,s_1} \des{c}{\vk,\vQ_j,l_2,\alpha_2,s_2}.
\end{aligned}
  \label{app:eqn:moire_momentum_space}
\end{equation}
The operator $\cre{c}{\vk,\vQ,l,\alpha,s}$ creates the continuum single-particle state associated with the refined plane-wave basis $\ket{\phi_{l\alpha s}(\vk,\vQ)}$ used in the projected DFT Hamiltonian in \cref{app:eqn:dft hamiltonian in plane wave basis}. Consequently, the coefficients $t^{n_x,n_y}_{l_1,\alpha_1,s_1;\,l_2,\alpha_2,s_2;\vG_M}$ in the analytical continuum Hamiltonian can be determined by matching the continuum expression for $h_{\vQ_i,\vQ_j}(\vk)$ to the numerically projected DFT Hamiltonian $h^{\mathrm{DFT}}_{\vQ_i,\vQ_j}(\vk)$. Once these coefficients are fixed, $h_{\vQ_i,\vQ_j}(\vk)$ defines an analytic continuum Hamiltonian that can be evaluated at arbitrary $\vk$.

We now constrain \(h_{\vQ_i,\vQ_j}(\vk)\) by the symmetries of the system, such as \(\{C_{3z},C_{2x},\mathcal T\}\) in twisted AB-stacked \ch{SnSe2} and \ch{ZrS2}. Under a symmetry operation \(g\), the continuum field operators transform as
\begin{equation}
\label{app:eqn:sym_action_moire_real}
g\,\cre{c}{l_1,\alpha_1,s_1}(\vec r)\,g^{-1}
=
\sum_{l_2,\alpha_2,s_2}
\left[D(g)\right]_{l_2\alpha_2 s_2;\,l_1\alpha_1 s_1}
\cre{c}{l_2,\alpha_2,s_2}(g\vec r),
\end{equation}
and therefore the momentum-space operators transform as
\begin{equation}
\label{app:eqn:sym_action_moire_mom}
g\,\cre{c}{\vk,\vQ,l_1,\alpha_1,s_1}\,g^{-1}
=
\sum_{l_2,\alpha_2,s_2}
\left[D(g)\right]_{l_2\alpha_2 s_2;\,l_1\alpha_1 s_1}
\cre{c}{g\vk,g\vQ,l_2,\alpha_2,s_2}.
\end{equation}
Here, \(D(g)\) is the representation matrix acting on the internal layer-orbital-spin space. Substituting \cref{app:eqn:sym_action_moire_mom} into \cref{app:eqn:moire_momentum_space}, we find
\begin{equation}
\label{app:eqn:sym_action_moire_mom_h}
h_{g\vQ_i,g\vQ_j}(g\vk)
=
D(g)\,
\left[h_{\vQ_i,\vQ_j}(\vk)\right]^{(*)}
D^\dagger(g),
\end{equation}
where \(^{(*)}\) denotes complex conjugation only when \(g\) is antiunitary.

The continuum Hamiltonian \(h(\vk)\) in \cref{app:eqn:moire_momentum_space} acts on the tensor-product space of the plane-wave sector labeled by \(\vQ\) and the internal layer-orbital-spin space. For each pair \((\vQ_i,\vQ_j)\), the block \(h_{\vQ_i,\vQ_j}(\vk)\) is a matrix acting on the internal space of dimension $N_{\mathrm{int}}=N_{\mathrm{layer}}N_{\mathrm{orbital}}N_s$. To systematically impose symmetry constraints, we construct a basis of symmetry-allowed terms for \(h(\vk)\) at fixed gradient order \(n=n_x+n_y\) and fixed moir\'e harmonic shell
\begin{equation}
\mathbf G_M^{(m)} \equiv \{\vG\in \mathbf G_M \mid |\vG|=|\vG_m|\},
\end{equation}
which is closed under point-group operations, since such operations preserve both the total polynomial degree \(n\) and the norm \(|\vG|\). \(|\vG_m|\) denotes the magnitude of the reciprocal vectors in the \(m\)-th moir\'e harmonic shell. This set is closed under point-group operations, since such operations preserve both the total polynomial degree \(n\) and the norm \(|\vG|\). At fixed \((n,m)\), the Hamiltonian is expanded in the tensor-product space
\begin{equation}
\mathcal V_{n,m}=\mathcal F_n\otimes \mathcal P \otimes \mathcal X_m,
\end{equation}
where the three factors are defined as follows:
\begin{itemize}
    \item \(\mathcal F_n\) is the \((n+1)\)-dimensional space of homogeneous polynomials of \(n\)-th gradient order in a two-dimensional momentum variable \(\vec p=(p_x,p_y)\),
    \begin{equation}
    \mathcal F_n=\left\{p_x^{n_x}p_y^{n_y}\,\middle|\, n_x,n_y\ge 0,\; n_x+n_y=n\right\}.
    \end{equation}
    In the Hamiltonian, \(\vec p\) is evaluated as \(\vk-\vQ_i\). For example, for \(n=2\),
    \begin{equation}
    \mathcal F_2=\{p_x^2,\;p_xp_y,\;p_y^2\}.
    \end{equation}

    \item \(\mathcal P\) is the space of Hermitian matrices acting on the internal layer-orbital-spin space,
    \begin{equation}
    \mathcal P=\{P_\mu\},
    \end{equation}
    whose dimension is \(N_{\mathrm{int}}^2\). For example, if spin is the only internal degree of freedom, \(\mathcal P\) is spanned by the four Pauli matrices.

    \item \(\mathcal X_m\) is the space spanned by the \((\vQ_i,\vQ_j)\)-space matrices associated with harmonics in the \(m\)-th shell. For each \(\vG\in \mathbf G_M^{(m)}\), we define
    \begin{equation}
    \label{app:eqn:def_XG}
    [X_{\vG}]_{\vQ_i,\vQ_j}\equiv \delta_{\vQ_i,\vQ_j+\vG}.
    \end{equation}
    The corresponding harmonic space is
    \begin{equation}
    \mathcal X_m=\{X_{\vG}\mid \vG\in \mathbf G_M^{(m)}\},
    \end{equation}
    whose dimension equals the number of reciprocal vectors in the \(m\)-th shell.
\end{itemize}
A generic term in the continuum Hamiltonian at fixed \((n,m)\) is therefore a linear combination of basis elements of the form
\begin{equation}
f_a(\vk)\,P_\mu\,X_{\vG},
\qquad
f_a(\vk)\in \mathcal F_n,\;\;
P_\mu\in \mathcal P,\;\;
X_{\vG}\in \mathcal X_m.
\end{equation}
The symmetry operation \(g\) acts separately on each factor space. On \(\mathcal F_n\), it acts by momentum transformation,
\begin{equation}
\hat F_g f_a(\vk)=f_a(g^{-1}\vk)=\sum_b f_b(\vk)\,\mathcal F_{ba}(g).
\end{equation}
where \(\mathcal F_{ba}(g)\) is the representation matrix of \(g\) on the space \(\mathcal F_n\) spanned by \(\{f_a(\vk)\}\). On \(\mathcal P\), it acts by conjugation with the internal representation,
\begin{equation}
\hat P_g P_\mu
=
D(g)\,P_\mu^{(*)}\,D^\dagger(g)
=
\sum_\nu P_\nu\,\mathcal P_{\nu\mu}(g),
\end{equation}
where \(^{(*)}\) again denotes complex conjugation only for antiunitary \(g\), and \(\mathcal P_{\nu\mu}(g)\) is the representation matrix of \(g\) on the space \(\mathcal P\) spanned by \(\{P_\mu\}\). On \(\mathcal X_m\), it acts by rotating the moir\'e reciprocal vector,
\begin{equation}
\hat X_g X_{\vG}=X_{g\vG}
=\sum_{\vG'\in \mathbf G_M^{(m)}} X_{\vG'}\,\mathcal X_{\vG',\vG}(g).
\end{equation}
where \(\mathcal X_{\vG',\vG}(g)\) is the representation matrix of \(g\) on the space \(\mathcal X_m\) spanned by \(\{X_{\vG}\}\). With these definitions, the symmetry condition in \cref{app:eqn:sym_action_moire_mom_h} is equivalent to requiring that the Hamiltonian matrix terms at fixed \((n,m)\) be invariant under the combined action
\begin{equation}
\hat F_g\otimes \hat P_g\otimes \hat X_g.
\end{equation}
In other words, the symmetry-allowed terms are precisely those basis combinations in \(\mathcal V_{n,m}\) that transform in the trivial representation of \(\hat F_g\otimes \hat P_g\otimes \hat X_g\).

The actions of point-group operations on \(\mathcal F_n\) and \(\mathcal X_m\) are fixed independently of the material. By contrast, the representation matrix \(D(g)\) acting on the internal layer-orbital-spin space depends on the chosen orbital basis in \ch{SnSe2} and \ch{ZrS2}. In the following section, we derive the corresponding representation matrices \(D(g)\), taking into account both the exact symmetries of the twisted bilayer and the additional constraints inherited from the untwisted configuration~\cite{cualuguaru2024new}, which we refer to as zero-twist symmetries.

\subsubsection{Zero-twist symmetry of $\Gamma$-valley from local stacking approximation}\label{app:sec:zero-twist symmetry}

We consider a family of untwisted bilayer Hamiltonians $\mathcal H(\Delta \vR)$, where the top and bottom layers are displaced by $+\Delta \vR/2$ and $-\Delta \vR/2$, respectively. In the AA-stacked geometry, the lattice site $(\vR,l)$ is located at
\begin{equation}
    \vec r^{\Delta \vR}_{l,\vR}
    =
    \vR+\frac{l}{2}\Delta \vR ,
\end{equation}
whereas in the AB-stacked geometry we use a convenient layer-dependent Bravais-label convention in which the lower layer is indexed by $-\vR$, so the site $(\vR,l)$ is located at
\begin{equation}
    \vec r^{\Delta \vR}_{l,\vR}
    =
    l\,\vR+\frac{l}{2}\Delta \vR ,
\end{equation}
with $l=\pm1$ labeling the two layers. We denote by $\cre{a}{\vR,l,\alpha,s}$ the corresponding lattice creation operator. The untwisted bilayer Hamiltonian can then be written as
\begin{equation}
    \mathcal H(\Delta \vR)
    =
    \sum_l \mathcal H_l^{\tSl}
    +
    \sum_{\substack{\vR_1,l_1,\alpha_1,s_1\\\vR_2,l_2,\alpha_2,s_2}}
    S_{l_1 \alpha_1 s_1; l_2 \alpha_2 s_2}
    \!\left(
        \Delta \vR,\,
        \vec r^{\Delta \vR}_{l_2,\vR_2}-\vec r^{\Delta \vR}_{l_1,\vR_1}
    \right)
    \cre{a}{\vR_1,l_1,\alpha_1,s_1}
    \des{a}{\vR_2,l_2,\alpha_2,s_2}.
    \label{app:eqn:gamma_untwisted_hamiltonian}
\end{equation}
Here
\begin{equation}
    \mathcal H_l^{\tSl}
    =
    \sum_{\vk}
    \left[h^{\sl}(\vk)\right]_{\alpha_1s_1; \alpha_2s_2}
    \cre{a}{\vk,l,\alpha_1,s_1}\des{a}{\vk,l,\alpha_2,s_2}
\end{equation}
is the single-layer Hamiltonian of layer $l$. The coefficient
$
S_{l_1 \alpha_1 s_1; l_2 \alpha_2 s_2}(\Delta \vR,\vec r)
$
is the stacking-induced single-particle matrix element between two lattice fermions separated by $\vec r$. For $l_1\neq l_2$ it describes interlayer tunneling, while for $l_1=l_2$ it describes stacking-induced intralayer
corrections.

Because $\mathcal H(\Delta \vR)$ retains the lattice translation symmetry of a monolayer, $S_{l_1 \alpha_1 s_1; l_2 \alpha_2 s_2}(\Delta \vR,\vec r)$ depends only on the relative coordinate $\vec r$, not on the absolute positions of the two sites. Moreover, translating one layer by a primitive lattice vector leaves the local stacking unchanged, so
\begin{equation}
    S_{l_1 \alpha_1 s_1; l_2 \alpha_2 s_2}(\Delta \vR+\vec a_i,\vec r)
    =
    S_{l_1 \alpha_1 s_1; l_2 \alpha_2 s_2}(\Delta \vR,\vec r),
    \qquad i=1,2.
    \label{app:eqn:gamma_S_periodicity}
\end{equation}
As in Ref.~\cite{cualuguaru2024new}, we assume that this kernel is short-ranged in its second argument,
\begin{equation}
    S_{l_1 \alpha_1 s_1; l_2 \alpha_2 s_2}(\Delta \vR,\vec r)\approx 0
    \qquad \text{for } |\vec r|\gg |\vec a_1|.
    \label{app:eqn:gamma_S_short_range}
\end{equation}
\cref{app:eqn:gamma_S_periodicity} allows us to Fourier expand $S$ in the interlayer displacement $\Delta \vR$, while its dependence on $\vec r$ can be represented in crystal momentum space. We introduce
\begin{equation}
    S_{l_1 \alpha_1 s_1; l_2 \alpha_2 s_2}(\Delta \vR,\vec r)
    =
    \frac{1}{N}\sum_{\vg,\vk}
    S_{l_1 \alpha_1 s_1; l_2 \alpha_2 s_2}(\vg,\vk)\,
    e^{-i\vg\cdot\Delta \vR}\,
    e^{-i\vk\cdot\vec r},
    \label{app:eqn:gamma_S_double_fourier}
\end{equation}
where $\vg$ is a reciprocal lattice vector of the monolayer crystal, $N$ is the number of monolayer unit cells and $\vk$ runs over monolayer Brillouin zone. The object that enters the $\Gamma$-valley is therefore simply
\begin{equation}
    S_{l_1 \alpha_1 s_1; l_2 \alpha_2 s_2}(\vg,\Gamma)
    \equiv
    S_{l_1 \alpha_1 s_1; l_2 \alpha_2 s_2}(\vg,\vec 0).
    \label{app:eqn:gamma_S_at_Gamma}
\end{equation}

We now turn to the twisted bilayer. As in Ref.~\cite{cualuguaru2024new}, the two twisted lattices are locally approximated by an untwisted bilayer with a slowly varying local displacement $\delta \vR(\vec r)$. Denoting the twisted-lattice fermion by $\cre{c}{\vR,l,\alpha,s}$, we write its physical position in the twisted lattice as $\vec r_{l,\vR}$, which replaces the uniform-displacement position $\vec r^{\Delta \vR}_{l,\vR}$ used in the untwisted case. The twisted Hamiltonian inherits the same two-piece structure as \cref{app:eqn:gamma_untwisted_hamiltonian}. The single-layer part carries over directly by replacing $a$ with $c$ in $\sum_l \mathcal H_l^{\tSl}$, while the stacking part is obtained by replacing the global displacement $\Delta \vR$ in \cref{app:eqn:gamma_untwisted_hamiltonian} with the local displacement $\delta \vR\bigl((\vec r_{l_1,\vR_1}+\vec r_{l_2,\vR_2})/2\bigr)$ evaluated at the midpoint of the two sites, a symmetric choice that treats both layers on the same footing. Because $S$ is short-ranged in its second argument [\cref{app:eqn:gamma_S_short_range}], only site pairs with $|\vec r_{l_2,\vR_2}-\vec r_{l_1,\vR_1}|\sim|\vec a_1|$ contribute, over which $\delta \vR(\vec r)$ is effectively constant, justifying the local-untwisted approximation. The stacking part of the twisted Hamiltonian is then
\begin{equation}
    \mathcal H^{\Gamma}_{\tMoire}
    \approx
    \sum_{\substack{\vR_1,l_1,\alpha_1,s_1\\\vR_2,l_2,\alpha_2,s_2}}
    S_{l_1 \alpha_1 s_1; l_2 \alpha_2 s_2}
    \!\left(
        \delta \vR\!\left(
            \frac{\vec r_{l_1,\vR_1}+\vec r_{l_2,\vR_2}}{2}
        \right),
        \vec r_{l_2,\vR_2}-\vec r_{l_1,\vR_1}
    \right)
    \cre{c}{\vR_1,l_1,\alpha_1,s_1}
    \des{c}{\vR_2,l_2,\alpha_2,s_2}.
    \label{app:eqn:gamma_twisted_lattice_hamiltonian}
\end{equation}
For the $\Gamma$ valley, the low-energy states sit near the $\Gamma$ point of the monolayer Brillouin zone. In this low-energy subspace, the lattice operator $\cre{c}{\vR,l,\alpha,s}$ is represented by a slowly varying continuum field $\cre{c}{l,\alpha,s}(\vec r)$ of \cref{app:sec:general continuum model} by the low-energy approximation
\begin{equation}
    \cre{c}{\vR,l,\alpha,s}
    \approx
    \sqrt{\Omega_0}\,\cre{c}{l,\alpha,s}(\vec r_{l,\vR})
    =
    \sqrt{\Omega_0}
    \int \dd[2]{r}\,
    \delta(\vec r_{l,\vR}-\vec r)\,
    \cre{c}{l,\alpha,s}(\vec r),
    \label{app:eqn:gamma_continuum_expansion}
\end{equation}
where $\Omega_0$ is the monolayer unit-cell area. We use the same symbol $c$ for the lattice and continuum operators, distinguished by their arguments: the lattice operator $\cre{c}{\vR,l,\alpha,s}$ is labelled by a discrete lattice vector $\vR$, while the continuum operator $\cre{c}{l,\alpha,s}(\vec r)$ is a function of
the continuous position $\vec r$.

We now reduce \cref{app:eqn:gamma_twisted_lattice_hamiltonian} to a continuum form following \cite{cualuguaru2024new}. First, the short-rangedness of $S$ in its second argument [\cref{app:eqn:gamma_S_short_range}] restricts the sum over $(\vR_1,\vR_2)$ in \cref{app:eqn:gamma_twisted_lattice_hamiltonian} to site pairs with $|\vec r_{l_2,\vR_2}-\vec r_{l_1,\vR_1}|\sim|\vec a_1|$, which is small compared to the scale $\Lambda^{-1}$ on which the continuum field varies. Substituting \cref{app:eqn:gamma_continuum_expansion} for both lattice operators yields a factor of $\Omega_0$, and by the slowly-varying assumption the two continuum fields can be evaluated at a common point $\vec r$. We finally arrive at
\begin{equation}
    \mathcal H^{\Gamma}_{\tMoire}
    \approx
    \sum_{\substack{l_1,\alpha_1,s_1 \\ l_2,\alpha_2,s_2}}
    \int \dd[2]{r}
    \sum_{\vR_2}
    S_{l_1 \alpha_1 s_1; l_2 \alpha_2 s_2}
    \!\left(\delta \vR(\vec r),\,\vec r_{l_2,\vR_2}-\vec r\right)
    \cre{c}{l_1,\alpha_1,s_1}(\vec r)\,
    \des{c}{l_2,\alpha_2,s_2}(\vec r).
    \label{app:eqn:gamma_moire_after_continuum_id}
\end{equation}
Second, we Fourier-expand $S$ via \cref{app:eqn:gamma_S_double_fourier}. The
remaining lattice sum takes the form
\begin{equation}
    \sum_{\vR_2} e^{-i\vk\cdot(\vec r_{l_2,\vR_2}-\vec r)}
    =
    e^{i\vk\cdot\vec r}
    \sum_{\vR_2} e^{-i\vk\cdot\vec r_{l_2,\vR_2}}.
\end{equation}
For fixed layer $l_2$, the position $\vec r_{l_2,\vR_2}$ is a linear function of $\vR_2$ plus a constant: in the AA convention $\vec r_{l_2,\vR_2}=\vR_2+\text{const.}$, while in the AB convention $\vec r_{l_2,\vR_2}=l_2\vR_2+\text{const.}$ The sum over $\vR_2$ therefore vanishes unless $\vk$ coincides with a reciprocal-lattice vector of the monolayer Bravais lattice. Since in \cref{app:eqn:gamma_S_double_fourier} we choose $\vk$ in the first monolayer Brillouin zone, the only such momentum is $\vk=\Gamma=\vec 0$. The $\Gamma$-valley projection thus keeps only this term, for which all layer-dependent offset phases are unity and the lattice sum reduces to $\sum_{\vR_2}1=N$. This collapses the double Fourier sum in \cref{app:eqn:gamma_S_double_fourier} to a single sum over $\vg$ with the $\vk=\Gamma$ coefficient $S_{l_1 \alpha_1 s_1; l_2 \alpha_2 s_2}(\vg,\Gamma)$ defined in \cref{app:eqn:gamma_S_at_Gamma}, and we obtain
\begin{equation}
    \mathcal H^{\Gamma}_{\tMoire}
    =
    \sum_{\substack{\vg\\l_1,\alpha_1,s_1,l_2,\alpha_2,s_2}}
    \int \dd[2]{r}\,
    S_{l_1 \alpha_1 s_1; l_2 \alpha_2 s_2}(\vg,\Gamma)\,
    \cre{c}{l_1,\alpha_1,s_1}(\vec r)\,
    \des{c}{l_2,\alpha_2,s_2}(\vec r)\,
    e^{-i\vg\cdot \delta \vR(\vec r)}.
    \label{app:eqn:gamma_moire_before_final_phase}
\end{equation}
Using the standard relation for the local displacement field in a twisted
bilayer,
    $e^{-i\vg\cdot\delta \vR(\vec r)}
    =
    e^{-2i\sin(\theta/2)(\vg\cross\hat{\vec z})\cdot\vec r}$,
we obtain
\begin{equation}
    \mathcal H^{\Gamma}_{\tMoire}
    =
    \sum_{\substack{\vg\\l_1,\alpha_1,s_1,l_2,\alpha_2,s_2}}
    \int \dd[2]{r}\,
    S_{l_1 \alpha_1 s_1; l_2 \alpha_2 s_2}(\vg,\Gamma)\,
    \cre{c}{l_1,\alpha_1,s_1}(\vec r)\,
    \des{c}{l_2,\alpha_2,s_2}(\vec r)\,
    e^{-2i\sin(\theta/2)(\vg\cross\hat{\vec z})\cdot\vec r}.
    \label{app:eqn:gamma_moire_hamiltonian}
\end{equation}
\cref{app:eqn:gamma_moire_hamiltonian} defines the real-space moir\'e
potential
\begin{equation}
    V^\Gamma_{l_1 \alpha_1 s_1; l_2 \alpha_2 s_2}(\vec r)
    =
    \sum_{\vg}
    S_{l_1 \alpha_1 s_1; l_2 \alpha_2 s_2}(\vg,\Gamma)\,
    e^{-2i\sin(\theta/2)(\vg\cross\hat{\vec z})\cdot\vec r}.
    \label{app:eqn:gamma_real_space_potential}
\end{equation}
The phases in \cref{app:eqn:gamma_real_space_potential} are labelled by a monolayer reciprocal vector $\vg$, while the natural Fourier labels in the twisted system are moir\'e reciprocal vectors. In the generic formulation, each layer carries its own valley-centre offset $\vq_l$ on top of the moir'e labelling. For the $\Gamma$-valley problem the monolayer valley sits at $\vec 0$ and is invariant under rotation, so $\vq_l=\vec 0$ in both layers and these offsets drop out. The monolayer reciprocal vector $\vg$ is then mapped directly onto a moir\'e reciprocal vector $\vG$ by the \emph{moir\'e reciprocal-lattice map}
\begin{equation}
    -2\sin(\theta/2)\,(\vg\cross\hat{\vec z}) = \vG,
    \label{app:eqn:gamma_moire_identification}
\end{equation}
which is a bijection between the monolayer and moir\'e reciprocal lattices.

Under this identification, the moir\'e potential admits the Fourier expansion
\begin{equation}
    V^\Gamma_{l_1 \alpha_1 s_1; l_2 \alpha_2 s_2}(\vec r)
    =
    \sum_{\vG}
    \left[T^\Gamma_{\vG}\right]_{l_1 \alpha_1 s_1; l_2 \alpha_2 s_2}
    e^{i\vG\cdot\vec r},
    \label{app:eqn:gamma_V_expand_in_T}
\end{equation}
where $[T^\Gamma_{\vG}]_{l_1 \alpha_1 s_1; l_2 \alpha_2 s_2}$ is the momentum-space matrix element coupling plane-wave components that differ by the moir\'e reciprocal vector $\vG$. Fourier inversion over a moir\'e unit cell of area $\Omega$ gives
\begin{equation}
    \left[T^\Gamma_{\vG}\right]_{l_1 \alpha_1 s_1; l_2 \alpha_2 s_2}
    =
    \frac{1}{\Omega}
    \int \dd[2]{r}\,
    V^\Gamma_{l_1 \alpha_1 s_1; l_2 \alpha_2 s_2}(\vec r)\,
    e^{-i\vG\cdot\vec r}.
    \label{app:eqn:gamma_T_definition}
\end{equation}
Substituting \cref{app:eqn:gamma_real_space_potential} into
\cref{app:eqn:gamma_T_definition} and inverting
\cref{app:eqn:gamma_moire_identification} as
$\vg=\vG\cross\hat{\vec z}/[2\sin(\theta/2)]$, we obtain
\begin{equation}
    \left[T^\Gamma_{\vG}\right]_{l_1 \alpha_1 s_1; l_2 \alpha_2 s_2}
    =
    S_{l_1 \alpha_1 s_1; l_2 \alpha_2 s_2}
    \!\left(
        \frac{\vG\cross\hat{\vec z}}{2\sin(\theta/2)},\,
        \Gamma
    \right).
    \label{app:eqn:gamma_T_S_relation}
\end{equation}
Thus the $\Gamma$-valley moir\'e matrix elements $T^\Gamma$ are in one-to-one correspondence with the Fourier components $S(\vg,\Gamma)$ of the untwisted stacking kernel.

To connect $T^\Gamma_{\vG}$ to the $(\vQ_1,\vQ_2)$ labelled form used elsewhere in the paper, we project the tunneling Hamiltonian onto the plane-wave mesh of \cref{app:eqn:fourier_trans_field_operator},
\begin{equation}
\begin{aligned}
        \mathcal H^\Gamma_{\tMoire}
    &=
    \sum_{\substack{l_1,\alpha_1,s_1 \\ l_2,\alpha_2,s_2}}
    \int \dd[2]{r}\,
    V^\Gamma_{l_1 \alpha_1 s_1; l_2 \alpha_2 s_2}(\vec r)\,
    \cre{c}{l_1,\alpha_1,s_1}(\vec r)\,
    \des{c}{l_2,\alpha_2,s_2}(\vec r)\\
    &=
    \sum_{\vk\in\mathrm{MBZ}}
    \sum_{\substack{\vQ_1,\vQ_2 \\ l_1,\alpha_1,s_1 \\ l_2,\alpha_2,s_2}}
    \left[T^\Gamma_{\vQ_1,\vQ_2}\right]_{l_1\alpha_1 s_1;l_2\alpha_2 s_2}
    \cre{c}{\vk,\vQ_1,l_1,\alpha_1,s_1}
    \des{c}{\vk,\vQ_2,l_2,\alpha_2,s_2},
    \label{app:eqn:gamma_H_moire_Q_basis}
\end{aligned}
\end{equation}
with
\begin{equation}
    \left[T^\Gamma_{\vQ_1,\vQ_2}\right]_{l_1\alpha_1 s_1;l_2\alpha_2 s_2}
    \;\equiv\;
    \left[T^\Gamma_{\vQ_1,\,\vQ_2+\vG}\right]_{l_1\alpha_1 s_1;l_2\alpha_2 s_2}
    \Big|_{\vG=\vQ_1-\vQ_2}
    =
    \left[T^\Gamma_{\vG=\vQ_1-\vQ_2}\right]_{l_1\alpha_1 s_1;l_2\alpha_2 s_2}.
    \label{app:eqn:gamma_T_Q_vs_G}
\end{equation}
This is the momentum-space tunneling block that enters the matrix-element definition $[h(\vk)_{\vQ_1,\vQ_2}]_{l_1\alpha_1 s_1;l_2\alpha_2 s_2}$ used elsewhere in the paper.

We now impose the zero-twist symmetry constraints by specialising the
general continuum symmetry relation derived earlier. For any symmetry
$g=\{p_g|\tau_g\}$ of the zero-displacement untwisted bilayer,
\cref{app:eqn:sym_action_moire_mom_h} states that the continuum Hamiltonian
matrix element satisfies
\begin{equation}
    h_{p_g\vQ_1,\,p_g\vQ_2}(p_g\vk)
    =
    D(g)\,[h_{\vQ_1,\vQ_2}(\vk)]^{(*)}\,D^\dagger(g),
\end{equation}
where $(*)$ denotes complex conjugation for antiunitary $g$. The moir\'e
tunneling block is the $\vk$-independent part of $h$ and equals
$T^\Gamma_{\vQ_1,\vQ_2}$, so evaluating this constraint on the tunneling
block gives
\begin{equation}
    T^\Gamma_{p_g\vQ_1,\,p_g\vQ_2}
    =
    D(g)\,[T^\Gamma_{\vQ_1,\vQ_2}]^{(*)}\,D^\dagger(g).
\end{equation}
By \cref{app:eqn:gamma_T_Q_vs_G}, $T^\Gamma_{\vQ_1,\vQ_2}$ depends only on
the difference $\vG\equiv\vQ_1-\vQ_2$, and
$p_g\vQ_1-p_g\vQ_2=p_g\vG$, so the constraint reduces to a condition on
the $\vG$-labelled coefficients,
\begin{equation}
    T^\Gamma_{p_g\vG}
    =
    D(g)\,[T^\Gamma_{\vG}]^{(*)}\,D^\dagger(g).
    \label{app:eqn:gamma_T_sym_constraint}
\end{equation}
Note that the fractional translation $\tau_g$ of a possibly nonsymmorphic
$g$ does not enter: at the lattice level it would appear through a Bloch
phase $e^{-i\vk\cdot\tau_g}$, which is trivial at the valley centre
$\vk=\Gamma=\vec 0$ on which the continuum theory is built.

We next factor $D(g)$ into its action on the layer index and on the
internal single-layer degrees of freedom $(\alpha,s)$,
\begin{equation}
    [D(g)]_{l_2 \alpha_2 s_2;\, l_1 \alpha_1 s_1}
    =
    \delta_{l_2,\,\epsilon_g l_1}\,[D^{\tSl}(g)]_{\alpha_2 s_2;\,\alpha_1 s_1},
\end{equation}
where $\epsilon_g=+1$ if $g$ preserves the layer index and $\epsilon_g=-1$
if it exchanges the two layers. In this notation
\cref{app:eqn:gamma_T_sym_constraint} reads
\begin{equation}
    \left[T^\Gamma_{p_g\vG}\right]_{\epsilon_g l_1,\alpha'_1 s'_1;\,\epsilon_g l_2,\alpha'_2 s'_2}
    =
    \sum_{\alpha_1,s_1,\alpha_2,s_2}
    \left[D^{\tSl}(g)\right]_{\alpha'_1 s'_1;\alpha_1 s_1}
    \left[T^\Gamma_{\vG}\right]^{(*)}_{l_1\alpha_1 s_1;\,l_2\alpha_2 s_2}
    \left[D^{\tSl}(g)\right]^*_{\alpha'_2 s'_2;\alpha_2 s_2}.
\end{equation}
Finally, substituting $T^\Gamma_{\vG}=S(\vg,\Gamma)$ from
\cref{app:eqn:gamma_T_S_relation}, we track how $p_g$ acts on the
microscopic reciprocal vector $\vg$ through the moir\'e map
\cref{app:eqn:gamma_moire_identification}. Applying $p_g$ to
$\vG=-2\sin(\theta/2)\,(\vg\cross\hat{\vec z})$ and using
$p_g\hat{\vec z}=\epsilon_g\hat{\vec z}$ (the point-group element flips
$\hat{\vec z}$ precisely when it exchanges the two layers), we find
\begin{equation}
    p_g\vG
    =
    -2\sin(\theta/2)\,(\epsilon_g p_g\vg\cross\hat{\vec z}),
\end{equation}
so $p_g\vG$ corresponds under the moir\'e map to the microscopic vector
$\epsilon_g p_g\vg$, i.e.,
$T^\Gamma_{p_g\vG}=S(\epsilon_g p_g\vg,\Gamma)$. This yields the
zero-twist symmetry constraint
\begin{equation}
    \sum_{\alpha_1,s_1,\alpha_2,s_2}
    \left[D^{\tSl}(g)\right]_{\alpha'_1 s'_1; \alpha_1 s_1}
    S^{(*)}_{l_1 \alpha_1 s_1; l_2 \alpha_2 s_2}(\vg,\Gamma)
    \left[D^{\tSl}(g)\right]^*_{\alpha'_2 s'_2; \alpha_2 s_2}
    =
    S_{\epsilon_g l_1, \alpha'_1 s'_1;\,\epsilon_g l_2, \alpha'_2 s'_2}
    (\epsilon_g p_g \vg,\Gamma),
    \label{app:eqn:gamma_general_symmetry_constraint}
\end{equation}
where the superscript $(*)$ means complex conjugation if $g$ is
antiunitary.

In the AA-stacked structure, the untwist symmetry absent in twist configuration is inversion $\mathcal I$. Since $p_{\mathcal I}\vg=-\vg$ and $\epsilon_{\mathcal I}=-1$, one has
$\epsilon_{\mathcal I}p_{\mathcal I}\vg=\vg$, and therefore
\begin{equation}
    \sum_{\alpha_1,s_1,\alpha_2,s_2}
    \left[D^{\tSl}(\mathcal I)\right]_{\alpha'_1 s'_1; \alpha_1 s_1}
    S^{\tAA}_{l_1 \alpha_1 s_1; l_2 \alpha_2 s_2}(\vg,\Gamma)
    \left[D^{\tSl}(\mathcal I)\right]^*_{\alpha'_2 s'_2; \alpha_2 s_2}
    =
    S^{\tAA}_{-l_1, \alpha'_1 s'_1;\,-l_2, \alpha'_2 s'_2}(\vg,\Gamma).
    \label{app:eqn:gamma_AA_constraint}
\end{equation}
In the AB-stacked structure, the untwist symmetry absent in twist configuration is $M_z$. Since $p_{M_z}\vg=\vg$ for an in-plane reciprocal vector and $\epsilon_{M_z}=-1$, one finds $\epsilon_{M_z}p_{M_z}\vg=-\vg$, so that
\begin{equation}
    \sum_{\alpha_1,s_1,\alpha_2,s_2}
    \left[D^{\tSl}(\mathcal M_z)\right]_{\alpha'_1 s'_1; \alpha_1 s_1}
    S^{\tAB}_{l_1 \alpha_1 s_1; l_2 \alpha_2 s_2}(\vg,\Gamma)
    \left[D^{\tSl}(\mathcal M_z)\right]^*_{\alpha'_2 s'_2; \alpha_2 s_2}
    =
    S^{\tAB}_{-l_1, \alpha'_1 s'_1;\,-l_2, \alpha'_2 s'_2}(-\vg,\Gamma).
    \label{app:eqn:gamma_AB_constraint}
\end{equation}
Consequently, the untwist $\mathcal I$ acts like $\mathcal M_z$ in twisted AA-stacking, while the untwist $\mathcal M_z$ acts like $\mathcal I$ in twisted AA-stacking.

\subsubsection{Exact and zero-twist symmetry constraints in twisted bilayer \ch{ZrS2}}\label{app:sec:zero-twist and exact symmetry in ZrS2}
When constructing the basis of continuum model for twisted AA/AB-stacked \ch{ZrS2}, we consider two layers, two orbitals (\textit{i.e.}, \(\alpha\in\{P,\bar{s}\}\)) and spin degree of freedom, as discussed in \cref{app:sec:general continuum model}. For twisted AA–\ch{ZrS2}, whose exact symmetries are $C_{3z}$, $C_{2x}$ and $\mathcal{T}$.
The $\mathcal{I}$ in untwist AA-stacked bilayer acts as ${\mathcal{M}}_z$ as derived in \cref{app:sec:zero-twist symmetry}.
Considering the spinful molecular orbital $P$ with $J_z=\pm\frac{3}{2}$, the representation matrices for both exact and zero-twist symmetries can be obtained by element-wise products (“$\circ$”) of the orbital and spin parts,
\begin{equation}\label{app:eqn: effectivep_translation_symm_AAZrS}
\begin{aligned}
&D^P(C_{3z})=e^{-i\frac{2\pi}{3}\sigma_z}\circ e^{-i\frac{\pi}{3}s_z}= -s_0 ,\quad D^P(C_{2x})=\sigma_x\circ -is_x=-is_x,\\
&D^P(\mathcal{T})=\sigma_x\circ -is_y\kappa = -is_y\kappa,\quad D^P(\mathcal{M}_z)=\sigma_0\circ -is_z = -is_z.
\end{aligned}
\end{equation}
where $\sigma_i$ and $s_i$ act in orbital and spin spaces, respectively, and $\kappa$ denotes complex conjugation. Similarly, the representation matrices for the spinful molecular orbital $\bar{s}$ with $J_z=\pm \frac{1}{2}$ are 
\begin{equation}\label{app:eqn: effectives_translation_symm_AAZrS}
\begin{aligned}
&D^{\bar{s}}(C_{3z})=e^{i\frac{2\pi}{3}\sigma_z}\circ e^{-i\frac{\pi}{3}s_z}= e^{i\frac{\pi}{3}s_z} ,\quad D^{\bar{s}}(C_{2x})=\sigma_x\circ -is_x=-is_x,\\
&D^{\bar{s}}(\mathcal{T})=\sigma_x\circ -is_y\kappa = -is_y\kappa,\quad D^{\bar{s}}(\mathcal{M}_z)=\sigma_0\circ -is_z = -is_z.
\end{aligned}
\end{equation}
We order the basis in the order of layer, orbital, and spin, 
\begin{equation}
\cre{c}{l,\alpha,s}=\{\cre{c}{t,P,\uparrow}, \cre{c}{t,P,\downarrow},\cre{c}{t,\bar{s},\uparrow}, \cre{c}{t,\bar{s},\downarrow},\cre{c}{b,P,\uparrow}, \cre{c}{b,P,\downarrow},\cre{c}{b,\bar{s},\uparrow}, \cre{c}{b,\bar{s},\downarrow}\},
\end{equation}
Therefore, the representation matrices for $\cre{c}{l,\alpha,s}$ are
\begin{equation}\label{app:eqn: effectiveps_translation_symm_AAZrS}
\begin{aligned}
&D^{\text{AA-\ch{ZrS2}}}(C_{3z})=\tau_0\otimes\left(-s_0\oplus e^{i\frac{\pi}{3}s_z}\right) ,\quad D^{\text{AA-\ch{ZrS2}}}(C_{2x})=\tau_x\otimes \left(-is_x\oplus -is_x\right)\\
&D^{\text{AA-\ch{ZrS2}}}(\mathcal{T})=\tau_0\otimes \left(-is_y\kappa\oplus -is_y\kappa\right),\quad D^{\text{AA-\ch{ZrS2}}}(\mathcal{M}_z)=\tau_x\otimes \left(-is_z\oplus -is_z\right).
\end{aligned}
\end{equation}
Here, $\tau_i$ (for $i=0,x,y,z$) are identity/Pauli matrices acting in the layer subspace. 

For AB-stacking case, the exact symmetries are $C_{3z}$, $C_{2y}$ and $\mathcal{T}$. $\mathcal{M}_z$ in untwist AB bilayer becomes zero-twist symmetry $\mathcal{I}$ in twisted configuration. Listing only what differs from the AA case, the representation matrices for the continuum basis $\cre{c}{l,\alpha,s}$ are
\begin{equation}\label{app:eqn: effectiveps_translation_symm_ABZrS}
\begin{aligned}
&D^{\text{AB-\ch{ZrS2}}}(C_{2y})=\tau_x\otimes \left(-is_y\oplus -is_y\right),\quad D^{\text{AB-\ch{ZrS2}}}(\mathcal{I})=\tau_x\otimes \left(s_0\oplus s_0\right).
\end{aligned}
\end{equation}

\subsubsection{Exact and zero-twist symmetry constraints in twisted bilayer \ch{SnSe2}}
The exact and zero-twist symmetries in twisted \ch{SnSe2} are identical to those in \ch{ZrS2}. However, in \ch{SnSe2}, we keep only the $\bar{P}$ orbital in our continuum model basis since the $s$ molecular orbital basis only contributes to deeper moir\'e bands as discussed in \cref{app:sec:projection method}. So we drop the orbital index $\alpha$ in \ch{SnSe2}. Besides, unlike \ch{ZrS2} where $P$ and $\bar s$ are inversion-odd,molecular orbitals $R$ in \ch{SnSe2} have even parity under inversion. Consequently, although the symmetry groups for the AA/AB-stacked \ch{SnSe2} and \ch{ZrS2} are the same, their corresponding representation matrices slightly differ. For twisted AA-stacked \ch{SnSe2}, the continuum basis $\cre{c}{l, s}$ is
\begin{equation}
\cre{c}{l,s}=\{\cre{c}{t,\uparrow}, \cre{c}{t,\downarrow},\cre{c}{b,\uparrow}, \cre{c}{b,\downarrow}\},
\end{equation}
the representation matrices reads,
\begin{equation}
\begin{aligned}
&D^{\text{AB-\ch{SnSe2}}}(C_{3z})=\tau_0\otimes s_0 ,\quad D^{\text{AB-\ch{SnSe2}}}(C_{2x})=\tau_x\otimes -is_x,\\
&D^{\text{AB-\ch{SnSe2}}}(\mathcal{T})=\tau_0\otimes -is_y\kappa,\quad D^{\text{AB-\ch{SnSe2}}}(\mathcal{M}_z)=\tau_x\otimes -is_z.
\end{aligned}
\end{equation}
In AB-stacked case, the representation matrices for $C_{2y}$ and $\mathcal{I}$ are,
\begin{equation}
\begin{aligned}
&D^{\text{AB-\ch{SnSe2}}}(C_{2y})=\tau_x\otimes is_y,\quad D^{\text{AB-\ch{SnSe2}}}(\mathcal{I})=\tau_x\otimes- is_0.
\end{aligned}
\end{equation}

\subsection{Continuum model extracted/fitted from \textit{ab initio} Hamiltonian}\label{app:sec:extracted continuum model}
In the above section, we outlined the symmetry constraints and representation matrices for the continuum basis $\cre{c}{l,\alpha,s}$ in twisted AA/AB-stacked \ch{ZrS2} and \ch{SnSe2}. By applying constraints of exact and/or zero-twist symmetries, we obtain the independent symmetry-allowed terms of $h_{\vQ_i, \vQ_j} (\vk)$. Since the projected DFT Hamiltonian ($h^{\mathrm{DFT}}(\vk)$ in \cref{app:eqn:dft hamiltonian in plane wave basis}) is equivalent to the symmetry-constrained continuum model in the continuum limit, we can parametrize the continuum Hamiltonian in terms of the independent symmetry-allowed terms and extract the numerical values of their coefficients from the projected DFT Hamiltonian $h^{\mathrm{DFT}}(\vk)$~\cite{cualuguaru2024new,ZHA24}.
We built two models of different complexity. First, we construct a continuum model with dozens to hundreds of parameters that reproduces both band dispersions and wavefunctions of the top six/two spinful bands in \ch{ZrS2}/\ch{SnSe2}. To assess the wavefunction agreement, we define the overlap between the DFT wavefunction $|\psi_{\text{dft}}\rangle$ and the continuum model wavefunction $|\psi_{\text{model}}\rangle$. Note that because the isolated low-energy bands are almost doubly degenerate, each pair of nearly-degenerate bands is grouped into one set. 
The overlap for $m$-set of bands at $\vec{k}$ is defined as follows,
\begin{equation}\label{app:eqn: definetion for overlap}
\begin{aligned}
    \mathcal{O}_m(\vec{k}) &= \sqrt{\frac{\operatorname{Tr}\left[P_{m,\text{DFT}}P_{m,\text{model}}\right]}{2}},\\
    P_{m,\text{DFT/model}}&=\sum_{i\in m}|\psi_{i,\text{DFT/model}}\rangle\langle \psi_{i,\text{DFT/model}}|.
\end{aligned}
\end{equation}
Here, $P_{m,\text{dft/model}}$ is the projector defined using the two degenerated bands in set $m$. The full models achieve overlaps above 99\% and 95\% for \ch{ZrS2} and \ch{SnSe2}, respectively.
Second, to gain analytical insight, we apply the step-wise regression procedure of Ref.~\cite{cualuguaru2024new} to reduce parameters of the full model and obtain a simplified model with less than 10 parameters (including effective mass).
We start from the continuum model with dozens to hundreds of parameters, and test the effect of removing each parameter individually, while keeping the others fixed. The parameter whose removal increases the loss the most is discarded, and the reduced model is then refit. Repeating this procedure produces a nested set of models with progressively fewer parameters. We then choose the simplest model that still achieves the desired accuracy.
The simplified model still matches the dispersions and wavefunctions of isolated bands, with overlaps exceeding 93\% and 75\% for \ch{ZrS2} and \ch{SnSe2}, respectively. In what follows, we detail both continuum models for each material.

\subsubsection{\ch{ZrS2}}\label{app:sec:full and simplified model for ZrS2}
\begin{figure}[!t]
    \centering
    \includegraphics[width=\textwidth]{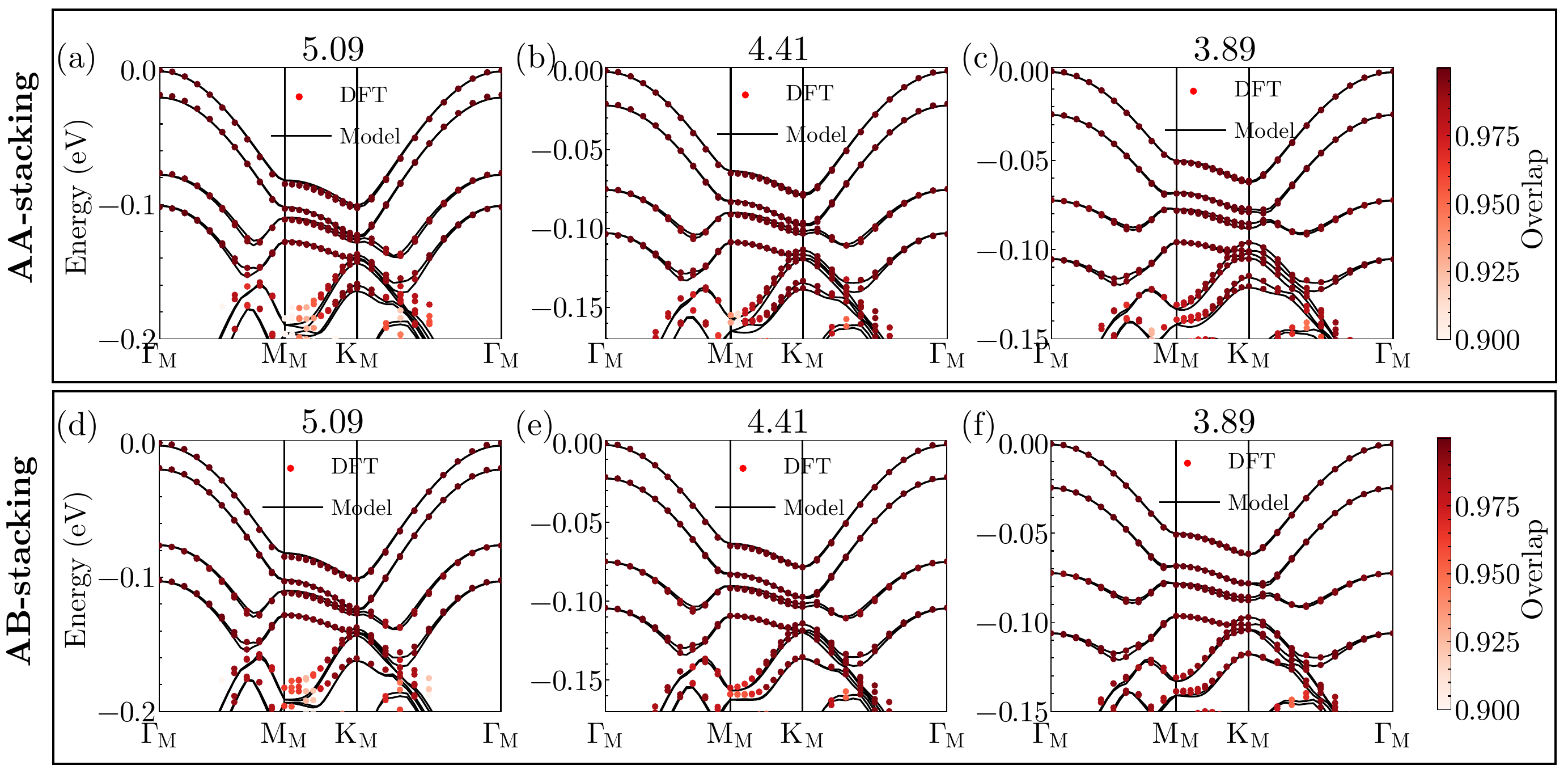}%
    \caption{The comparison of band structure between the first-harmonic continuum model and DFT for twisted \ch{ZrS2}. (a-c) Band structure for twisted AA-stacked \ch{ZrS2} from \SI{5.09}{\degree} to \SI{3.89}{\degree}. (d-f) Band structure for twisted AB-stacked \ch{ZrS2} from \SI{5.09}{\degree} to \SI{3.89}{\degree}. The black line is the result of the first-harmonic continuum model, and the red dots are DFT results with color intensity reflecting the overlap between the DFT and continuum-model wavefunctions.
    }
    \label{app:fig:ZrS2-first-harmonic-model-band}
\end{figure}

For \ch{ZrS2} system,  we consider up to first harmonic (\(\left|\vQ_i - \vQ_j\right| = \left|b_{M1}\right|\)) for $\vk$-independent terms ( $n_x+n_y= 0$), and $\vQ_i=\vQ_j$ for the $\vk$-dependent terms (\ie, $1\leq n_x+n_y\leq2$). Because the model contains two layers, two orbitals ($P$ and $\bar{s}$ molecular orbitals) and two spin flavors, the first-harmonic continuum Hamiltonians constrained by the exact symmetries involve 65 and 61 independent terms for twisted AA-stacked and AB-stacked \ch{ZrS2}, respectively, as listed in \cref{app:tab:continuum_parameters for first harmonic AA ZrS2,app:tab:continuum_parameters for first harmonic AB ZrS2}. We extract the parameters by linear least-squares fitting and compare the full continuum model with the projected DFT Hamiltonian in \cref{app:fig:ZrS2-first-harmonic-model-band} from twist angles \SI{5.09}{\degree} to \SI{3.89}{\degree} (AA on the top row, AB on the bottom row). The dispersions match closely and the wavefunction overlap exceeds 99\% for the top three sets of bands. Besides, the $C_3$ eigenvalues of the continuum model at $C_3$ invariant high symmetry points agree with the DFT results shown in \cref{app:tab:spin_chern_number_AA_ZrS2} and \cref{app:tab:spin_chern_number_AB_ZrS2}.

\begin{figure}[!t]
    \centering
    \includegraphics[width=\textwidth]{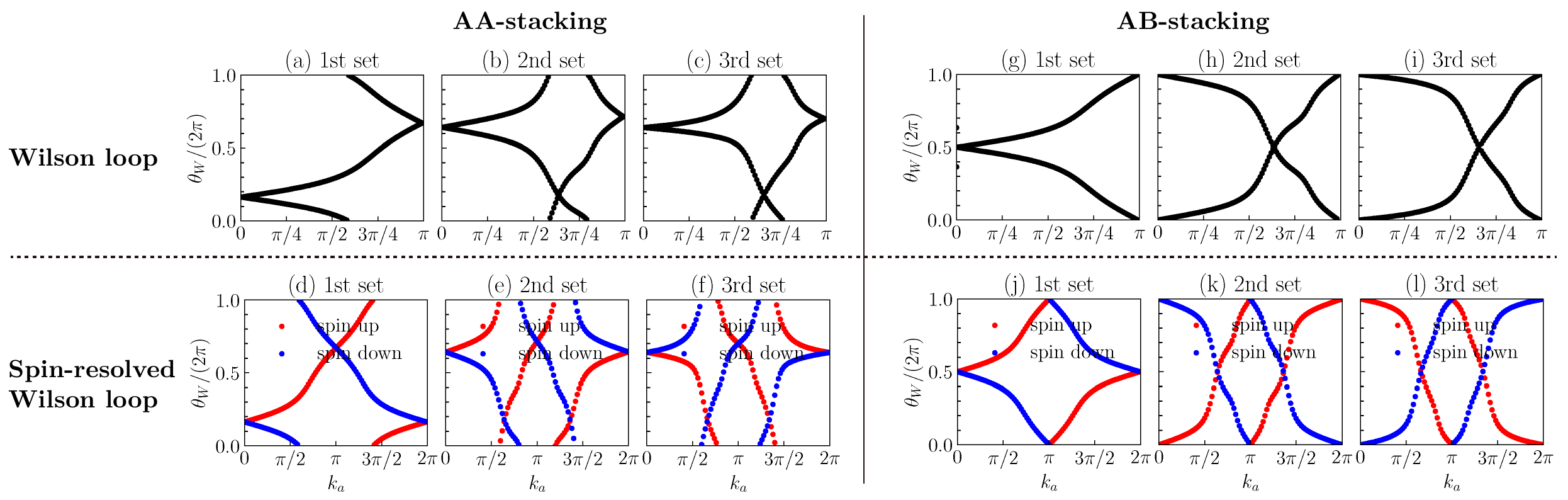}
     \subfloat{\label{app:fig:ZrS2-first-harmonic-model-wcc:a}}%
     \subfloat{\label{app:fig:ZrS2-first-harmonic-model-wcc:b}}%
     \subfloat{\label{app:fig:ZrS2-first-harmonic-model-wcc:c}}%
     \subfloat{\label{app:fig:ZrS2-first-harmonic-model-wcc:d}}%
     \subfloat{\label{app:fig:ZrS2-first-harmonic-model-wcc:e}}%
     \subfloat{\label{app:fig:ZrS2-first-harmonic-model-wcc:f}}%
     \subfloat{\label{app:fig:ZrS2-first-harmonic-model-wcc:g}}%
     \subfloat{\label{app:fig:ZrS2-first-harmonic-model-wcc:h}}%
     \subfloat{\label{app:fig:ZrS2-first-harmonic-model-wcc:i}}%
     \subfloat{\label{app:fig:ZrS2-first-harmonic-model-wcc:j}}%
     \subfloat{\label{app:fig:ZrS2-first-harmonic-model-wcc:k}}%
     \subfloat{\label{app:fig:ZrS2-first-harmonic-model-wcc:l}}%
    \caption{Wilson loop of first-harmonic continuum model for twisted \ch{ZrS2} at \SI{3.89}{\degree}. (a-f) are the Wilson loop for AA-stacked \ch{ZrS2}. (a-c) are the Wilson loops of the first, second, and third sets of bands. (d-f) are the spin-projected Wilson loop of the topmost set of bands. (g-l) are the Wilson loop for AB-stacked \ch{ZrS2}. (g-i) are the Wilson loops of the first, second, and third sets of bands. (j-l) are the spin-projected Wilson loop of the topmost set of bands. In the spin-projected Wilson loop spectrum, spinor-up and spinor-down components are denoted with red and blue dots, respectively. }
    \label{app:fig:ZrS2-first-harmonic-model-wcc}
\end{figure}

For the first-harmonic model with 65 and 61 parameters for twisted AA- and AB-stacked \ch{ZrS2}, we calculate the Wilson loops for the top three sets of bands  at \SI{3.89}{\degree}. As shown in \cref{app:fig:ZrS2-first-harmonic-model-wcc:a,app:fig:ZrS2-first-harmonic-model-wcc:g}, the first set of bands has a nontrivial winding number and exhibits a quantum spin Hall (QSH) state for both stackings. On the contrary, both the second and third sets of band have trivial $\mathbb{Z}_2$ indices as shown in \cref{app:fig:ZrS2-first-harmonic-model-wcc:b,app:fig:ZrS2-first-harmonic-model-wcc:c,app:fig:ZrS2-first-harmonic-model-wcc:h,app:fig:ZrS2-first-harmonic-model-wcc:i}. 
We quantify the breaking of spin-\Uone symmetry in the continuum Hamiltonian $h(\mathbf k)$ by
\begin{equation}
\Delta_H(\mathbf k)=
\frac{\|[h(\mathbf k),s_z]\|_F}
{2\left\|h(\mathbf k)-\dfrac{\Tr[h(\mathbf k)]}{N}\mathbb I\right\|_F},
\end{equation}
where $N$ is the dimension of $h(\mathbf k)$. For both twisted AA- and AB-stacked \ch{ZrS2}, the maximum value of $\Delta_H(\mathbf k)$ is below $2\%$. This indicates that spin-\Uone symmetry is only weakly broken, so that spin-projected geometric quantities are still well defined to a good approximation.
For the $n$-th isolated two-band set, we define the projector
\begin{equation}
P_n(\mathbf k)=\sum_{\alpha=1}^{2}|u_{n\alpha\mathbf k}\rangle\langle u_{n\alpha\mathbf k}|,
\end{equation}
where $|u_{n\alpha\mathbf k}\rangle$ are the Bloch eigenstates spanning that isolated set. We then diagonalize the projected spin operator
\begin{equation}
P_n(\mathbf k)s_zP_n(\mathbf k)\,|u^{s}_{n\mathbf k}\rangle
=
\lambda_n^s(\mathbf k)\,|u^{s}_{n\mathbf k}\rangle,
\qquad s=\uparrow,\downarrow,
\end{equation}
and use the resulting spin-adapted states $|u^{s}_{n\mathbf k}\rangle$ to define the spin-projected Berry connection
\begin{equation}
\mathcal A^{s}_{n,i}(\mathbf{k})
=
i\left\langle
u^{s}_{n\mathbf{k}}
\middle|
\partial_{k_i}
u^{s}_{n\mathbf{k}}
\right\rangle .
\end{equation}
The corresponding spin-projected Wilson-loop phase is
\begin{equation}\label{app:eqn:spin-projected Wilson loop}
\theta_n^s(k_a)
=
\mathrm{Im}\,\log
\left[
\exp\!\left(
i\oint dk_b\, \mathcal A^s_{n,b}(k_a,k_b)
\right)
\right].
\end{equation}
The spin-projected Wilson loops for the topmost band set, shown in \cref{app:fig:ZrS2-first-harmonic-model-wcc:d,app:fig:ZrS2-first-harmonic-model-wcc:j}, yield a spin Chern number $|C_s|=1$ for both AA and AB stackings. For the second and third band sets, shown in \cref{app:fig:ZrS2-first-harmonic-model-wcc:e,app:fig:ZrS2-first-harmonic-model-wcc:f,app:fig:ZrS2-first-harmonic-model-wcc:k,app:fig:ZrS2-first-harmonic-model-wcc:l}, the spin-projected Wilson loops yield $|C_s|=2$, even though the corresponding $\mathbb{Z}_2$ indices are trivial.

\begin{figure}[!t]
    \centering
    \includegraphics[width=\textwidth]{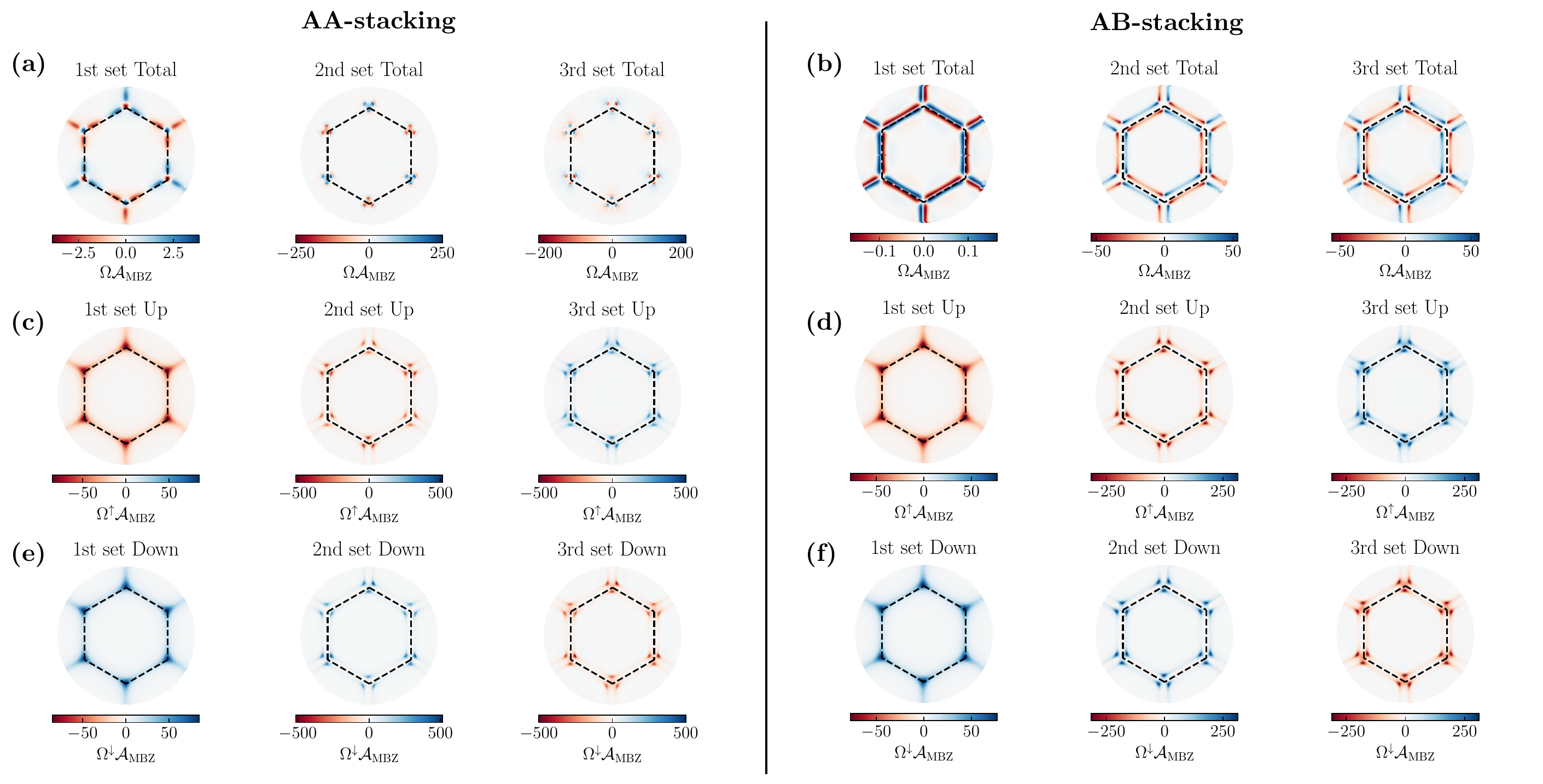}
    \subfloat{\label{app:fig:ZrS2-first-harmonic-model-berrycurvature:a}}
    \subfloat{\label{app:fig:ZrS2-first-harmonic-model-berrycurvature:b}}
    \subfloat{\label{app:fig:ZrS2-first-harmonic-model-berrycurvature:c}}
    \subfloat{\label{app:fig:ZrS2-first-harmonic-model-berrycurvature:d}}
    \subfloat{\label{app:fig:ZrS2-first-harmonic-model-berrycurvature:e}}
    \subfloat{\label{app:fig:ZrS2-first-harmonic-model-berrycurvature:f}}
    \caption{Berry curvature distribution of the first-harmonic continuum model for twisted \ch{ZrS2} at a twist angle of \SI{3.89}{\degree}. (a,b) Ordinary berry curvature for the first, second, and third sets of bands in twisted AA and AB-stacked \ch{ZrS2}. (d-f) Spin-projected Berry curvature for the first, second, and third sets of bands in AA and AB-stacked \ch{ZrS2}. Because $\tilde{C}_{2z}/{C_{2z}}$ is only an approximate symmetry in twisted AA/AB-stacked \ch{ZrS2}, the Ordinary Berry curvature is not zero but much smaller compared to the spin-projected Berry curvature.
    }
    \label{app:fig:ZrS2-first-harmonic-model-berrycurvature}
\end{figure}

Similarly, the spin-projected Berry curvature is defined as
\begin{equation}
\Omega_n^{s}(\mathbf{k})
=
\partial_{k_x}\mathcal A^s_{n,y}(\mathbf{k})
-
\partial_{k_y}\mathcal A^s_{n,x}(\mathbf{k}).
\end{equation}
The Berry-curvature distributions $\Omega_n(\mathbf{k})$ of the top three isolated band sets in the first-harmonic continuum model for both stackings are shown in \cref{app:fig:ZrS2-first-harmonic-model-berrycurvature}. For easier comparison across different twist angles, we plot the dimensionless quantity $\Omega_n \mathcal A_{\rm MBZ}$, where $\mathcal A_{\rm MBZ}$ is the area of the first moir\'e Brillouin zone. Because ${C_{2z}\mathcal T}$/$C_{2z}\mathcal T$ is only an approximate symmetry in first-harmonic model of AA/AB-stacked \ch{ZrS2}, the ordinary Berry curvature is nonzero but suppressed when compared with the spin-projected Berry curvature $\Omega_n^s(\mathbf{k})$.

\begin{figure}[!t]
    \centering
    \includegraphics[width=0.8\textwidth]{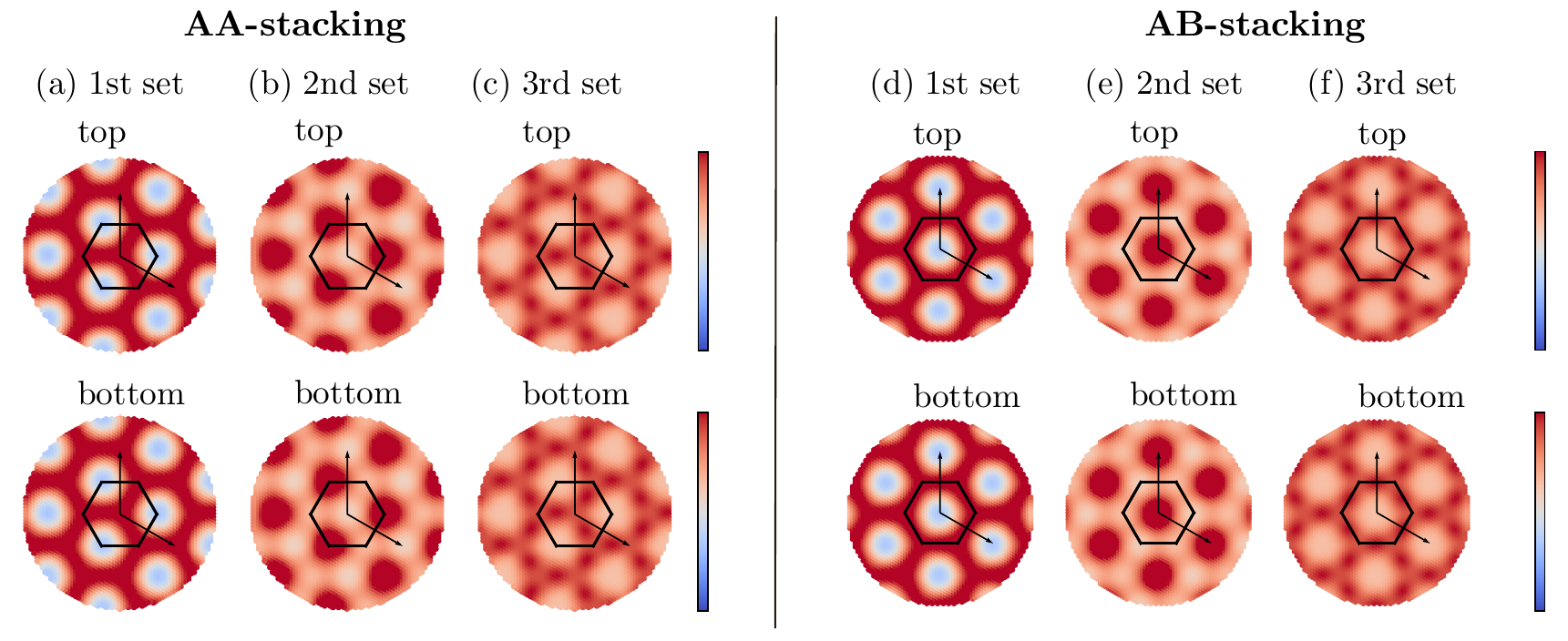}
     \subfloat{\label{app:fig:ZrS2-first-harmonic-model-ldos:a}}%
     \subfloat{\label{app:fig:ZrS2-first-harmonic-model-ldos:b}}%
     \subfloat{\label{app:fig:ZrS2-first-harmonic-model-ldos:c}}%
     \subfloat{\label{app:fig:ZrS2-first-harmonic-model-ldos:d}}%
     \subfloat{\label{app:fig:ZrS2-first-harmonic-model-ldos:e}}%
     \subfloat{\label{app:fig:ZrS2-first-harmonic-model-ldos:f}}%
    \caption{Local density of states (LDOS) computed from the first-harmonic continuum model for twisted \ch{ZrS2} at a twist angle of \SI{3.89}{\degree}. (a-c) LDOS for the first, second, and third sets of bands in AA-stacked \ch{ZrS2}. (d-f) LDOS for the first, second, and third sets of bands in AB-stacked \ch{ZrS2}. The first row corresponds to the LDOS of the top layer, while the second row shows the LDOS of the bottom layer. LDOS intensity is represented by color, ranging from blue (weak) to red (intense).
    }
    \label{app:fig:ZrS2-first-harmonic-model-ldos}
\end{figure}

We next examine the real-space charge distribution of the isolated bands. The local density of states (LDOS) for the top three sets in both AA- and AB-stacked twisted bilayer \ch{ZrS2} are shown in \cref{app:fig:ZrS2-first-harmonic-model-ldos}. The hexagon marks the moir\'e Wigner–Seitz cell, the two arrows indicate the moir\'e lattice vectors, and color encodes intensity from blue (low) to red (high).  The LDOS distribution exhibit the crystalline symmetry for both stackings:  (a) within each layer, the LDOS in both AA-stacked and AB-stacked \ch{ZrS2} show the $C_{3z}$ symmetry, (b) the top layer (first row) is related to the bottom layer (second row) by the exact symmetry $C_{2x}$ and $C_{2y}$ in AA-stacked and AB-stacked configuration, respectively. In \cref{app:fig:ZrS2-first-harmonic-model-ldos:a}, the LDOS of the topmost set of band is relatively extended with minimum at the $1b$ Wyckoff position. The second set of band is maximized at $1b$ Wyckoff position and forms a triangular lattice as illustrated in \cref{app:fig:ZrS2-first-harmonic-model-ldos:b}. The third set exhibits a kagome pattern with sites in the middle between the $1a$ and $1c$ Wyckoff positions, as depicted in \cref{app:fig:ZrS2-first-harmonic-model-ldos:c}. Furthermore, the AA and AB LDOS are related by a rigid fractional moir\'e translation: shifting the AB LDOS by $\tfrac{2}{3}\vec a_{M,1}+\tfrac{1}{3}\vec a_{M,2}$ maps the patterns of the top three sets in AB onto those in AA as shown by \cref{app:fig:ZrS2-first-harmonic-model-ldos:d,app:fig:ZrS2-first-harmonic-model-ldos:e,app:fig:ZrS2-first-harmonic-model-ldos:f}.  Finally, the LDOS patterns display approximate twofold symmetry about $z$. More specifically, LDOS in AA are nearly invariant under $C_{2z}$ operation with in-plane fractional translation, \textit{i.e.}, $\tilde C_{2z}=\{C_{2z}\,|\,(\frac{2}{3},\,\frac{1}{3})\}$, whereas in AB they are nearly invariant under $C_{2z}$. Neither of them is an exact symmetry of the system.

\begin{figure}[!t]
    \centering
    \includegraphics[width=\textwidth]{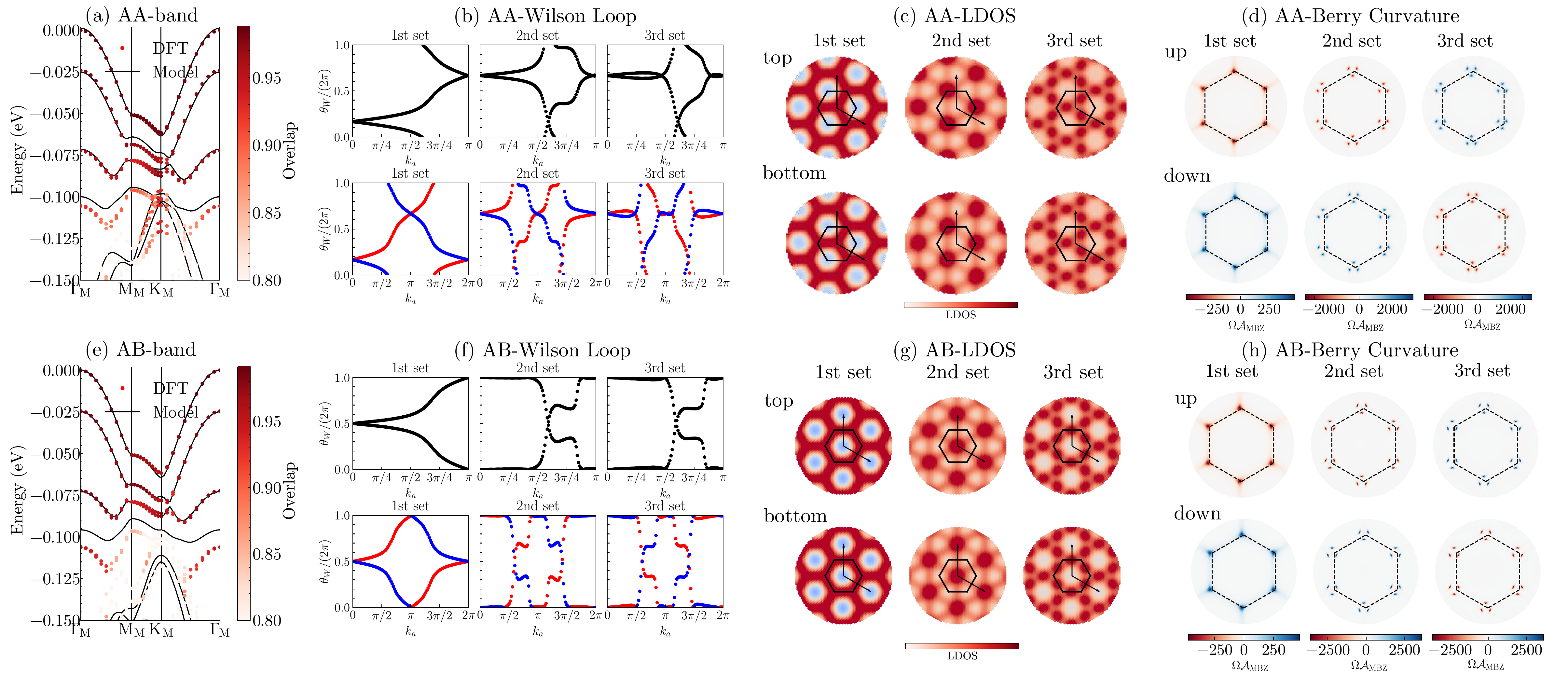}
     \subfloat{\label{app:fig:Fig-ZrS2-simplified-model:a}}%
     \subfloat{\label{app:fig:Fig-ZrS2-simplified-model:b}}%
     \subfloat{\label{app:fig:Fig-ZrS2-simplified-model:c}}%
     \subfloat{\label{app:fig:Fig-ZrS2-simplified-model:d}}%
     \subfloat{\label{app:fig:Fig-ZrS2-simplified-model:e}}%
     \subfloat{\label{app:fig:Fig-ZrS2-simplified-model:f}}%
     \subfloat{\label{app:fig:Fig-ZrS2-simplified-model:g}}%
     \subfloat{\label{app:fig:Fig-ZrS2-simplified-model:h}}%
    \caption{Electronic structure and topology from the simplified model for twisted bilayer \ch{ZrS2} at \SI{3.89}{\degree}. Panels (a–d) show AA-stacked and panels (e–h) show AB-stacked. (a,e) Band structures with black lines for the simplified continuum model, colored dots for DFT, and dot intensity indicating the wavefunction overlap between DFT and simplified model. (b,f) Wilson loops for the top three isolated band sets, Regular on the upper panels and spin-projected on the lower panels. (c,g) Layer-resolved LDOS for the top three band sets. (d,h) spin-projected Berry curvature for the top three band sets.
    }
    \label{app:fig:Fig-ZrS2-simplified-model}
\end{figure}

To clarify the relation between AA and AB stackings and the approximate symmetries suggested by the LDOS, we reduce the number of parameters by applying the step-wise regression to the first harmonic model at \SI{3.89}{\degree}~\cite{cualuguaru2024new}. The simplified models of AA-stacked and AB-stacked \ch{ZrS2} at \SI{3.89}{\degree} have the same analytical expression as follows but different numerical values for parameters,
\begin{equation}
\label{app:eqn:simplified model for AA/AB-ZrS2}
\begin{aligned}
h^{\text{AA/AB}} (\vec{r}) &=-\frac{ \hbar^2\boldsymbol{\nabla}^2}{2m_1^*}\tau_0\otimes\Gamma_{00}-\frac{ \hbar^2}{2m^*_2}\tau_0\otimes\left[(\nabla_x^2-\nabla_y^2)\Gamma_{x0}+2\nabla_x\nabla_y\Gamma_{yz}\right]\\
&+V_1\tau_0\otimes\Gamma_{z0}
+V_2\tau_z\otimes\sum_{\alpha=\pm 1}e^{i\alpha \phi_1}\left[\Gamma_{yz}e^{i\alpha\boldsymbol{g}_1\cdot\vec{r}}+\left(\frac{\sqrt{3}}{2}\Gamma_{x0}-\frac{1}{2}\Gamma_{yz}\right)e^{i\alpha\boldsymbol{g}_2\cdot\vec{r}}-\left(\frac{\sqrt{3}}{2}\Gamma_{x0}+\frac{1}{2}\Gamma_{yz}\right)e^{i\alpha\boldsymbol{g}_3\cdot\vec{r}}\right]\\
&+w_1\tau_x\otimes\Gamma_{00}+ w_2\tau_x\otimes\Gamma_{00}\sum_{i=j}^3\sum_{\alpha=\pm 1} e^{i\alpha \left(\phi_2+\boldsymbol{g}_j\cdot \vec{r}\right)},\\
\end{aligned}
\end{equation}
Here, we use $V_i$ and $w_i$ to denote the interlayer and intralayer coupling parameters, respectively. We define $\Gamma_{ij}=\sigma_i\otimes s_j$, where $\sigma_i$ and $s_i$ ($i=0,x,y,z$) are the identity/Pauli matrices acting in the molecular-orbital and spin spaces. The first-harmonic moir\'e reciprocal vectors are $\boldsymbol{g}_i=C_{3z}^{i-1}\vec{b}_{M,1}$ with $i=1,2,3$. From the simplified model, we can notice there's an emergent symmetry $\tilde{C}_{2z}/C_{2z}$ 
The numerical values of all parameters are listed in \cref{app:tab:ZrS2_simplified_model_para_values}. Moreover, $\frac{1}{m_{\pm}}=\frac{1}{m_1^*}\pm\frac{1}{m_2^*}$ describes the effective masses of the two molecular-orbital branches: the \(J_z=\pm \frac{3}{2}\) branch, corresponding to $P$ molecular orbital with \(\bar{\Gamma}_{6}\bar{\Gamma}_{7}\), and the \(J_z=\pm \frac{1}{2}\) branch, corresponding to $\bar{s}$ molecular orbital with \(\bar{\Gamma}_{9}\). As shown in \cref{app:fig:monolayer-band:f}, these two branches have different effective masses. To assess the fidelity of the simplified model, we compare its dispersion with DFT and compute the overlap with DFT wavefunctions in \cref{app:fig:Fig-ZrS2-simplified-model:a,app:fig:Fig-ZrS2-simplified-model:e}. For the top three band sets, the overlap exceeds 93\% along the high symmetry line. We also plot the regular and spin-projected Wilson loop for the simplified models in \cref{app:fig:Fig-ZrS2-simplified-model:b,app:fig:Fig-ZrS2-simplified-model:f}, where the winding numbers match those of the full model, with slight spectral differences arising from the small gap between band sets around $K_M$. The LDOS of the simplified model in \cref{app:fig:Fig-ZrS2-simplified-model:c,app:fig:Fig-ZrS2-simplified-model:g} exhibits the $\tilde{C}_{2z}/C_{2z}$ symmetry more prominently. The spin-projected Berry curvature for the top three sets (\cref{app:fig:Fig-ZrS2-simplified-model:d,app:fig:Fig-ZrS2-simplified-model:h}) is concentrated near $K_M$, and the two spin channels have equal magnitude and opposite sign.

\begin{table}[h!]
\centering
\begin{tabular}{c|c|c|c|c|c|c|c|c}
\hline
Parameters & $m_1^*$ $(m_e)$ & $m_2^*$ $(m_e)$ & $V_1$ (meV)& $V_2$ (meV) & $\phi_1$ ($\degree$) & $w_1$ (meV) &$w_2$ (meV) & $\phi_2$ ($\degree$)  \\
\hline
AA & -0.2052 & -0.3823 & 43.68 & 30.80& 59.90 &11.56 & 15.40  & 59.99  \\
\hline
AB& -0.2376 & -0.7387 & 43.23 & -35.66&0& 12.08 &-15.84& 0  \\
\hline
\end{tabular}
\caption{Table of parameters for simplified continuum model for AA- and AB-stacking twisted \ch{ZrS2} at \SI{3.89}{\degree}. Note that $\phi_i=0$ in AB-stacking is required by $C_{2y}$.}
\label{app:tab:ZrS2_simplified_model_para_values}
\end{table}
\begin{figure}[!t]
    \centering
    \includegraphics[width=\textwidth]{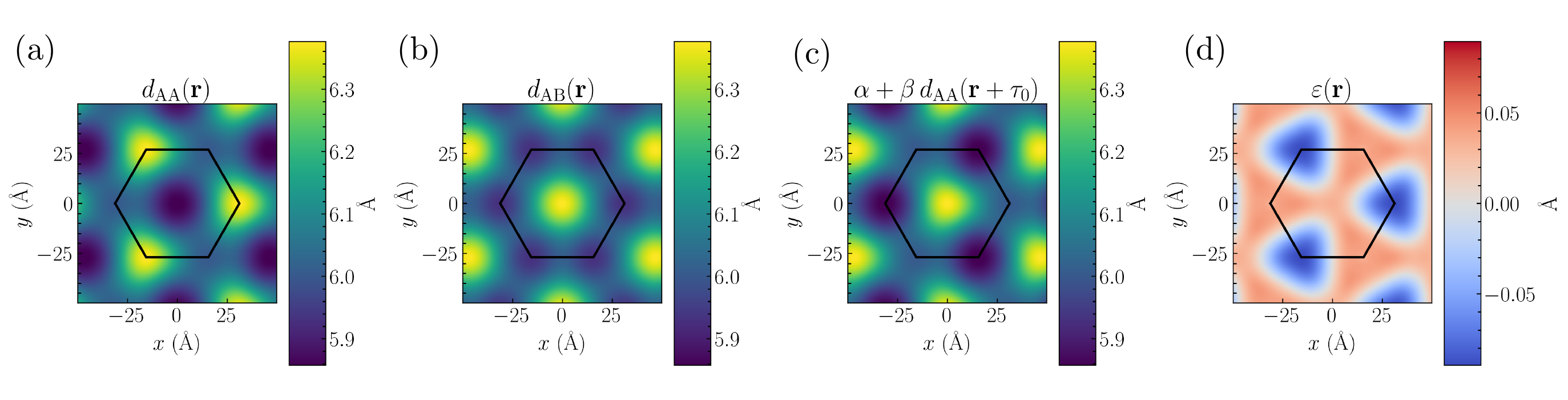}
    \subfloat{\label{app:fig:ZrS2-ILD:a}}
    \subfloat{\label{app:fig:ZrS2-ILD:b}}
    \subfloat{\label{app:fig:ZrS2-ILD:c}}
    \subfloat{\label{app:fig:ZrS2-ILD:d}}
    \caption{ Interlayer-distance fields for twisted AA- and AB-stacked \ch{ZrS2}. (a) \(d_{\mathrm{AA}}(\mathbf{r})\) (b) \(d_{\mathrm{AB}} (\mathbf{r})\) (c) the fitted translated field \(\alpha+\beta\,d_{\mathrm{AA}}(\mathbf{r}+\bm{\tau}_0)\) and (d) the residual   \(\varepsilon(\mathbf{r})\). Here \(\bm{\tau}_0=\tfrac{2}{3}\mathbf a_{M,1}+\tfrac{1}{3}\mathbf a_{M,2}\), and the black hexagon denotes the moir\'e Wigner-Seitz cell. The agreement between the AB field and the translated AA field shows that the two corrugation patterns are approximately related by an emergent fractional translation. }
    \label{app:fig:ZrS2-ILD}
\end{figure}
The analytical form of the simplified continuum model clarifies these observations. From \cref{app:eqn:simplified model for AA/AB-ZrS2}, the AA and AB models share the same functional structure. If we round the parameters in \cref{app:tab:ZrS2_simplified_model_para_values} so that $\phi^{AA}_{i}=4\pi/3$, $V^{AA}_i=V^{AB}_i$, and $w^{AA}_i=w^{AB}_i$, the moir\'e potential of twisted AA and AB-stacked \ch{ZrS2} is related by fractional moir\'e lattice translation $T_{\tau_0}$ with $\tau_0=\tfrac{2}{3}\vec a_{M,1}+\tfrac{1}{3}\vec a_{M,2}$. This is because under the fractional translation $\mathbf r\to \mathbf r+\frac23\mathbf a_{M1}+\frac13\mathbf a_{M2}$, the three first-harmonic factors acquire the same phase $e^{i\alpha 4\pi/3}$, due to $\mathbf g_1\cdot\mathbf r_0=\mathbf g_2\cdot\mathbf r_0= \mathbf g_3\cdot\mathbf r_0= 4\pi/3 \mod 2\pi$. This explains the similarity of LDOS of the top three band sets in AA and AB that only differ by $\tau_0$. Moreover, the combined symmetry $\tilde C_{2z}\mathcal T$ and $C_{2z}\mathcal T$ in twisted AA and AB stacked \ch{ZrS2} together with the spin-\Uone symmetry enforce the doubly degenerate bands as shown in \cref{app:fig:Fig-ZrS2-simplified-model}), and the equal-magnitude but opposite-sign spin-projected Berry curvatures as shown in \cref{app:fig:Fig-ZrS2-simplified-model}).

The emergent relation between the simplified AA and AB models can be traced to the local-stacking dependence of the interlayer corrugation. In the convention where the AB structure is compared to a translated AA pattern, we fit the interlayer-distance field as
\begin{equation}
d_{\mathrm{AB}}(\mathbf r)\approx \alpha+\beta\, d_{\mathrm{AA}}(\mathbf r+\bm{\tau}_0),
\qquad
\bm{\tau}_0=\frac{2}{3}\mathbf a_{M,1}+\frac{1}{3}\mathbf a_{M,2}.
\end{equation}
Here $\alpha$ accounts for the difference in average interlayer spacing, while $\beta$ captures a small renormalization of the corrugation amplitude. To quantify the remaining mismatch after removing the optimal translation, offset, and amplitude rescaling, we define the residual field
\begin{equation}
  \varepsilon(\mathbf r)=d_{\mathrm{AB}}(\mathbf r)-\bigl[\alpha+\beta\, d_{\mathrm{AA}}(\mathbf r+\bm{\tau}_0)\bigr],
\end{equation}
and evaluate its root-mean-square value on the moir\'e sampling grid,
\begin{equation}
  \varepsilon_{\mathrm{rms}}
  =
  \sqrt{\frac{1}{N}\sum_{i=1}^{N}\varepsilon(\mathbf r_i)^2 }.
\end{equation}
We also report the Pearson correlation coefficient between the target field $d_{\mathrm{AB}}(\mathbf r)$ and the fitted field
  \(
  d_{\mathrm{fit}}(\mathbf r)=\alpha+\beta\, d_{\mathrm{AA}}(\mathbf r+\bm{\tau}_0)
  \),
defined as
\begin{equation}\label{app:eqn:pearson correlation}
  \rho
  =
  \frac{
  \sum_{i=1}^{N}
  \left[d_{\mathrm{AB}}(\mathbf r_i)-\overline{d}_{\mathrm{AB}}\right]
  \left[d_{\mathrm{fit}}(\mathbf r_i)-\overline{d}_{\mathrm{fit}}\right]
  }{
  \sqrt{
  \sum_{i=1}^{N}\left[d_{\mathrm{AB}}(\mathbf r_i)-\overline{d}_{\mathrm{AB}}\right]^2
  }
  \sqrt{
  \sum_{i=1}^{N}\left[d_{\mathrm{fit}}(\mathbf r_i)-\overline{d}_{\mathrm{fit}}\right]^2
  }
  },
  \end{equation}
where the overline denotes the average over the moir\'e sampling grid. A value of $\rho$ close to unity indicates that the translated and affinely rescaled AA field reproduces the spatial pattern of the AB field very well, even if a small residual mismatch remains.
For the relaxed \ch{ZrS2} structures studied here, we obtain $\bm{\tau}_0\approx(2/3,\,1/3)$ in fractional moir\'e coordinates, $\alpha\approx 0.851~\text{\AA}$, $\beta\approx 0.861$, and $\varepsilon_{\mathrm{rms}}\approx 0.040~\text{\AA}$, with correlation $\approx 0.937$. This shows that the AA and AB structures share essentially the same leading moir\'e corrugation texture up to a shifted origin, while the residual field measures the remaining higher-harmonic and relaxation-induced corrections, so the $\bm{\tau}_0$ relation is emergent rather than exact.

\subsubsection{AB-stacked \ch{SnSe2}}
\begin{figure}[!t]
    \centering
    \includegraphics[width=\textwidth]{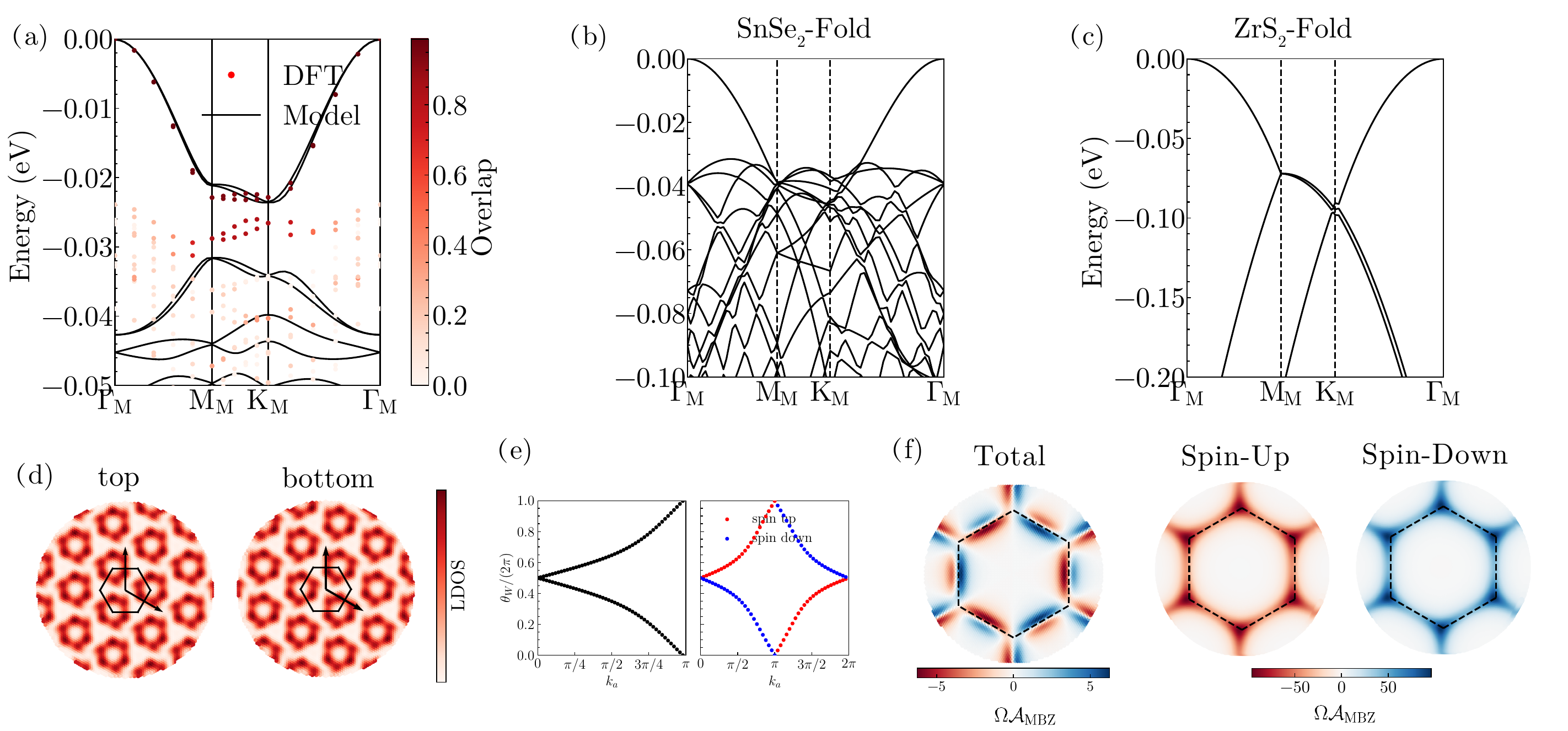}
     \subfloat{\label{app:fig:Fig-SnSe-fullmodel:a}}%
     \subfloat{\label{app:fig:Fig-SnSe-fullmodel:b}}%
     \subfloat{\label{app:fig:Fig-SnSe-fullmodel:c}}%
     \subfloat{\label{app:fig:Fig-SnSe-fullmodel:d}}%
     \subfloat{\label{app:fig:Fig-SnSe-fullmodel:e}}%
     \subfloat{\label{app:fig:Fig-SnSe-fullmodel:f}}%
    \caption{Band structure and topological properties of second-harmonic continuum model for twisted AB-stacked \ch{SnSe2} at \SI{4.41}{\degree}. (a) Band structures with black lines for the full continuum model, colored dots for DFT, and dot intensity indicating the wavefunction overlap between DFT and full model. (b) Folded band structure of \ch{SnSe2} obtained by folding monolayer topmost band into moir\'e BZ at \SI{4.41}{\degree}, indicating the NHSP valley contributes significantly to the low-energy moir\'e bands. (c) Folded band structure of \ch{ZrS2} obtained by folding monolayer topmost band into moir\'e BZ at \SI{3.89}{\degree}. (d) Layer-resolved LDOS for the topmost band set. (e) Wilson loops for the topmost isolated band set. Regular on the left panel and spin-projected on the right panel.  (f) Regular (left) and spin-projected (middle and right) Berry curvature for the topmost band set.
    }
    \label{app:fig:Fig-SnSe-fullmodel}
\end{figure}

For the numerically exact model of twisted AB-stacked \ch{SnSe2} at \SI{4.41}{\degree}, we include up to the second harmonic for the $\mathbf k$-independent and linear terms ($0\le n_x+n_y\le 1,\ \lvert\vQ_i-\vQ_j\rvert\le \lvert\mathbf b_{M1}+\mathbf b_{M2}\rvert$) and restrict to $\vQ_i=\vQ_j$ for the quadratic terms ($n_x+n_y=2$). Imposing the exact symmetries yields 61 independent terms as listed in \cref{app:tab:continuum_parameters for second harmonic AB SnSe2}. We fit their coefficients to the reduced DFT Hamiltonian and compare the resulting continuum model with DFT in \cref{app:fig:Fig-SnSe-fullmodel:a}. The top two spinful bands of the model track the DFT dispersions, and the minimum wavefunction overlap exceeds 93\%. The $C_{3}$ eigenvalues at $C_{3}$-invariant momenta also match the DFT results in \cref{app:tab:SnSe2-C3eigenvalue}. Note that we restrict to the top two spinful bands when extracting parameters from DFT reduced Hamiltonian. Higher-energy moir\'e bands receive sizable weight from a monolayer subvalley near $\Gamma$ along the $\Gamma\!-\!M$ line, \ie, a non–high-symmetry-point (NHSP)s valley (See \cref{app:fig:untwist-bilayer:b,app:fig:untwist-bilayer:e}). Because this NHSP valley lies close in energy to $\Gamma$-valley, folding the monolayer bands into the moir\'e Brillouin zone at \SI{4.41}{\degree} produces extra low-energy bands in \ch{SnSe2} that are absent in folded \ch{ZrS2} at \SI{3.89}{\degree} (\cref{app:fig:Fig-SnSe-fullmodel:b,app:fig:Fig-SnSe-fullmodel:c}). These additional bands originate from the NHSP subvalley, drive strong deviations from a simple parabolic kinetic approximation, and make stable parameter extraction difficult if more bands are included.

We evaluate the LDOS of the topmost band set for the full model (\cref{app:fig:Fig-SnSe-fullmodel:d}). The top- and bottom-layer LDOS form rings around the $1b$ and $1c$ positions, respectively. They are related by $C_{2y}$ and also exhibit an approximate inversion ($\mathcal I$) symmetry in addition to the exact generators. To determine the topology, we compute the Wilson loop (the left of \cref{app:fig:Fig-SnSe-fullmodel:e}), which shows a nontrivial winding as in AB–\ch{ZrS2}. Given the approximate spin-\Uone symmetry present both in the projected DFT Hamiltonian and full continuum model, we also compute spin-projected Wilson loops (the right of \cref{app:fig:Fig-SnSe-fullmodel:e}). The spin-up and spin-down sectors have winding numbers $+1$ and $-1$, consistent with the $C_3$ eigenvalues in \cref{app:tab:SnSe2-C3eigenvalue}. The spin-projected Berry curvature for the top band set (middle and right of \cref{app:fig:Fig-SnSe-fullmodel:f}) is concentrated near $K_M$ line and nearly cancels between spins, as seen in the total Berry curvature (left of \cref{app:fig:Fig-SnSe-fullmodel:f}).

\begin{figure}[!t]
    \centering
    \includegraphics[width=\textwidth]{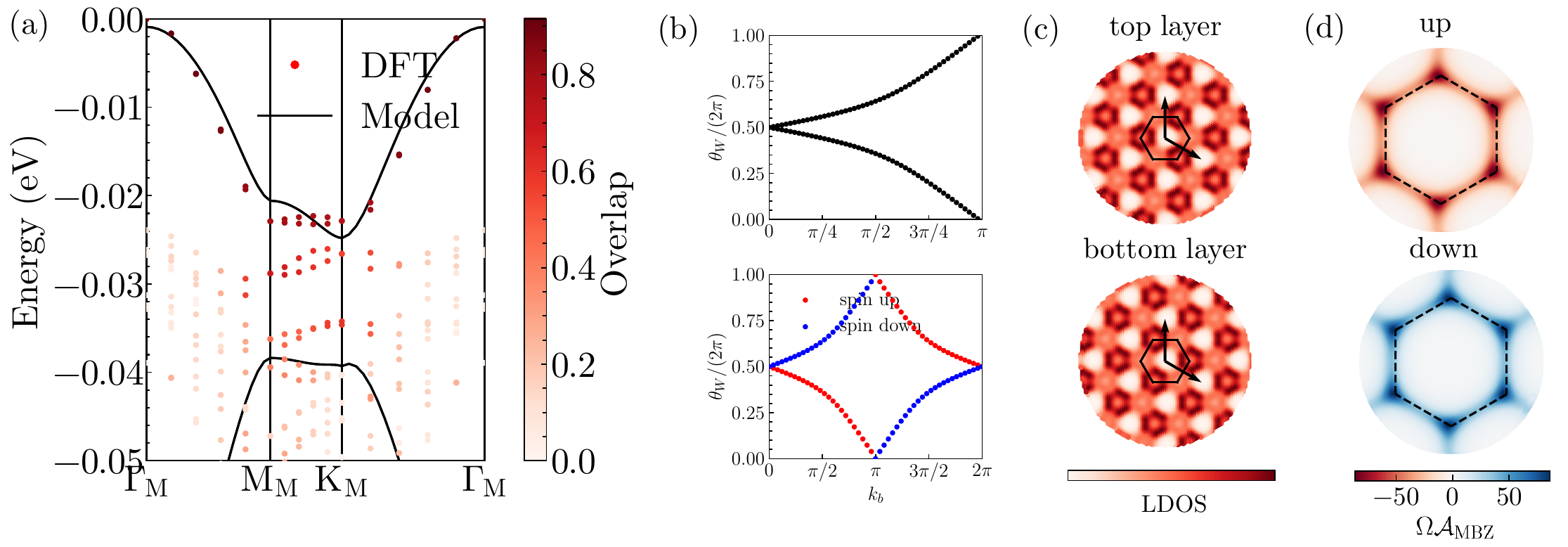}
     \subfloat{\label{app:fig:Fig-SnSe2-simplified-model:a}}%
     \subfloat{\label{app:fig:Fig-SnSe2-simplified-model:b}}%
     \subfloat{\label{app:fig:Fig-SnSe2-simplified-model:c}}%
     \subfloat{\label{app:fig:Fig-SnSe2-simplified-model:d}}%
    \caption{Band structure and topological properties of simplified continuum model in \cref{app:eqn:simplified model for AB-SnSe2} for twisted AB-stacked \ch{SnSe2} at \SI{4.41}{\degree}. (a) Band structures with black lines for the simplified continuum model, colored dots for DFT, and dot intensity indicating the wavefunction overlap between DFT and simplified model. (b) Wilson loops for the top band set, with regular on the upper panel and spin-projected on the lower panel. (c) Layer-resolved LDOS for the top three band set. (d) spin-projected Berry curvature for the top band set.
    }
    \label{app:fig:Fig-SnSe2-simplified-model}
\end{figure}

To gain a deep understanding of the system, we reduce the number of parameters and obtain a simplified analytic model. Inspecting the projected DFT Hamiltonian, we find the zero-twist inversion symmetry holds to within about 1\% for the top band set, which explains the additional $\mathcal I$ observed in the LDOS of the full model. We therefore impose $\mathcal I$ together with the approximate spin-\Uone symmetry, and drop the $\vk$-dependent moir\'e potential terms. With these simplifications, the continuum model for \ch{SnSe2} takes the form
\begin{equation}
\label{app:eqn:simplified model for AB-SnSe2}
\begin{aligned}
h^{\text{AB}} (\vec{r}) &=-\frac{ \hbar^2 \boldsymbol{\nabla}^2}{2m_1^*}\tau_0\otimes s_{0}+V_1\tau_0\otimes s_0\sum_{j=1}^3\sum_{\alpha=\pm 1} e^{i\alpha\boldsymbol{g}_j\cdot\vec{r}}\\
&+i\alpha V_2 \tau_z\otimes s_0\sum_{j=1}^3\sum_{\alpha=\pm 1} e^{i\alpha\boldsymbol{g}_j\cdot\vec{r}}+V_3\tau_0\otimes s_0\sum_{j=1}^3\sum_{\alpha=\pm 1} e^{i\alpha\boldsymbol{g}_{2j}\cdot\vec{r}}\\
&-\frac{ \hbar^2\boldsymbol{\nabla}^2}{2m^*_2}\tau_x\otimes s_0+w_1\tau_x\otimes s_{0}+w_2\tau_x\otimes s_{0}\sum_{j=1}^3\sum_{\alpha=\pm 1} e^{i\alpha\boldsymbol{g}_j\cdot\vec{r}}\\
&+w_3\tau_x\otimes s_0\sum_{j=1}^3\sum_{\alpha=\pm 1} e^{i\alpha\boldsymbol{g}_{2j}\cdot\vec{r}}+i\alpha w_4\tau_y\otimes s_z\sum_{j=1}^3\sum_{\alpha=\pm 1} e^{i\alpha\boldsymbol{g}_{2j}\cdot\vec{r}}.
\end{aligned}
\end{equation}
Here, we use $V_i$ and $w_i$ to denote the interlayer and intralayer coupling parameters, respectively. $\tau_i$ and $s_i$ ($i=0,x,y,z$) are identity/Pauli matrices acting in layer and spin subspaces, respectively.  The first- and second-harmonic moir\'e reciprocal vectors are $\boldsymbol{g}_i=C_{3z}^{i-1}\vec{b}_{M,1}$ and $\boldsymbol{g}_{2i}=C_{3z}^{i-1}(\vec{b}_{M,1}+\vec{b}_{M,2})$. The numerical values of the parameters are listed in \cref{app:tab:SnSe2_simplified_model_para_values}. Because of the nearby monolayer NHSP valley, the $\vk$-dependent moir\'e terms are essential in \ch{SnSe2}. Omitting them causes substantial renormalization of the fitted parameters in the simplified model compared to the full continuum model. To assess the fidelity of the simplified model, we compare the dispersion with DFT and compute wavefunction overlap. As shown in \cref{app:fig:Fig-SnSe2-simplified-model:a}, the top band set reproduces the overall dispersion, and the overlaps exceed 75\% along high-symmetry lines. The regular and spin-projected Wilson loops are shown in \cref{app:fig:Fig-SnSe2-simplified-model:b}, which match those of the full model. \cref{app:fig:Fig-SnSe2-simplified-model:c} show the LDOS of the simplified model, which differs slightly in detail but retains the ring-like weight around the $1b$ and $1c$ positions. Besides, LDOS of top and bottom layers are related by both the exact symmetry $C_{2y}$ and zero-twist symmetry $\mathcal I$. The spin-projected Berry curvature for the topmost band set (\cref{app:fig:Fig-SnSe2-simplified-model:d}) has equal magnitude and opposite sign in the two spin channels. The approximate combined $\mathcal I\mathcal T$ symmetry accounts for both the near double degeneracy of the DFT isolated band set and the nearly vanishing total Berry curvature.
\begin{table}[h!]
\centering
\begin{tabular}{c|c|c|c|c|c|c|c|c|c|}
\hline
Parameters & $m_1^*$ $(m_e)$ & $m_2^*$ $(m_e)$ & $V_1$ (eV)& $V_2$ (eV) & $V_3$ (eV) & $w_1$ (eV) &$w_2$ (eV) & $w_3$ (eV) & $w_4$ (eV)  \\
\hline
AB& -0.3207 & -0.6434 & -0.115 & 0.387 &-1.704 & -4.141 &-0.104&-1.678 &0.088  \\
\hline
\end{tabular}
\caption{Table of parameters for simplified contiuum model for AB-stacking twisted \ch{SnSe2} at \SI{4.41}{\degree}.}
\label{app:tab:SnSe2_simplified_model_para_values}
\end{table}

\section{Topological property of $\Gamma$-valley systems}\label{app:sec: analysis for topology of Gamma valley}

The isolated topological moir\'e bands observed in twisted bilayer \ch{SnSe2} and \ch{ZrS2} motivate us to examine the origin of their topology. We start from the monolayer band structure and identify which features are prerequisites for nontrivial moiré topology. Two closely related symmetry-based analysis studies were developed in Ref.~\cite{liu2025symmetry,crepel2025efficient}, where the topology of the low-energy moir\'e bands is inferred from the atomic symmetry group, the monolayer IRREP at $\Gamma$, and the moir\'e symmetry group. In our case, the presence of spin-\Uone symmetry in moir\'e system inherited from the monolayer provides additional information but also imposes an additional constraint. One the one hand, spin-\Uone symmetry allows us to distinguish the stable topology from the fragile topology through the $C_{3z}$ symmetry indicator~\cite{fang2012bulk}. On the other hand, the spin-\Uone symmetry applies additional constraints to the conditions of isolated topological moir\'e band as shown below.

\subsection{Approximate spin-\Uone symmetry in $D_{3d}$/$D_{3h}$ monolayer $\Gamma$-valley}\label{app:sec:monolayer origin of topology}
\begin{figure}[!t]
    \centering
    \includegraphics[width=0.5\textwidth]{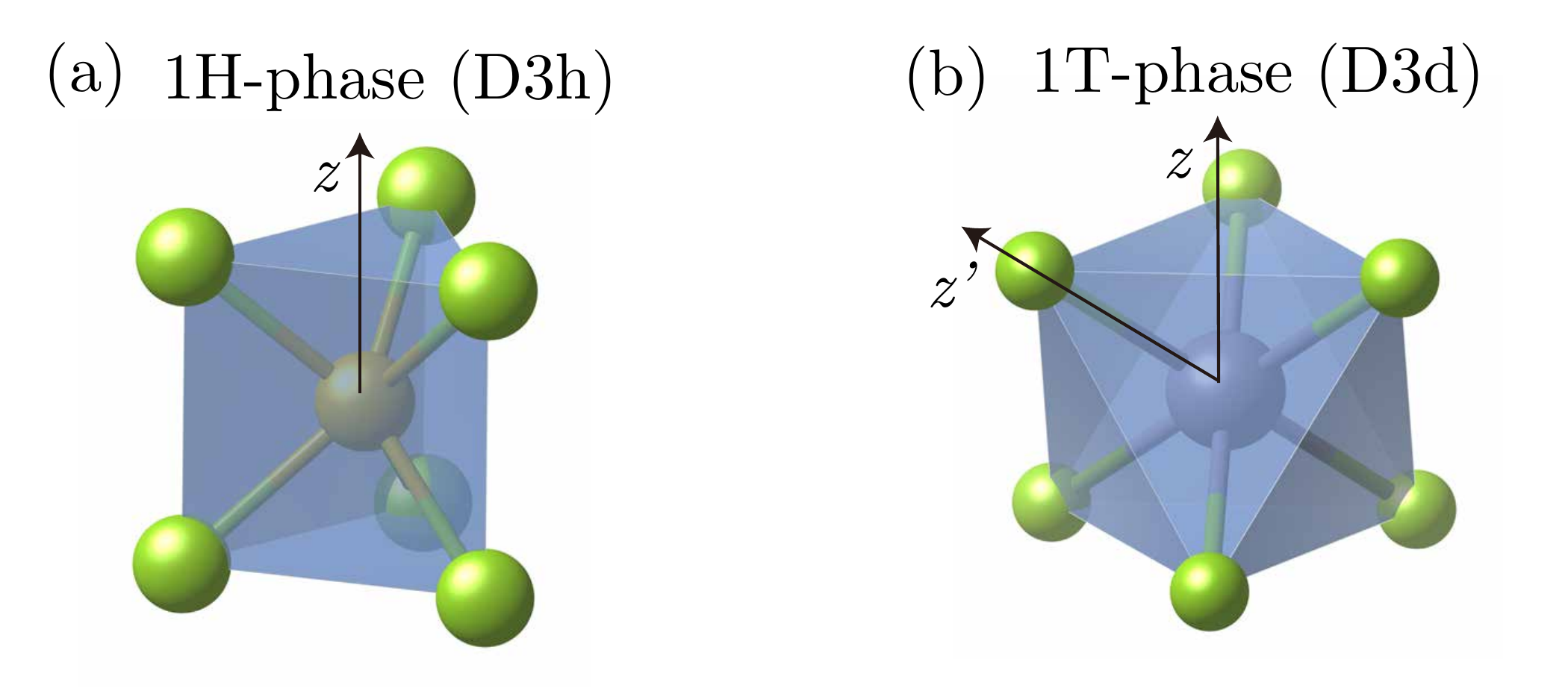}
     \subfloat{\label{app:fig:Fig-1H1T-crystal:a}}%
     \subfloat{\label{app:fig:Fig-1H1T-crystal:b}}%
    \caption{Monolayer crystal structure of (a) 1H-phase and (b) 1T-phase TMDs.
    }
    \label{app:fig:Fig-1H1T-crystal}
\end{figure}

In this section, we explore the origin of approximate spin-\Uone symmetry and spin-$\mathrm{SU}(2)$ symmetry observed in $\Gamma$-valley moir\'e bands based on twisted bilayer/multilayer 1H-phase \ch{MoTe2}\cite{JIA24,qi2025chern} and twisted bilayer 1T-\ch{ZrS2}/\ch{SnSe2}, respectively. 

We first analyze the orbital component dominating band edge at monolayer $\Gamma$ in these three materials, which can be inferred from crystal-field splitting together with the metal electron configuration. In 1H-phase as shown in \cref{app:fig:Fig-1H1T-crystal:a}, the metal atoms reside in a trigonal-prismatic crystal field, causing the $d$ orbitals to split into a singlet $a_1'$ ($d_{z^2}$) and two doublets, namely $e'$ ($d_{x^2-y^2}$, $d_{xy}$) and $e''$ ($d_{xz}$, $d_{yz}$)~\cite{chhowalla2013chemistry}. Because orbitals with lobes pointing along the metal–chalcogen bonds are pushed to higher energies by Coulomb repulsion, the energy ordering is $E_{e''}>E_{e'}>E_{a_1'}$. Because 1H-\ch{MoTe2} is composed of group-VI transition metals, where Mo donates 4 electrons and has effectively $d^2$ shell, only the lowest $a'$-state is occupied while the other states are unfilled. Consequently, in \ch{MoTe2} (and other group-VI TMDs), the valence band edge at $\Gamma$ are typically dominated by the $d_{z^2}$ orbital, while conduction band edge at $\Gamma$ have predominantly $e'$-type character hybridized with in-plane chalcogen $p$ orbitals~\cite{li2016strong,kormanyos2015k,liu2013three}. In 1T phase as shown in \cref{app:fig:Fig-1H1T-crystal:b}, the metal atoms occupy octahedral sites. In the local octahedral frame the $d$ orbitals split into $t_{2g}$ ($d_{x'y'}$, $d_{x'z'}$, $d_{y'z'}$) and $e_g$ ($d_{z'^2}$, $d_{x'^2-y'^2}$) manifolds, where $E_{e_g}>E_{t_{2g}}$. Note that we use $x'y'z'$, where $z'$ is pointing from the metal atom to one of the chalcogen atom, to distinguish the $xyz$ used in the following global framework. In the ideal 1T-phase monolayer with trigonal distortion, the $t_{2g}$ manifold further rehybridizes into an $a_{1g}$ singlet and an $e_g^\pi$ doublet, which in the global hexagonal frame can be viewed as $d_{z^2}$-like and $(d_{x^2-y^2}, d_{xy})$-like states, respectively~\cite{mattheiss1973band,chhowalla2013chemistry}. For 1T-phase composed by the group-IV metal such as \ch{ZrS2}, the metal is nominally in a $d^0$ configuration. Therefore, the valence bands is dominated by filled chalcogen $p$ orbitals and the conduction band is dominated by empty metal $d$-orbitals~\cite{zhuang2013computational,lau2019electronic}. Note that \ch{SnSe2} is composed of the group-14 main-group element Sn, which is formally in a $4d^{10}$ configuration. The filled $4d$ shell lies deep in energy and is chemically inert. Consequently, the conduction band states are dominated by antibonding hybridization between Sn $5s$ and Se $4p$ states, while the valence-band edge retain predominantly  chalcogen $p$ character. In both \ch{ZrS2} and \ch{SnSe2}, the valence band edge at $\Gamma$ is then dominated by a $(p_x,p_y)$ doublet as shown in \cref{app:fig:monolayer-bands}.

The little group of the monolayer $\Gamma$ point is $D_{3h}$ (generated by $C_{3z}$, $C_{2x}$, $\mathcal M_z$, and $\mathcal T$) for the 1H phase, and $D_{3d}$ (generated by $C_{3z}$, $C_{2x}$, $\mathcal I$, and $\mathcal T$) for the 1T phase. For the two point groups, the IRREPs in the non-SOC case and their orbital basis functions are listed in \cref{app:tab:combined_point_groups}. From the discussion above, band-edge states at $\Gamma$ dominated by out-of-plane orbitals (\eg, $d_{z^2}$ in valence band edge of \ch{MoTe2}) transform as one-dimensional IRREPs, whereas states dominated by in-plane orbitals (\eg, $(p_x,p_y)$ in \ch{ZrS2} and \ch{SnSe2}) transform as two-dimensional IRREPs. We consider the atomic SOC interaction,
\begin{equation}
    H_{\mathrm{SOC}} = \lambda\,\vec{L}\cdot\vec{S}
    = \lambda\left[L_z S_z + \frac{1}{2}\big(L_+ S_- + L_- S_+\big)\right],
\end{equation}
with $L_{\pm}=L_x\pm iL_y$.  For bands transforming as 1D IRREPs, \textit{e.g.,} valence band edge at $\Gamma$ in 1H-\ch{MoTe2}, the dominant orbital component is $d_{z^2}$ has $m_l = 0$. In this case $L_z|m_l=0\rangle = 0$, and the ladder operators $L_{\pm}$ connect the low-energy orbital to higher-lying $|m_l=\pm 1\rangle$ states, so SOC only enters the low-energy subspace in second order. As a result, the low-energy states retain an approximate spin-$\mathrm{SU}(2)$ symmetry.

By contrast, orbitals forming 2D IRREPs have nonzero magnetic quantum numbers ($m_l=\pm1$ for $(p_x,p_y)$ and $m_l=\pm 2$ for $(d_{x^2-y^2}, d_{xy})$). In the $D_{3d}$ case relevant to 1T-TMDs, the low-energy doublet at $\Gamma$ is contributed by either the chalcogen $(p_x,p_y)$ orbitals in the valence band or the metal $(d_{x^2-y^2}, d_{xy})$ orbitals in the conduction band. Projecting $H_{\mathrm{SOC}}$ onto this isolated two-dimensional IRREP without SOC, only the diagonal $L_z S_z$ term survives to leading order and yields an Ising-type SOC that splits. In $D_{3h}$, where inversion symmetry is absent, metal $(d_{x^2-y^2}, d_{xy})$ and chalcogen $(p_x,p_y)$ basis states can mix within the same 2D IRREP. Nonetheless, the horizontal mirror $\mathcal M_z$ constrains the projected SOC on $E'$-IRREP manifold to be Ising-like and supports an emergent spin-\Uone conservation in the low-energy continuum model. Crucially, these $\mathrm{SU}(2)$-breaking (often effectively \Uone-preserving) SOC terms are inherited by the moir\'e continuum model and can facilitate the formation of isolated topological bands upon twisting.

From this analysis, we propose that the presence of a two-dimensional IRREP without SOC at $\Gamma$ is a key ingredient for realizing nontrivial $\Gamma$-valley moir\'e bands. Because monolayer and untwisted bilayer calculations are significantly less computationally expensive than fully relaxed twisted-bilayer simulations, this criterion provides a practical pre-screening tool to identify candidate materials capable of hosting topological moir\'e physics.

\begin{table}[htbp]
\centering
\begin{minipage}[t]{0.48\textwidth}
    \centering
    \vspace{5pt}
    \begin{tabular}{l c c c c l}
    \toprule
    $D_{3d}$ (1T-phase) & $E$ & $C_{3z}$ & $C_{2x}$ & $\mathcal I$ & Basis ($s,p,d$) \\
    \midrule
    $A_{1g}$ & $1$ & $1$ & $1$ & $1$ & $s$, $d_{z^2}$ \\
    $A_{2g}$ & $1$ & $1$ & $-1$ & $1$ &  \\
    $E_g$    & $2$ & $-1$ & $0$ & $2$ & $(d_{x^2-y^2}, d_{xy})$,  $(d_{xz}, d_{yz})$ \\
    % \midrule
    $A_{1u}$ & $1$ & $1$ & $1$ & $-1$ & \\
    $A_{2u}$ & $1$ & $1$ & $-1$ & $-1$ & $p_z$ \\
    $E_u$    & $2$ & $-1$ & $0$ & $-2$ & $(p_x, p_y)$ \\
    \bottomrule
    \end{tabular}
\end{minipage}
\hfill
\begin{minipage}[t]{0.48\textwidth}
    \centering
    \vspace{5pt}
    
    \begin{tabular}{l c c c c l}
    \toprule
    $D_{3h}$ (1H-phase) & $E$ & $C_{3z}$ & $C_{2x}$ & $\sigma_h$ & Basis ($s,p,d$) \\
    \midrule
    $A'_1$ & $1$ & $1$ & $1$ & $1$ & $s$, $d_{z^2}$ \\
    $A'_2$ & $1$ & $1$ & $-1$ & $1$ & \\
    $E'$   & $2$ & $-1$ & $0$ & $2$ & $(p_x, p_y)$,  $(d_{x^2-y^2}, d_{xy})$ \\
    $A''_1$& $1$ & $1$ & $1$ & $-1$ &\\
    $A''_2$& $1$ & $1$ & $-1$ & $-1$ & $p_z$ \\
    $E''$  & $2$ & $-1$ & $0$ & $-2$ & $(d_{xz}, d_{yz})$ \\
    \bottomrule
    \end{tabular}
\end{minipage}
\caption{Character tables for $D_{3d}$ (1T-phase) and $D_{3h}$ (1H-phase) point groups with orbital basis functions.}
\label{app:tab:combined_point_groups}
\end{table}

\subsection{\Uone-symmetric $\Gamma$-valley moire systems with different symmetry space groups}\label{app:sec: moire systems with different symmetry space groups}
Based on the above discussion, we now analyze the topology of a $\Gamma$-valley moir\'e system with spin-\Uone inherited from the monolayer. The twisted bilayer belongs to the layer group $p321$ or $p312$ with $\mathcal T$, depending on whether it is in the AA- or AB-stacked configuration. A generic twisted multilayer configuration belongs to the layer group $p3$. In the presence of spin-\Uone symmetry, the topology of an isolated band can be diagnosed from the spin Chern number $C_s$ using \cref{app:eqn:spin chern number formula}, which is determined by the $C_3$ eigenvalues $\xi_{\vec{k},s}$ at $\vec{k}\in\{\Gamma_M,K_M,K'_M\}$ with $s=\uparrow,\downarrow$. We assume that turning on the moir\'e potential does not result in level crossing at $\Gamma_M$, so the Kramers pair at moir\'e $\Gamma_M$ nearest to the charge neutrality is adiabatically connected to the one of the untwisted multilayer at $\Gamma$. If this local Kramers doublet has quantum number $J_z=\pm m_J$ for the total angular momentum, its $C_3$ eigenvalue are $\xi_{\Gamma_M,\uparrow}=e^{- i\frac{2\pi}{3}m_J}$ and $\xi_{\Gamma_M,\downarrow}=e^{i\frac{2\pi}{3}m_J}$. In the following sections, we discuss the possible value of $C_3$ eigenvalues at at $K_M$ and $K'_M$ under different conditions for different layer groups, and hence determine the topology of the system. Finally, because of the strong interlayer coupling in $\Gamma$-valley systems, we transform from the top/bottom layer basis, denoted by $l$, to the layer-hybridized bonding/antibonding basis, denoted by $\lambda=\pm1$.

\subsubsection{$\Gamma$-valley moire system I: layer group $p3$ }\label{app:subsec:triple model}
We begin with the moir\'e system that belongs layer group $p3$ with $\mathcal T$. Since we are interested in the isolated band set closest to the charge neutrality, we truncate to the three nearest $\vec Q$-sites around $K_M$ and $K'_M$,
\begin{equation}\label{app:eqn:triple Q basis}
\begin{aligned}
    \vQ_{K_M}^{(1)} = \vec{b}_{M,1},\qquad &\vQ_{K_M}^{(2)} = \vec{b}_{M,2},\qquad \vQ_{K_M}^{(3)} = \vec{0},\\
    \vQ_{K'_M}^{(1)} = -\vec{b}_{M,1},\qquad &\vQ_{K'_M}^{(2)} = -\vec{b}_{M,2},\qquad \vQ_{K'_M}^{(3)} = \vec{0}.\\
\end{aligned}
\end{equation}
For simplicity, we focus on the \textit{single-branch and single-orbital} limit in this section. In this limit, the coupling between different layer-hybridized branches can be neglected, and the low-energy subspace is dominated by either the bonding/antibonding branch of one effective orbital with $J_z=\pm m_J$. We may therefore suppress the branch ($\lambda$) and orbital ($\alpha$) indices. The basis of $h(K_M)$ and $h(K'_M)$ is then
\begin{equation}
\begin{aligned}
    &\{\cre{c}{\vQ^{(1)},\uparrow}, \cre{c}{\vQ^{(2)},\uparrow},\cre{c}{\vQ^{(3)},\uparrow},\cre{c}{\vQ^{(1)},\downarrow}, \cre{c}{\vQ^{(2)},\downarrow},\cre{c}{\vQ^{(3)},\downarrow}\} \quad \text{For } K_M,\\
    &\{\cre{c}{\vQ'^{(1)},\uparrow}, \cre{c}{\vQ'^{(2)},\uparrow},\cre{c}{\vQ'^{(3)},\uparrow},\cre{c}{\vQ'^{(1)},\downarrow}, \cre{c}{\vQ'^{(2)},\downarrow},\cre{c}{\vQ'^{(3)},\downarrow}\}  \quad \text{For } K'_M.\\
\end{aligned}
\end{equation}
For brevity, we write $\vQ_{K_M}^{(i)}\to\vQ^{(i)}$ and $\vQ_{K'_M}^{(i)}\to\vQ'^{(i)}$. Consequently, the three $\vQ^{(i)}$ form a $C_3$ ``triple'' and generate three states at $K_M$ in each spin sector with eigenvalues$\xi_{K_M,\uparrow}=e^{- i\frac{2\pi}{3}(m_J+\Pi_z)}$ and $\xi_{K_M,\downarrow}=e^{-i\frac{2\pi}{3}(-m_J+\Pi_z)}$, where $\Pi_z\in\{-1,0,1\}$ denotes the $\vQ$-lattice angular momentum. 

\begin{figure}[!t]
    \centering
    \includegraphics[width=\textwidth]{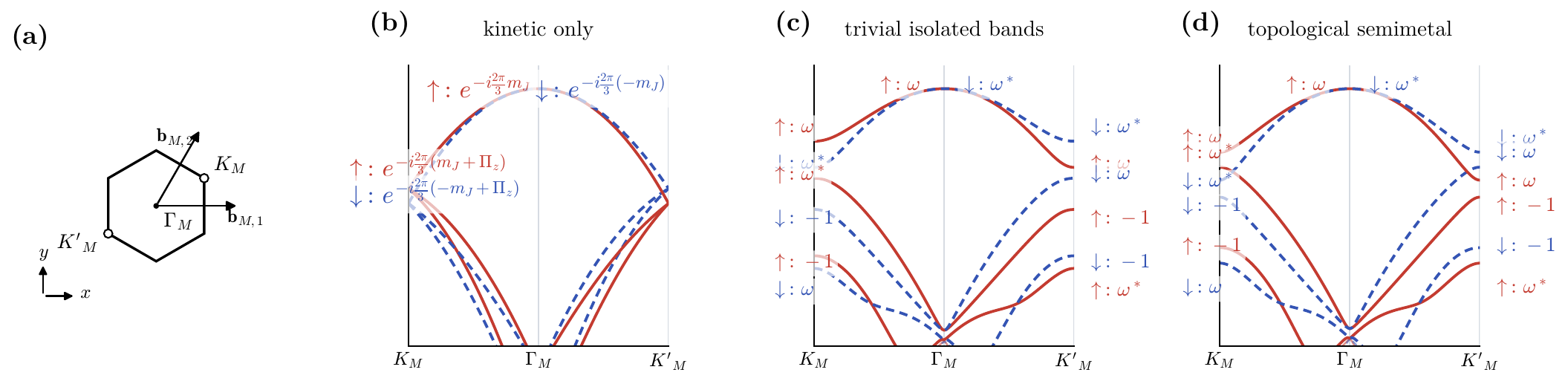}
    \subfloat{\label{app:fig:fig-P3-SBSO:a}}
    \subfloat{\label{app:fig:fig-P3-SBSO:b}}
    \subfloat{\label{app:fig:fig-P3-SBSO:c}}
    \subfloat{\label{app:fig:fig-P3-SBSO:d}}
    \caption{First-harmonic continuum model of layer group $p3$ in the \emph{single-orbital-single-branch} limit. (a) Moir\'e Brillouin zone. (b) Band structure from the kinetic term \cref{app:eqn:P3-SBSO-kinetic} only, with $m=-0.2$ and $d_1=d_2=0.3$. (c) Trivial isolated bands of the continuum Hamiltonian \cref{app:eqn:moire_momentum_space_bonding_basis}, with $m=-0.2$, $d_1=d_2=0.6$, $V_1=0.2$, and $V_2=-0.2$. (d) Topological semimetal of the same continuum Hamiltonian, with $m=-0.2$, $d_1=d_2=0.6$, $V_1=0.15$, and $V_2=-0.2$. Spin-up and spin-down bands are shown by red solid and blue dashed curves, respectively. The displayed $C_3$ eigenvalues are shown in matching colors. In panels (c) and (d), they are obtained by diagonalizing the $C_3$ representation matrix in the numerically degenerate subspaces and are identified using $m_J=-\frac{1}{2}$, consistent with \cref{app:eqn:triple model for P3,app:eqn:ordering at K_M for P3,app:eqn:ordering at K'_M for P3}. }
    \label{app:fig:fig-P3-SBSO}
\end{figure}

We now derive how the corresponding states at $K_M$ and $K'_M$ are related. We decompose Hamiltonian into a $\vk$-dependent $\vQ$-diagonal part $W_{\vQ}(\vk)$ and a $\vk$-independent moir\'e-potential part $T_{\vQ_i,\vQ_j}$,
\begin{equation}\label{app:eqn:moire_momentum_space_bonding_basis}
    \left[h_{\vQ_i,\vQ_j}(\vk)\right]_{s}= \left[W_{\vQ_i}(\vk)\right]_{s}\delta_{\vQ_i,\vQ_j}+\left[T_{\vQ_i,\vQ_j}\right]_{s}.
\end{equation}
Spin-\Uone symmetry forbids inter-spin mixing, so the spin label $s$ only distinguishes the two decoupled spin sectors. We first turn off the moir\'e potential and analyze $W_{\vQ}(\vk)$. Keeping the symmetry-allowed terms up to cubic order in momentum,
\begin{equation}\label{app:eqn:P3-SBSO-kinetic}
    \left[W(\vec{k})\right]_{\vQ}=\frac{(\vec{k}-\vQ)^2}{2m}s_0+\frac{d_1}{2}\left[(\vec{k}_+-\vec{Q}_+)^3+(\vec{k}_--\vec{Q}_-)^3\right]s_z-i\frac{d_2}{2}\left[(\vec{k}_+-\vec{Q}_+)^3-(\vec{k}_--\vec{Q}_-)^3\right]s_z,
\end{equation}
where $\vec{k}_\pm=k_x\pm ik_y$ and $\vQ_\pm=Q_x\pm iQ_y$. As shown in \cref{app:fig:fig-P3-SBSO:a}, within a fixed spin sector the three nearest $\vQ$-sites at $K_M$ (and likewise at $K'_M$) are degenerate before the moir\'e potential is turned on. The two spin sectors are generally split in energy by the spin-dependent terms in \cref{app:eqn:P3-SBSO-kinetic}.

Next we include the moir\'e potential. Since the matrix element $[T_{\vQ_i,\vQ_j}]_s$ depends only on the momentum transfer $\vQ_i-\vQ_j$, and because the $\vQ$ meshes around $K_M$ and $K'_M$ are related by inversion of momentum, the hopping from $\vQ^{(i)}_{K_M}$ to $\vQ^{(i+1\,\mathrm{mod}\,3)}_{K_M}$ is equal to the hopping from $\vQ'^{(i+1\,\mathrm{mod}\,3)}_{K'_M}$ to $\vQ'^{(i)}_{K'_M}$, as shown in \cref{fig:fig-SBSO-SB-SO}:
\begin{equation}\label{app:eqn:relation between K and Kprim in SBSO}
    [T_{\vQ^{(i)},\vQ^{(i+1\,\mathrm{mod}\,3)}}]_s
    =
    [T_{\vQ'^{(i+1\,\mathrm{mod}\,3)},\vQ'^{(i)}}]_s.
\end{equation}
By Hermiticity,
\begin{equation}\label{app:eqn:hermitioan in SBSO}
    [T_{\vQ'^{(i+1\,\mathrm{mod}\,3)},\vQ'^{(i)}}]_s
    =
    [T_{\vQ'^{(i)},\vQ'^{(i+1\,\mathrm{mod}\,3)}}]_s^{*}.
\end{equation}
Therefore, within a fixed spin sector, the moir\'e potentials at $K_M$ is the complex conjugate of that at $K_M$,
\begin{equation}
    [T_{\vQ^{(i)},\vQ^{(i+1\,\mathrm{mod}\,3)}}]_s
    =
    [T_{\vQ'^{(i)},\vQ'^{(i+1\,\mathrm{mod}\,3)}}]_s^{*}.
\end{equation}
The same conclusion continues to hold on any $\vQ$ mesh, provided the meshes around $K_M$ and $K'_M$ are exchanged by momentum inversion and each remains $C_{3z}$-invariant about its valley center.

Restricting to the three nearest $\vQ$-sites, we may write the effective Hamiltonians as
\begin{equation}\label{app:eqn:triple model for P3}
[h(K_M)]_s=\left(\begin{array}{ccc}
     \varepsilon_{1,s} & t & t^*  \\
     t^*&  \varepsilon_{1,s} & t\\
     t &t^* &  \varepsilon_{1,s}
\end{array}\right),\qquad 
[h(K'_M)]_s=\left(\begin{array}{ccc}
     \varepsilon_{2,s} & t^* & t  \\
     t&  \varepsilon_{2,s} & t^*\\
     t^* &t &  \varepsilon_{2,s}
\end{array}\right),
\end{equation}
where the onsite energies at $K_M$ and $K'_M$ may in general differ. The representation matrix of $C_{3z}$ on the three-site $\vQ$ subspace is
\begin{equation}
    D_{s}(C_{3z})=e^{-i\frac{2\pi}{3}\sigma m_J}\left(\begin{array}{ccc}
     0 & 1 & 0  \\
     0&  0 & 1\\
     1 &0 &  0
\end{array}\right),
\end{equation}
where $\sigma=+$ when $s=\uparrow$ and $\sigma=-$ when $s=\downarrow$. Diagonalizing \cref{app:eqn:triple model for P3}, the threefold multiplet at $K_M$ splits into
\begin{equation}\label{app:eqn:ordering at K_M for P3}
\begin{aligned}
    &E_{1,s}(K_M)= \varepsilon_{1,s}+2\Re[t],\qquad \xi_{K_M,s}^{(1)}=e^{-i \frac{2\pi}{3}\sigma m_J},\\
    &E_{2,s}(K_M)= \varepsilon_{1,s}-\Re[t]-\sqrt{3}\Im[t],\qquad \xi_{K_M,s}^{(2)}=e^{-i\frac{2\pi}{3} (\sigma m_J-1)},\\
    &E_{3,s}(K_M)= \varepsilon_{1,s}-\Re[t]+\sqrt{3}\Im[t],\qquad \xi_{K_M,s}^{(3)}=e^{-i\frac{2\pi}{3} ( \sigma m_J+1)},
\end{aligned}
\end{equation}
while for $K'_M$, we have 
\begin{equation}\label{app:eqn:ordering at K'_M for P3}
\begin{aligned}
    &E_{1,s}(K'_M)= \varepsilon_{2,s}+2\Re[t],\qquad \xi_{K'_M,s}^{(1)}=e^{-i \frac{2\pi}{3}\sigma m_J},\\
    &E_{2,s}(K'_M)= \varepsilon_{2,s}-\Re[t]-\sqrt{3}\Im[t],\qquad \xi_{K'_M,s}^{(2)}=e^{-i\frac{2\pi}{3} (\sigma m_J+1)},\\
    &E_{3,s}(K'_M)= \varepsilon_{2,s}-\Re[t]+\sqrt{3}\Im[t],\qquad \xi_{K'_M,s}^{(3)}=e^{-i\frac{2\pi}{3} (\sigma m_J-1)}.
\end{aligned}
\end{equation}
Hence, within the same spin sector, the $j$-th states at $K_M$ and $K'_M$ always carry opposite $\vQ$-lattice angular momentum, and therefore 
\begin{equation}\label{app:eqn:constraints from SBSO}
\xi^{(j)}_{K_M,s}\xi^{(j)}_{K'_M,s}\large|_{j=1}^3=e^{-i\frac{4\pi}{3}\sigma m_J}.
\end{equation}
Equivalently, the product $\xi^{(j)}_{K_M,s}\xi^{(j)}_{K'_M,s}$ is independent of the lattice angular momentum and depends only on $m_J$. Physically, this happens because the hoppings on the $\{\vQ^{(i)}\}$ mesh around $K_M$ and on the $\{\vQ'^{(i)}\}$ mesh around $K'_M$ circulate in opposite directions, so the two valleys carry opposite $\vQ$-lattice angular momentum.

For the isolated set, the two states at $K_M$ must come from different spin sectors. Otherwise, one of them would have to connect to a lower Kramers pair at $\Gamma_M$, and the set could not be isolated. Within each spin sector, the isolated set at $K_M$ and $K'_M$ therefore corresponds to the same split branch $j$, as shown in \cref{app:fig:fig-P3-SBSO:b,app:fig:fig-P3-SBSO:c}. Using \cref{app:eqn:constraints from SBSO} together with $\xi_{\Gamma_M,s}=e^{-i\frac{2\pi}{3}\sigma m_J}$, the $C_3$ indicator formula gives
\begin{equation}
    C_s=\left[3\left(\sigma m_J+\frac{1}{2}\right)\right]\bmod 3=0,
\end{equation}
where $m_J$ for a spinful orbital is always a half-integer. Hence, in the \emph{single-orbital-single-branch} limit, the $C_3$ symmetry indicator is always trivial for the topmost isolated band set.

For the numerical illustration in \cref{app:fig:fig-P3-SBSO}, we evaluate the band structure using the full continuum Hamiltonian in \cref{app:eqn:moire_momentum_space_bonding_basis}. Up to first harmonic, the moir\'e potential is taken to be
\begin{equation}\label{app:eqn:P3-SBSO-moire potential}
    T_{\vQ,\vQ'}=\sum_{j=1}^3\left(V_1\delta_{\vQ,\vQ'\pm\boldsymbol{g}_j}\mp iV_2\delta_{\vQ,\vQ'\pm\boldsymbol{g}_j}\right)s_0.
\end{equation}
The $C_3$ eigenvalues displayed in \cref{app:fig:fig-P3-SBSO:c,app:fig:fig-P3-SBSO:d} are obtained by diagonalizing the $C_3$ representation matrix on the numerically degenerate subspaces of the continuum Hamiltonian and then identifying the result using $m_J=-\frac{1}{2}$ as an example.

\subsubsection{$\Gamma$-valley moire system II: layer group $p321$}\label{app:sec:p3211' moire system}

\begin{figure}[!t]
    \centering
    \includegraphics[width=\textwidth]{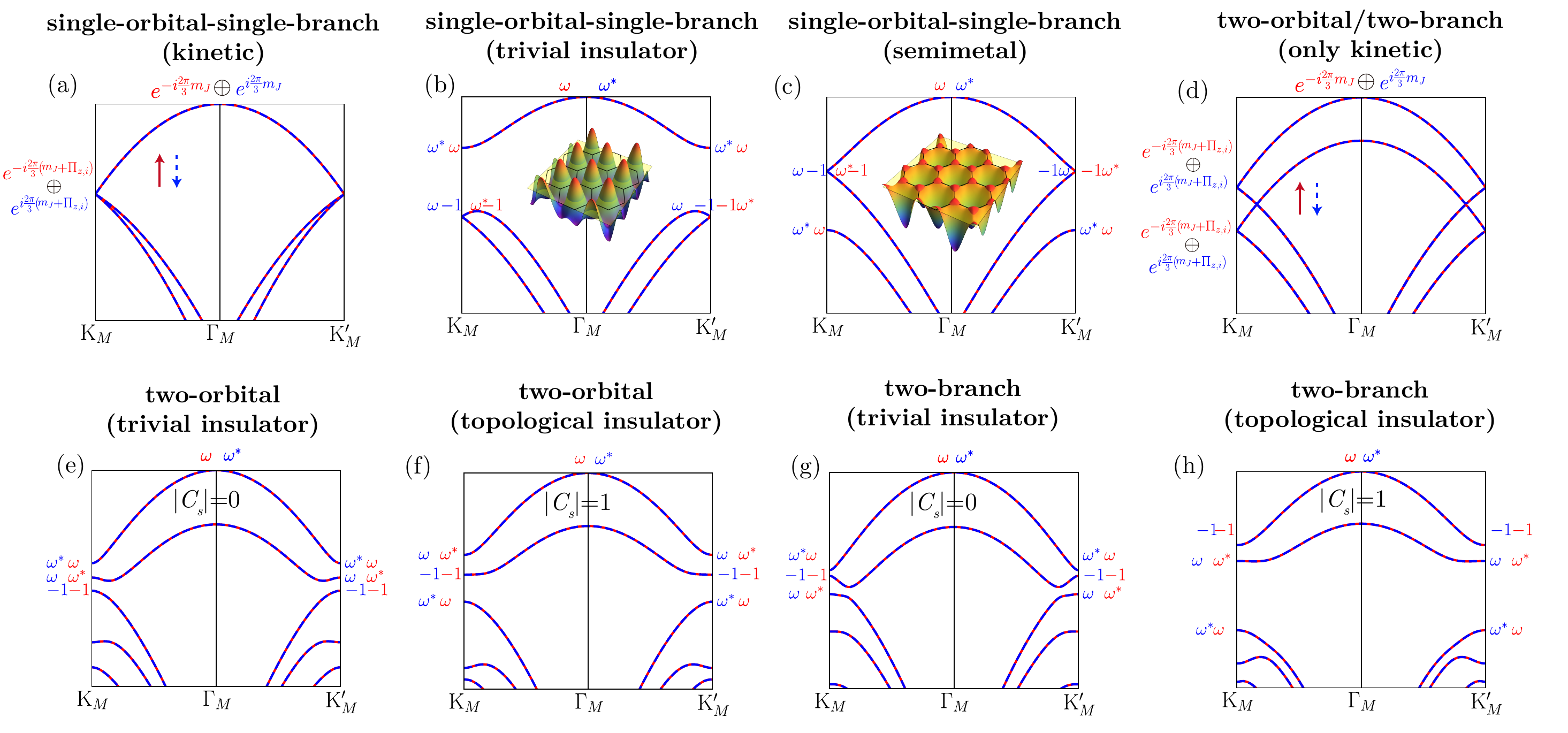}
     \subfloat{\label{app:fig:P321-SBSO:a}}
     \subfloat{\label{app:fig:P321-SBSO:b}}
     \subfloat{\label{app:fig:P321-SBSO:c}}
     \subfloat{\label{app:fig:P321-SBSO:d}}
     \subfloat{\label{app:fig:P321-SBSO:e}}
     \subfloat{\label{app:fig:P321-SBSO:f}}
     \subfloat{\label{app:fig:P321-SBSO:g}}
     \subfloat{\label{app:fig:P321-SBSO:h}}
    \caption{
    First harmonic continuum model of layer group $p321$ in (a-c) \emph{single-orbital-single-branch} limit, (d-f) single branch (SB) limit and (g,h) single orbital (SO) limit. (a) Consider only the kinetic part with $m=-0.2$ and $d_1=0.2$. (b) Trivial isolated bands when including moir\'e potential with $V_1=0.2$. (c) Topological semimetal when including moir\'e potential with $V_1=-0.2$. (d) Consider only the kinetic part with $a_1=0.2$, $m_1=-0.2$ in \cref{app:eqn:P321 SB limit} or $a'_1=0.2$, $m'_1=-0.2$ in \cref{app:eqn:P321 SO limit}. (e) Trivial insulating band structure when including moir\'e potential with $c_1=-c_2=0.1$ in SB limit. (f) Topological insulating band structure when including moir\'e potential with $c_1=c_2=-0.1$ in SB limit. (g) Trivial insulating band structure with $a'_1=0.2$, $m'_1=-0.2$, $b'_8=-0.4$,  $c'_1=0.1$ and $c'_8=-0.2$ in SO limit. (h) Topological insulating band structure with $a'_1=0.2$, $m'_1=-0.2$, $b'_8=-0.4$,  $c'_1=-0.1$ and $c'_8=-0.2$ in SO limit. Energy bands with spin-up and spin-down are denoted with red solid and blue dashed lines, respectively. The $C_3$ eigenvalue for spin-up and spin-down states is written in red and blue, respectively. Without loss of generality, we take $m_J=-\frac{1}{2}$ when labeling the $C_3$ eigenvalue. Insets in (b, c) are the corresponding moir\'e potential in \emph{single-orbital-single-branch} limit.}
    \label{app:fig:fig-P321-SBSO}
\end{figure}

Following the discussion of the layer group $p3$, we now turn to the spin-\Uone-symmetric moir\'e system with layer group $p321$, generated by $\{C_{3z},C_{2y}\}$, and again begin from the \emph{single-orbital-single-branch} limit. We first switch off the moir\'e potential and retain terms up to $O(\mathbf{k}^4)$ for $\left[W(\vec{k})\right]_{\vQ}$,
\begin{equation}\label{app:eqn:P321-SBSO-kinetic}
    \left[W(\vec{k})\right]_{\vQ}=\frac{(\vec{k}-\vQ)^2}{2m}s_0+\frac{d_1}{2}\left[(\vec{k}_+-\vec{Q}_+)^3+(\vec{k}_--\vec{Q}_-)^3\right]s_z,
\end{equation}
 Compared to \cref{app:eqn:P3-SBSO-kinetic}, $d_2$ term is forbidden by $C_{2y}$. Due to the presence of $C_{2y}$ and spin-\Uone symmetry, the dispersion along $K'_M-\Gamma_M-K_M$ is spin-degenerate, and the top six spinful states at $K_M/K'_M$ are degenerate as shown in \cref{app:fig:P321-SBSO:a}. Now we turn on the moir\'e potential that splits the six-fold degenerate state. Without loss of generality, we consider moir\'e potential up to first harmonic, 
 \begin{equation}\label{app:eqn:P321-SBSO-moire potential}
    T_{\vQ,\vQ'} =\sum_{j=1}^3V_1\delta_{\vQ,\vQ'\pm\boldsymbol{g}_j}s_0.
\end{equation} 
In the basis of the three nearest $\vQ$-sites within the same spin sector [\cref{app:eqn:triple Q basis}], $[h(K_M)]_s$ and $[h(K'_M)]_s$ reads,
\begin{equation}
    h_s(K_M)=h_s(K'_M)=\left(\begin{array}{ccc}
        \varepsilon &  V_1& V_1\\
        V_1 &\varepsilon &V_1\\
        V_1 & V_1 &\varepsilon
    \end{array}\right).
\end{equation}
In this three-site model, there is an emergent symmetry $\bar C_{2y}$ represented by
\begin{equation}
    D_s(\bar{C}_{2y})= \left(\begin{array}{ccc}
     0 & 0 & 1  \\
     0&  1 & 0\\
     1 &0 &  0
\end{array}\right).
\end{equation}
$\bar C_{2y}$ acts within a fixed spin sector and therefore differs from the ordinary $C_{2y}$ operation, which flips spin.
Consequently, the sixfold-degenerate state splits into a twofold-degenerate spinful state
\begin{equation}\label{app:eqn:double degenetate in P321 SBSO}
E_{1,s}=\varepsilon+2V_1,
\qquad
\xi_{s}^{(1)}=e^{-i\frac{2\pi}{3}\sigma m_J},
\qquad s=\uparrow,\downarrow,
\end{equation}
and a fourfold-degenerate spinful state
\begin{equation}\label{app:eqn:four-fold degenetate in P321 SBSO}
E_{2,s}=E_{3,s}=\varepsilon-V_1,
\qquad
\xi_{s}^{(2,3)}=
e^{-i\frac{2\pi}{3}(\sigma m_J-1)}
\oplus
e^{-i\frac{2\pi}{3}(\sigma m_J+1)},
\qquad s=\uparrow,\downarrow,
\end{equation}
as shown in \cref{app:fig:P321-SBSO:b,app:fig:P321-SBSO:c}. The four-fold degeneracy at $K_M$ and $K'_M$ is protected jointly by $C_{2y}$ and the emergent $\bar C_{2y}$ in the presence of spin-\Uone. Take $K_M$ as an example, the twofold degeneracy between the two spin sectors is enforced by the exact $C_{2y}$ together with spin-\Uone, which relates the state with $\xi_{K_M,s}=e^{-i\frac{2\pi}{3}(\sigma m_J-1)}$ to the opposite-spin state with $\xi_{K_M,\bar s}=e^{-i\frac{2\pi}{3}(\bar\sigma m_J+1)}$, where $\bar s$ and $\bar \sigma=-\sigma$ denote the opposite spin sector. The additional doubling within each spin sector is enforced by the emergent $\bar C_{2y}$ symmetry, which acts within a fixed spin sector and relates the two states with $\vQ$-lattice angular momentum $\Pi_z=\pm1$, namely the pair with $C_3$ eigenvalues $e^{-i\frac{2\pi}{3}(\sigma m_J-1)}$ and $e^{-i\frac{2\pi}{3}(\sigma m_J+1)}$.

The existence of the fourfold degenerate state at $K_M$ and $K'_M$ does not rely on truncating the moir\'e potential to first harmonics, but persists for arbitrary harmonics in the \emph{single-orbital-single-branch} limit because the constraint from \cref{app:eqn:constraints from SBSO} fixes the same-spin relation between $K_M$ and $K'_M$. Concretely, suppose a spinful doublet at $K_M$ has the $C_3$ content
\begin{equation}\label{app:eqn:P321 doublet 1 in SBSO limit}
\xi^{(j)}_{K_M,s}\oplus \xi^{(j)}_{K_M,\bar s}
=
e^{-i\frac{2\pi}{3}(\sigma m_J+1)}
\oplus
e^{-i\frac{2\pi}{3}(\bar\sigma m_J-1)},
\end{equation}
then \cref{app:eqn:constraints from SBSO} requires the corresponding same-spin state at $K'_M$ to carry
\begin{equation}\label{app:eqn:P321 SBSO constraints in SBSO limit}
\xi^{(j)}_{K'_M,s}\oplus \xi^{(j)}_{K'_M,\bar s}
=
e^{-i\frac{2\pi}{3}(\sigma m_J-1)}
\oplus
e^{-i\frac{2\pi}{3}(\bar\sigma m_J+1)}.
\end{equation}
Because $C_{2y}$ still enforces spin degeneracy at $K'_M$, \cref{app:eqn:P321 SBSO constraints in SBSO limit} is itself a spinful doublet. Applying time-reversal symmetry $\mathcal T$ maps this doublet at $K'_M$ back to a second spinful doublet at $K_M$ with the same energy,
\begin{equation}\label{app:eqn:P321 doublet 2 in SBSO limit}
\xi^{(j')}_{K_M,s}\oplus \xi^{(j')}_{K_M,\bar s}
=
e^{-i\frac{2\pi}{3}(\sigma m_J-1)}
\oplus
e^{-i\frac{2\pi}{3}(\bar\sigma m_J+1)}.
\end{equation}
Therefore, the two spinful doublets of $j$-th state \cref{app:eqn:P321 doublet 1 in SBSO limit} and of $j'$-th state in \cref{app:eqn:P321 doublet 2 in SBSO limit}
must be degenerate, and together form a fourfold degenerate state at $K_M$.

As shown in \cref{app:eqn:four-fold degenetate in P321 SBSO,app:eqn:double degenetate in P321 SBSO}, the ordering between the split twofold-degenerate state and the fourfold-degenerate state is determined by the sign of $V_1$. When $V_1<0$, the fourfold-degenerate state lies higher in energy, whereas when $V_1>0$, the twofold-degenerate state lies higher. This can be understood from the real-space structure of the moir\'e potential together with the elementary band representations (EBRs) of layer group $p321$ (No.~68) with $\mathcal T$, listed in \cref{app:tab:P321_EBR_and_irreps}.
For the first-harmonic moir\'e potential in \cref{app:eqn:P321-SBSO-moire potential}, when $V_1<0$, the potential is maximized at the $2c$ position. The topmost band set can then be understood as placing an $s$-like ($p$-like) orbital with $|m_J|=1/2$ ($|m_J|=3/2$) at $2c$, which yields two copies of a spinless Dirac band as shown in \cref{app:fig:P321-SBSO:c}. The corresponding band representation is $\ce{^1\mathrm{\bar{E}}}\ce{^2\mathrm{\bar{E}}}@2c$ ($\mathrm{\bar{E}}\mathrm{\bar{E}}@2c$), and the fourfold-degenerate state at $K_M$ carries $\xi_{K_M,s}=e^{-i\frac{2\pi}{3}(\sigma m_J-1)}
\oplus
e^{-i\frac{2\pi}{3}(\sigma m_J+1)}$.
By contrast, when $V_1>0$, the potential is maximized at the $1a$ position. The topmost band set can be induced by placing a $s$-like ($p$-like) orbital with $|m_J|=1/2$ ($|m_J|=3/2$) at $1a$, which gives us a trivial isolate band as shown in \cref{app:fig:P321-SBSO:b}. The corresponding band representation is $\mathrm{\bar{E}}_1@1a$ ($\ce{^1\mathrm{\bar{E}}}\ce{^2\mathrm{\bar{E}}}@1a$), where a twofold-degenerate state at $K_M$ carries
$\xi_{K_M,s}=e^{-i\frac{2\pi}{3}\sigma m_J}$. 
In moir\'e systems, we assume that the strength of the moir\'e potential can be roughly reflected by the spatial variation of the interlayer hybridization. Regions with smaller interlayer distance typically have stronger interlayer hybridization and hence correspond to maxima of the effective moir\'e potential. For twisted bilayer AB-stacked \ch{ZrS2} and \ch{SnSe2}, which belong to layer group $p321$, the interlayer distance is minimized at the $2c$ position as shown in Fig.S11 and Fig.S16 in Ref.\cite{cualuguaru2024new}. Consequently, in both materials the topmost band set is expected to originate from the Dirac-like band.

\begin{table}[!h]
\centering
\scriptsize
\renewcommand{\arraystretch}{1.4}
\setlength{\tabcolsep}{4pt}
\begin{tabular}{|c|c|c|c|c|c|}
\hline
\multicolumn{6}{|c|}{\textbf{(a) Elementary band representations}} \\ \hline
Wyckoff pos. & \multicolumn{2}{c|}{1a $(0,0)$} & \multicolumn{3}{c|}{2c $(1/3,2/3)+(2/3,1/3)$} \\ \hline
Site sym. & \multicolumn{2}{c|}{32} & \multicolumn{3}{c|}{3} \\ \hline
BR
& $\ce{^1\mathrm{\bar{E}}}\ce{^2\mathrm{\bar{E}}}$
& $\mathrm{\bar{E}}_1$
& \multicolumn{2}{c|}{$\mathrm{\bar{E}}\mathrm{\bar{E}}$}
& $\ce{^1\mathrm{\bar{E}}}\ce{^2\mathrm{\bar{E}}}$ \\ \hline
$\Gamma~(0,0)$
& $\bar{\Gamma}_4\bar{\Gamma}_5(2)$
& $\bar{\Gamma}_6(2)$
& \multicolumn{2}{c|}{$2\bar{\Gamma}_4\bar{\Gamma}_5(2)$}
& $2\bar{\Gamma}_6(2)$ \\ \hline
$K~(1/3,1/3)$
& $\bar{K}_4(1)\oplus \bar{K}_5(1)$
& $\bar{K}_6(2)$
& \multicolumn{2}{c|}{$2\bar{K}_6(2)$}
& $\bar{K}_4(1)\oplus \bar{K}_5(1)\oplus 2\bar{K}_6$ \\ \hline
\multicolumn{6}{c}{} \\[-2pt]
\hline
\multicolumn{6}{|c|}{\textbf{(b) Matrix representations of irreps at high-symmetry momenta}} \\ \hline
 &
  $\bar{\Gamma}_4\bar{\Gamma}_5$ &
  $\bar{\Gamma}_6$ &
  $\bar{K}_4$ &
  $\bar{K}_5$ &
  $\bar{K}_6$ \\ \hline
$E$
& $\begin{pmatrix}1&0\\0&1\end{pmatrix}$
& $\begin{pmatrix}1&0\\0&1\end{pmatrix}$
& $1$
& $1$
& $\begin{pmatrix}1&0\\0&1\end{pmatrix}$ \\ \hline
$C_{3z}$
& $\begin{pmatrix}-1&0\\0&-1\end{pmatrix}$
& $\begin{pmatrix}e^{-i\pi/3}&0\\0&e^{i\pi/3}\end{pmatrix}$
& $-1$
& $-1$
& $\begin{pmatrix}e^{-i\pi/3}&0\\0&e^{i\pi/3}\end{pmatrix}$ \\ \hline
$C_{2y}$
& $\begin{pmatrix}-i&0\\0&i\end{pmatrix}$
& $\begin{pmatrix}0&e^{-i\pi/3}\\ e^{-i2\pi/3}&0\end{pmatrix}$
& $-i$
& $i$
& $\begin{pmatrix}0&e^{-i\pi/3}\\ e^{-i2\pi/3}&0\end{pmatrix}$ \\ \hline
\end{tabular}
\caption{Elementary band representations and matrix representations of irreducible representations at high-symmetry momenta for layer group $p321$ (No.~68) with time-reversal symmetry.}
\label{app:tab:P321_EBR_and_irreps}
\end{table}

So far the discussion has been restricted to the \emph{single-orbital-single-branch} limit, where every isolated band set is topologically trivial and equivalent to an EBR. In realistic systems this limit is not always sufficient, and one may need to include both layer-hybridized branches or more than one orbital. Such inter-branch or inter-orbital coupling acts as an effective ``SOC'' term that breaks the emergent $\bar C_{2y}$ symmetry. We first consider the single-branch limit containing two effective orbitals. Without loss of generality, we assume the two effective orbitals originate from monolayer $(p_{x},p_y)$ and are characterized by total angular momentum $J_z=\pm m_J$ with $m_J=-\frac{1}{2}$ (\ie, $p_{-,\uparrow}$ and $p_{+,\downarrow}$) and $m_J=\frac{3}{2}$ (\ie, $p_{+,\uparrow}$ and $p_{-,\downarrow}$), respectively. With the basis 
\begin{equation}
    \{\cre{c}{\vk,\vQ,p_-,\uparrow},\cre{c}{\vk,\vQ,p_+,\downarrow},\cre{c}{\vk,\vQ,p_+,\uparrow},\cre{c}{\vk,\vQ,p_-,\downarrow}\},
\end{equation}
the exact symmetries of layer group $p321$ has the following representation matrices
\begin{equation}
\begin{aligned}
&D({C}_{3z})= \begin{pmatrix}
e^{-i\pi/3} & 0 & 0 & 0\\
0 & e^{i\pi/3} & 0 & 0\\
0 & 0 & -1 & 0\\
0 & 0 & 0 & -1
\end{pmatrix},\quad D({C}_{2y})= \begin{pmatrix}
0 & 1 & 0 & 0\\
-1 & 0 & 0 & 0\\
0 & 0 & 0 & 1\\
0 & 0 & -1 & 0
\end{pmatrix},\quad D(\mathcal T)=\begin{pmatrix}
0 & -1 & 0 & 0\\
1 & 0 & 0 & 0\\
0 & 0 & 0 & -1\\
0 & 0 & 1 & 0
\end{pmatrix},
\end{aligned}
\end{equation}
Under the constraints of these exact symmetries together with spin-\Uone, the $\vk$-dependent part up to $O(\vk^4)$ and moir\'e potential part up to first harmonic continuum model take the following form,
\begin{equation}\label{app:eqn:P321 SB limit}
\begin{aligned}
    \left[W(\vec{k})\right]^{\text{two-orb}}_{\vQ,\vQ}=&a_1\Gamma_{z0}+\frac{(\vec k-\vQ)^2}{2m_1}\Gamma_{00}-\frac{\left[(\vk_+-\vQ_+)^2\Gamma_++(\vk_--\vQ_-)^2\Gamma_-\right]}{4m_2}+\frac{(\vec k-\vQ)^2}{2m_3}\Gamma_{z0}\\
    &-\frac{1}{2b_1}\left[(\vec{k}_+-\vec{Q}_+)^3+(\vec{k}_--\vec{Q}_-)^3\right]\Gamma_{0z}-\frac{1}{2b_3}\left[(\vec{k}_+-\vec{Q}_+)^3+(\vec{k}_--\vec{Q}_-)^3\right]\Gamma_{zz}\\
    &+\frac{1}{2b_5}(\vec k-\vQ)^2\left[(\vk_+-\vQ_+)\tilde{\Gamma}_- +(\vk_--\vQ_-)\tilde{\Gamma}_+ \right],\\
   T_{\vQ,\vQ'}^{\text{two-orb}}=&\sum_{\alpha=\pm 1}\sum_{j=1}^3 \delta_{\vQ,\vQ'+\alpha \boldsymbol{g}_j} \left[ c_1 \Gamma_{00} - c_2 \left( z_j \Gamma_- + z_j^* \Gamma_+ \right)+c_3\Gamma_{z0}+\alpha c_8  (z_j^* \Gamma_+ - z_j \Gamma_-)  \right].\\
\end{aligned}
\end{equation}
Here $\Gamma_{ij}=\sigma_i\otimes s_j$, where $\sigma_i$ and $s_i$ are Pauli matrices acting on orbital space (distinguishing $J_z=\pm 1/2$ and $J_z=\pm 3/2$) and spin space, respectively. We also use $\Gamma_{\pm}=\Gamma_{x0}\pm i\Gamma_{yz}$, $\tilde{\Gamma}_{\pm}=\Gamma_{xz}\pm i\Gamma_{yz}$, and $z_j=e^{i\frac{2\pi}{3}(j-1)}$. The terms $\{m_2,b_5,c_2,c_8\}$ in \cref{app:eqn:P321 SB limit} break the emergent $\bar C_{2y}$ symmetry, which protects the four-fold degeneracy in the \emph{single-orbital-single-branch} limit.

Similarly, one may retain both layer-hybridized bonding and antibonding branches while staying in the single-orbital limit. We take the orbital with $J_z=\pm\frac{1}{2}$ (\ie, an $s$-like orbital). With the basis 
\begin{equation}
    \{\cre{c}{\vk,\vQ,\text{antibonding}, \uparrow},\cre{c}{\vk,\vQ,\text{antibonding},\downarrow},\cre{c}{\vk,\vQ,\text{bonding}, \uparrow},\cre{c}{\vk,\vQ,\text{bonding},\downarrow}\},
\end{equation}
the exact symmetries of layer group $p321$ has the following representation matrices
\begin{equation}
\begin{aligned}
&D({C}_{3z})= \tau_0\otimes e^{i\frac{\pi}{3}s_z},\quad D({C}_{2y})= -\tau_z\otimes is_y,\quad D(\mathcal T)=\tau_0\otimes -is_y.
\end{aligned}
\end{equation}
Here, the Pauli matrices $\tau_i$ and $s_i$ act on the branch degrees of freedom (bonding/antibonding) and spin space, respectively.
In this case, the $\vk$-dependent part up to $O(\vk^4)$ and moir\'e potential part up to first harmonic take the following form,
\begin{equation}\label{app:eqn:P321 SO limit}
\begin{aligned}
\left[W(\vec{k})\right]^{\text{two-branch}}_{\vQ,\vQ}=&a'_1\tau_z\otimes s_0+a'_2\tau_y\otimes s_z+\frac{(\vec k-\vQ)^2}{2m'_1}\tau_0\otimes s_0+\frac{(\vec k-\vQ)^2}{2m'_2}\tau_z\otimes s_0+\frac{(\vec k-\vQ)^2}{2m'_3}\tau_y\otimes s_z\\
    &+\left[(\vec{k}_+-\vec{Q}_+)^3+(\vec{k}_--\vec{Q}_-)^3\right]\left(\frac{1}{2b'_1}\tau_0\otimes s_z+\frac{1}{2b'_3}\tau_z\otimes s_z+\frac{1}{2b'_5}\tau_y\otimes s_0\right)\\
    &+\frac{i}{2b'_8}\left[(\vec{k}_+-\vec{Q}_+)^3-(\vec{k}_--\vec{Q}_-)^3\right]\tau_x\otimes s_z,\\
   T^{\text{two-branch}}_{\vQ,\vQ'}=&\sum_{\alpha=\pm 1}\sum_{j=1}^3 \delta_{\vQ,\vQ'+\alpha \boldsymbol{g}_j} \left( c'_1 \tau_0\otimes s_0 + c'_2 \tau_z\otimes s_0+c_3'\tau_y\otimes s_z+i\alpha c'_8  \tau_x\otimes s_0 \right).\\
\end{aligned}
\end{equation}
In this two-branch model, the terms $\{m'_3,b'_8,c'_3\}$ in \cref{app:eqn:P321 SO limit} break the emergent $\bar C_{2y}$ symmetry with $D(\bar{C}_{2y})=-\tau_z\otimes s_0$.

Thus, once either the two-orbital terms in \cref{app:eqn:P321 SB limit} or the two-branch terms in \cref{app:eqn:P321 SO limit} are included, the emergent $\bar C_{2y}$ symmetry is lost while the exact $C_{2y}$ and $\mathcal T$ symmetries remain. Consequently, the additional same-spin doubling is lifted, but the exact $C_{2y}$ symmetry still enforces spin degeneracy at each valley. Therefore the fourfold multiplet of the SBSO limit splits into two spinful doublets, namely \cref{app:eqn:P321 doublet 1 in SBSO limit} and \cref{app:eqn:P321 doublet 2 in SBSO limit}. Time-reversal symmetry $\mathcal T$ maps each spinful doublet at $K_M$ to a spinful doublet at $K'_M$ with the same unordered $C_3$ content. Hence an isolated split doublet can only have one of the following two $C_3$ patterns:
\begin{equation}
\begin{aligned}
\xi^{(j)}_{K_M,s}\oplus \xi^{(j)}_{K_M,\bar s}
&=
e^{-i\frac{2\pi}{3}(\sigma m_J+1)}
\oplus
e^{-i\frac{2\pi}{3}(\bar\sigma m_J-1)}
\;\Longleftrightarrow\;
\xi^{(j)}_{K'_M,s}\oplus \xi^{(j)}_{K'_M,\bar s}
=
e^{-i\frac{2\pi}{3}(\sigma m_J+1)}
\oplus
e^{-i\frac{2\pi}{3}(\bar\sigma m_J-1)},\\
\xi^{(j')}_{K_M,s}\oplus \xi^{(j')}_{K_M,\bar s}
&=
e^{-i\frac{2\pi}{3}(\sigma m_J-1)}
\oplus
e^{-i\frac{2\pi}{3}(\bar\sigma m_J+1)}
\;\Longleftrightarrow\;
\xi^{(j')}_{K'_M,s}\oplus \xi^{(j')}_{K'_M,\bar s}
=
e^{-i\frac{2\pi}{3}(\sigma m_J-1)}
\oplus
e^{-i\frac{2\pi}{3}(\bar\sigma m_J+1)}.
\end{aligned}
\end{equation}
For materials such as twisted bilayer AB-stacked \ch{ZrS2} and \ch{SnSe2}, in which the moir\'e potential is maximized at the honeycomb sites, the topmost valence-band set is therefore expected to originate from the split Dirac-like band. Since $\xi_{\Gamma_M,s}\oplus \xi_{\Gamma_M,\bar s}=e^{-i\frac{2\pi}{3}\sigma m_J}\oplus e^{-i\frac{2\pi}{3}\bar\sigma m_J}$, the topmost isolated band set can have the following irrep content:
\begin{equation}\label{app:eqn:split of 2D EBR}
\begin{aligned}
&J_z=\pm\frac{3}{2}: \qquad \{\bar{\Gamma}_4\bar{\Gamma}_5,\bar K_6\}, \\
&J_z=\pm\frac{1}{2}: \qquad \{\bar{\Gamma}_6,\bar K_6\}
\quad\text{or}\quad
\{\bar{\Gamma}_6,\bar K_4\bar K_5\}.
\end{aligned}
\end{equation}
The nonzero indicator now follows directly from the spin-resolved $C_3$ eigenvalues. For the first split doublet and a fixed spin sector $s$, the $C_3$ indicator formula gives
\begin{equation}
    e^{-i\frac{2\pi}{3}C_s}
    =
    -\,\xi_{\Gamma_M,s}\,\xi^{(j)}_{K_M,s}\,\xi^{(j)}_{K'_M,s}
    =
    -\,e^{-i2\pi\sigma m_J}e^{-i\frac{4\pi}{3}}
    =
    e^{-i\frac{4\pi}{3}},
\end{equation}
where we used that $m_J$ is half-integer, so $e^{-i2\pi\sigma m_J}=-1$. The second split doublet similarly gives
\begin{equation}
    e^{-i\frac{2\pi}{3}C_s}=e^{i\frac{4\pi}{3}}.
\end{equation}
Hence every isolated split doublet satisfies $|C_s|=1 \bmod 3$.

We now verify this analytic conclusion numerically using the two toy models defined in \cref{app:eqn:P321 SO limit,app:eqn:P321 SB limit}. For both models, the band structure in the absence of the moir\'e potential is shown in \cref{app:fig:P321-SBSO:d}. Qualitatively, it can be viewed as two copies of the \emph{single-orbital-single-branch} case shown in \cref{app:fig:P321-SBSO:a}, giving rise to the sixfold-degenerate state at $K_M$ and $K'_M$. When the moir\'e potential is maximized on the honeycomb sites, this sixfold state splits into three doubly degenerate bands, as shown in \cref{app:fig:P321-SBSO:f,app:fig:P321-SBSO:h}, and the topmost isolated band set carrying nonzero spin Chern number is topologically nontrivial. By contrast, when the moir\'e potential is maximized on the triangular sites, the topmost isolated band set is topologically trivial and admits an atomic EBR description, as shown in \cref{app:fig:P321-SBSO:e,app:fig:P321-SBSO:g}. The parameter values (${m_i,a_i,b_i,c_i}$) in \cref{app:eqn:P321 SO limit} and (${m'_i,a'_i,b'_i,c'_i}$) in \cref{app:eqn:P321 SB limit} for each subfigure are listed in the corresponding caption. Thus, our analytic results in the previous paragraph are verified by the numerical models.

\subsubsection{$\Gamma$-valley moire system III: layer group $p312$}\label{app:sec:p3121' moire system}
\begin{figure}[!t]
    \centering
    \includegraphics[width=\textwidth]{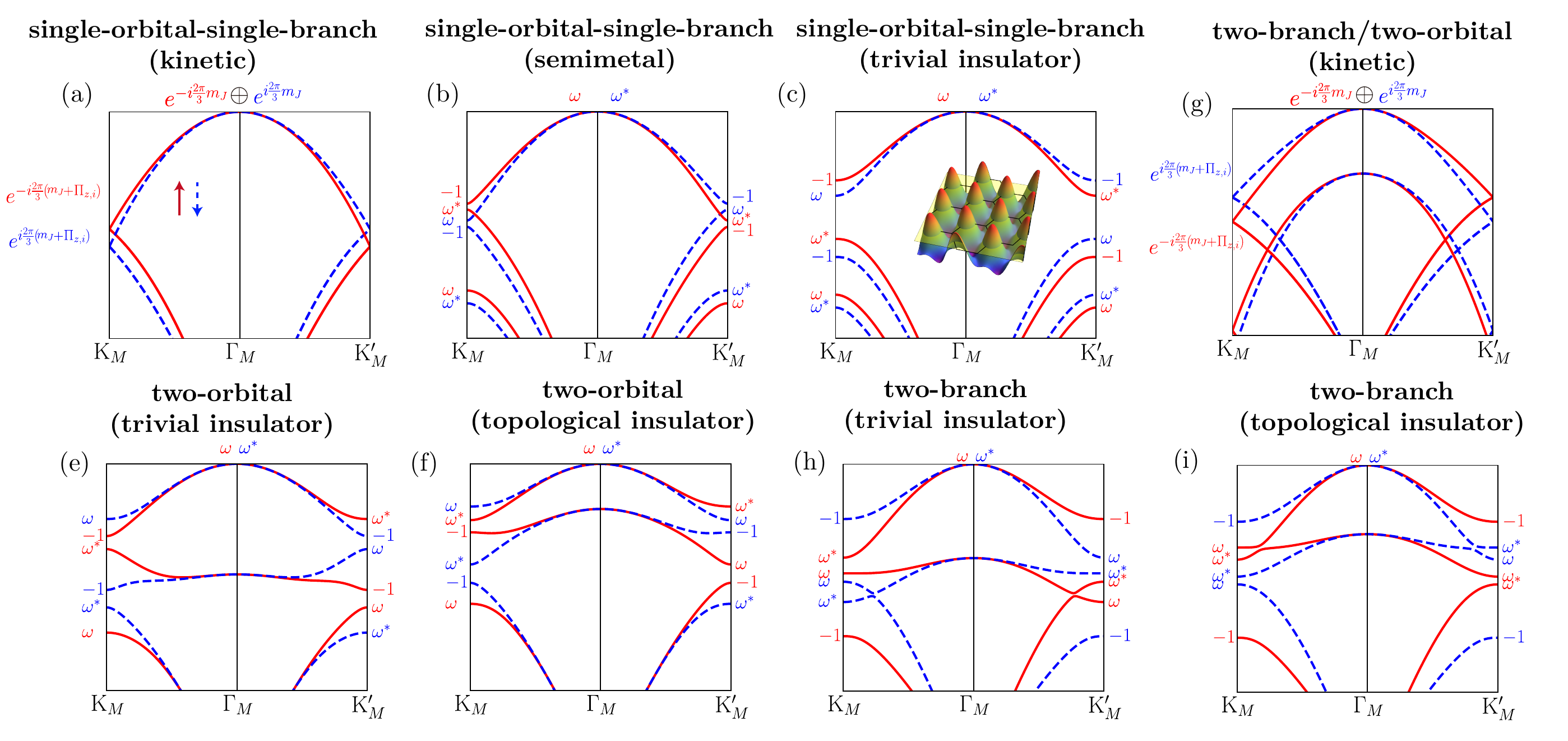}
     \subfloat{\label{app:fig:P312-SBSO:a}}
     \subfloat{\label{app:fig:P312-SBSO:b}}
     \subfloat{\label{app:fig:P312-SBSO:c}}
     \subfloat{\label{app:fig:P312-SBSO:d}}
     \subfloat{\label{app:fig:P312-SBSO:e}}
     \subfloat{\label{app:fig:P312-SBSO:f}}
     \subfloat{\label{app:fig:P312-SBSO:g}}
     \subfloat{\label{app:fig:P312-SBSO:h}}
     \subfloat{\label{app:fig:P312-SBSO:i}}
    \caption{First harmonic continuum model of layer group $p312$ in (a-c) \emph{single-orbital-single-branch} limit, (d-f) single branch (SB) limit and (g,h) single orbital (SO) limit. (a) Consider only the kinetic part with $m=-0.2$ and $d_2=0.3$. (b) Metallic bands when including moir\'e potential with $V_1=-0.2$ and $V_2=0.01$. (c) Trivial insulating bands when including moir\'e potential with $V_1=-0.2$ and $V_2=0.1$. (d) Consider only the kinetic part with $a'_1=0.14$ and $m'_1=-0.4$ and $b_4'=-0.4$ in the SO limit. (h) Trivial insulating band structure with $c'_1=-0.01$, $c'_2=-0.2$ and $c'_5=-0.04$ in SO limit. (e) Trivial insulating band structure when including moir\'e potential with $c_1=c_2=-0.2$ and $c_5=0.1$ in SB limit. (g) Topological insulating band structure when including moir\'e potential with $c_1=c_2=-0.2$, $c_5=0.1$, and modifying $a_1$ as $a_1=0.2$ in the SB limit.  (h) Topological insulating band structure with $c'_1=-0.01$, $c'_2=-0.2$ and $c'_5=-0.04$, and modifying $a'_1$ as $a'_1=0.1$ in SO limit. Energy bands with spin-up and spin-down are denoted with red solid and blue dashed lines, respectively. The $C_3$ eigenvalue for spin-up and spin-down states is written in red and blue, respectively. We take $m_J=-\frac{1}{2}$ when the moir\'e potential is included for better illustration. Insets in (c) are the corresponding moir\'e potential in \emph{single-orbital-single-branch} limit.}
    \label{app:fig:fig-P312-SBSO}
\end{figure}

In this section, we discuss the spin-\Uone-symmetric moir\'e system with time-reversal symmetric layer group $p312$, generated by $\{C_{3z},C_{2x}\}$. 
The following paragraphs are organized as follows. First, we show that the topmost isolate band in the \emph{single-orbital-single-branch} limit is required to be trivial as in the layer group $p3$ case. Second, we show that the simple relation \cref{app:eqn:constraints from SBSO} in the \emph{single-orbital-single-branch} limit is lost once two orbitals/branches are included, derive the general condition under which the topmost isolated band carries a nonzero $C_{3z}$ indicator, and identify a band inversion between two blocks with different \(\Pi_z\) as the route to tune the indicator. Finally, we use the first-harmonic model to demonstrate numerically the phase transition between a trivial and a topological isolated band.

We begin with the \emph{single-orbital-single-branch} limit. For the $\vk$-dependent part, we consider up to $O(\vec{k}^4)$, 
\begin{equation}
    \left[W(\vec{k})\right]_{\vQ}=\frac{(\vec{k}-\vQ)^2}{2m}s_0-i\frac{d_2}{2}\left[(\vec{k}_+-\vec{Q}_+)^3-(\vec{k}_--\vec{Q}_-)^3\right]s_z,
\end{equation}
the dispersion without moir\'e potential is shown in \cref{app:fig:P312-SBSO:a} with $m=-0.2$ and $d_2=0.3$. Similar to layer group $p3$ case in \cref{app:subsec:triple model}, states within the same spin sector related by $C_3$ are degenerate at $K_M$ and $K'_M$, while states with opposite spins are separated due to spin splitting. The symmetry-allowed moir\'e potential up to first harmonic reads,
\begin{equation}
    T_{\vQ,\vQ'} =\sum_{j=1}^3(V_1\delta_{\vQ,\vQ'\pm\boldsymbol{g}_j}\mp iV_2\delta_{\vQ,\vQ'\pm\boldsymbol{g}_j})s_0.
\end{equation} 
When the moir\'e potential is included, the Hamiltonian at $K_M$ and $K'_M$, written in the basis of the three nearest $\vQ$-sites, has the same form as \cref{app:eqn:triple model for P3}. Consequently, the three-fold degenerate state within the same spin sector at $K_M$ and $K'_M$ splits into three singlets unless fine tuned. Moreover, both $\mathcal T$ and $C_{2x}$ relate states at $K_M$ and $K'_M$ belonging to opposite spin sectors, and reverse the $\vQ$-lattice angular momentum $\Pi_z\in\{0,\pm1\}$ appearing in the $C_{3z}$ eigenvalues,
\begin{equation}
\begin{aligned}
\xi^{(j)}_{K_M,s}
&=
e^{-i\frac{2\pi}{3}(\sigma m_J+\Pi_z)}
\;\Longleftrightarrow\;
\xi^{(j)}_{K'_M,\bar s}
=
e^{-i\frac{2\pi}{3}(\bar\sigma m_J-\Pi_z)}.
\end{aligned}
\end{equation}
As in the layer group $p3$ case in \cref{app:subsec:triple model}, the topmost isolated band set can be separated from the lower bands only when the highest states at $K_M$ and $K'_M$ belong to different spin sectors, by the same band-connectivity argument, as shown in \cref{app:fig:P312-SBSO:c}. Otherwise, the topmost states cross the lower bands from the opposite spin sector, as shown in \cref{app:fig:P312-SBSO:b}.
Combining \cref{app:eqn:constraints from SBSO} (enforced by the $\vk$-independent moir\'e potential in the \emph{single-orbital-single-branch} limit) with the $\mathcal T$/$C_{2x}$-relation above fixes the topmost states at $K_M$ in the two spin sectors to share the same $\Pi_z$. Consequently, the isolated band set, consisting of one band from each spin sector, has the following $C_{3z}$ eigenvalues at $K_M$,
\begin{equation}
\begin{aligned}
\xi^{(j)}_{K_M,s}
&=
e^{-i\frac{2\pi}{3}(\sigma m_J+\Pi_z)}
,\qquad 
\xi^{(j)}_{K_M,\bar s}
=
e^{-i\frac{2\pi}{3}(\bar\sigma m_J+\Pi_z)}.
\end{aligned}
\end{equation}
As a result, this isolated band set carries a trivial $C_{3z}$ indicator. Besides, its possible irreps at $\Gamma_M$ and $K_M$ are
\begin{equation}
\begin{aligned}
&J_z=\pm\frac{3}{2}: \qquad \{\bar{\Gamma}_4\bar{\Gamma}_5,2\bar K_4\}\quad \text{or}\quad \{\bar{\Gamma}_4\bar{\Gamma}_5,2\bar K_5\}\quad \text{or}\quad \{\bar{\Gamma}_4\bar{\Gamma}_5,2\bar K_6\} ,\\
&J_z=\pm\frac{1}{2}: \qquad \{\bar{\Gamma}_6,\bar K_5\oplus\bar K_6\}
\quad\text{or}\quad
\{\bar{\Gamma}_6,\bar K_4\oplus\bar K_6\}\quad\text{or}\quad
\{\bar{\Gamma}_6,\bar K_4\oplus\bar K_5\},
\end{aligned}
\end{equation}
all of which are induced from EBRs of layer group $p312$, as listed in \cref{app:tab:P312_EBR_and_irreps}. Therefore, the isolated band set in layer group $p312$ is topologically trivial in the \emph{single-orbital-single-branch} limit.

\begin{table}[!h]
\centering
\scriptsize
\renewcommand{\arraystretch}{1.4}
\setlength{\tabcolsep}{4pt}

\begin{tabular}{|c|c|c|c|c|c|c|}
\hline
\multicolumn{7}{|c|}{\textbf{(a) Elementary band representations}} \\ \hline
Wyckoff pos. & \multicolumn{2}{c|}{1a} & \multicolumn{2}{c|}{1b} & \multicolumn{2}{c|}{1c} \\ \hline
Site sym. & \multicolumn{2}{c|}{32} & \multicolumn{2}{c|}{32} & \multicolumn{2}{c|}{32} \\ \hline
BR
& $\ce{^1\mathrm{\bar E}}\ce{^2\mathrm{\bar E}}$
& $\mathrm{\bar E}_1$
& $\ce{^1\mathrm{\bar E}}\ce{^2\mathrm{\bar E}}$
& $\mathrm{\bar E}_1$
& $\ce{^1\mathrm{\bar E}}\ce{^2\mathrm{\bar E}}$
& $\mathrm{\bar E}_1$ \\ \hline
$\Gamma~(0,0)$
& $\bar{\Gamma}_4\bar{\Gamma}_5(2)$
& $\bar{\Gamma}_6(2)$
& $\bar{\Gamma}_4\bar{\Gamma}_5(2)$
& $\bar{\Gamma}_6(2)$
& $\bar{\Gamma}_4\bar{\Gamma}_5(2)$
& $\bar{\Gamma}_6(2)$ \\ \hline
$K~(1/3,1/3)$
& $2\bar{K}_4(1)$
& $\bar{K}_5(1)\oplus \bar{K}_6(1)$
& $2\bar{K}_5(1)$
& $\bar{K}_4(1)\oplus \bar{K}_6(1)$
& $2\bar{K}_6(1)$
& $\bar{K}_4(1)\oplus \bar{K}_5(1)$ \\ \hline
\end{tabular}

\vspace{6pt}

\begin{tabular}{|c|c|c|c|c|c|}
\hline
\multicolumn{6}{|c|}{\textbf{(b) Matrix representations of irreps at high-symmetry momenta}} \\ \hline
 &
 $\bar{\Gamma}_4\bar{\Gamma}_5$ &
 $\bar{\Gamma}_6$ &
 $\bar{K}_4$ &
 $\bar{K}_5$ &
 $\bar{K}_6$ \\ \hline
$E$
& $\begin{pmatrix}1&0\\0&1\end{pmatrix}$
& $\begin{pmatrix}1&0\\0&1\end{pmatrix}$
& $1$
& $1$
& $1$ \\ \hline
$C_{3z}$
& $\begin{pmatrix}-1&0\\0&-1\end{pmatrix}$
& $\begin{pmatrix}e^{-i\pi/3}&0\\0&e^{i\pi/3}\end{pmatrix}$
& $-1$
& $e^{-i\pi/3}$
& $e^{i\pi/3}$ \\ \hline
$C_{2x}$
& $\begin{pmatrix}-i&0\\0&i\end{pmatrix}$
& $\begin{pmatrix}0&e^{-i\pi/3}\\ e^{i2\pi/3}&0\end{pmatrix}$
& ---
& ---
& --- \\ \hline
\end{tabular}
\caption{Elementary band representations and matrix representations of irreducible representations at high-symmetry momenta for layer group $P312$ (No. 67) with time-reversal symmetry.}
\label{app:tab:P312_EBR_and_irreps}
\end{table}

We now relax the \emph{single-orbital-single-branch} limit. We consider the following two cases
\begin{enumerate}
    \item \emph{two-orbital-single-branch}: a single bonding/antibonding branch with two orbitals carrying angular momentum $J_z=\pm m_{J,1}$ and $J_z=\pm m_{J,2}$,
    \item \emph{single-orbital-two-branch}: a single orbital carrying $J_z=\pm m_{J,1}$ with both bonding and antibonding branches,
\end{enumerate}
where the moir\'e potential part reads $[T_{\vQ_i,\vQ_j}]_{\alpha ,\alpha'}^s$. Here we focus on a single spin sector $s$ due to spin-\Uone. For brevity, in case (1), $\alpha$ labels the orbital, while in case (2), $\alpha$ labels the branch. Similar to \cref{app:eqn:relation between K and Kprim in SBSO,app:eqn:hermitioan in SBSO}, the sole dependence of $[T_{\vQ_i,\vQ_j}]_{\alpha ,\alpha'}^s$ on $\vQ_i-\vQ_j$ leads to
\begin{equation}
    [T_{\vQ^{(i)},\vQ^{(i+1\,\mathrm{mod}\,3)}}]^s_{\alpha,\alpha'}
    =
    [T_{\vQ'^{(i+1\,\mathrm{mod}\,3)},\vQ'^{(i)}}]^s_{\alpha,\alpha'}.
\end{equation}
By Hermiticity,
\begin{equation}\label{app:eqn:hermitioan}
    [T_{\vQ'^{(i+1\,\mathrm{mod}\,3)},\vQ'^{(i)}}]^s_{\alpha,\alpha'}
    =
    \left([T_{\vQ'^{(i)},\vQ'^{(i+1\,\mathrm{mod}\,3)}}]^s_{\alpha',\alpha}\right)^{*}.
\end{equation}
Consequently, moir\'e potential at $K_M$ and $K'_M$ are no longer related by a simple conjugation,
\begin{equation}\label{app:eqn:relation between K and Kprim}
    [T_{\vQ^{(i)},\vQ^{(i+1\,\mathrm{mod}\,3)}}]^s_{\alpha,\alpha'}
    =
    \left([T_{\vQ'^{(i)},\vQ'^{(i+1\,\mathrm{mod}\,3)}}]^s_{\alpha',\alpha}\right)^{*}.
\end{equation}
We denote the \(2\times2\) orbital-space matrix by by $[T_s]_{\alpha,\alpha'}\equiv [T_{\vQ^{(1)},\vQ^{(2)}}]^s_{\alpha,\alpha'}$, then \cref{app:eqn:relation between K and Kprim} implies that the corresponding hopping matrix at \(K'_M\) is \(T_s^\dagger\). Therefore, in the basis
\(\{
|\vQ^{(1)},\alpha,s\rangle,
|\vQ^{(2)},\alpha,s \rangle,
|\vQ^{(3)},\alpha,s\rangle
\}\) for $K_M$ and basis
\(\{
|\vQ'^{(1)},\alpha,s\rangle,
|\vQ'^{(2)},\alpha,s\rangle,
|\vQ'^{(3)},\alpha,s\rangle
\}\) for $K'_M$,
the Hamiltonians at \(K_M\) and \(K'_M\) take the block-circulant form
\begin{equation}\label{app:eqn:two-orbital-triple-model}
[h(K_M)]_s=
\begin{pmatrix}
\mathcal E_{1,s} & T_s & T_s^\dagger\\
T_s^\dagger & \mathcal E_{1,s} & T_s\\
T_s & T_s^\dagger & \mathcal E_{1,s}
\end{pmatrix},
\qquad
[h(K'_M)]_s=
\begin{pmatrix}
\mathcal E_{2,s} & T_s^\dagger & T_s\\
T_s & \mathcal E_{2,s} & T_s^\dagger\\
T_s^\dagger & T_s & \mathcal E_{2,s}
\end{pmatrix},
\end{equation}
where \(\mathcal E_{1,s}\) and \(\mathcal E_{2,s}\) are Hermitian \(2\times2\) onsite matrices in orbital space. In the \(C_3\)-adapted basis
\begin{equation}
    |\alpha,\Pi_z,s\rangle
    =
    \frac{1}{\sqrt 3}\sum_{j=1}^{3}\Omega^{-\Pi_z(j-1)}|\vQ^{(j)},\alpha,s\rangle,
    \qquad \Omega=e^{i2\pi/3},
\end{equation}
and analogously at $K'_M$ with $\vQ^{(j)}$ replaced by $\vQ'^{(j)}$, the total \(C_{3z}\) eigenvalue is
\begin{equation}
    \xi_{\alpha,\Pi_z,s}
    =
    e^{-i\frac{2\pi}{3}(\sigma m_{J,\alpha}+\Pi_z)}.
\end{equation}
The matrix element between two such states can be nonzero only if their $C_{3z}$ eigenvalues are equal. Hence the Hamiltonian decomposes into three two-dimensional subspaces
\begin{equation}
    h_s(K_M)=\bigoplus_{\Pi_z=0,\pm1} h_{\Pi_z,s}(K_M),
    \qquad
    h_s(K'_M)=\bigoplus_{\Pi_z=0,\pm1} h_{\Pi_z,s}(K'_M),
\end{equation}
In the \emph{two-orbital-single-branch} case with \(m_{J,1}=-\frac12\) and \(m_{J,2}=\frac32\), the two-dimensional subspace labeled by \(\Pi_z\) is spanned by $\Big\{|1,\Pi_z,s\rangle,\,|2,\Pi_z+\sigma,s\rangle\Big\}$, where the second index is understood mod 3. This is because
\(|1,\Pi_z,s\rangle\) and \(|2,\Pi_z+\sigma,s\rangle\) carry the same total angular momentum,
\begin{equation}
    \sigma m_{J,1}+\Pi_z= \left(\sigma m_{J,2}+\Pi_z+\sigma\right) \mod 3.
\end{equation}
In the \emph{single-orbital-two-branch} case, the two-dimensional subspace labeled by \(\Pi_z\) is spanned by $\Big\{|1,\Pi_z,s\rangle,\,|2,\Pi_z,s\rangle\Big\}$ with the total angular momentum,
\begin{equation}
    \sigma m_{J,1}+\Pi_z.
\end{equation}
In both cases, because \(h_{\Pi_z,s}(K_M)\) and \(h_{\Pi_z,s}(K'_M)\) are now independent $2\times2$ matrices rather than scalars, the topmost states at $K_M$ in spin sectors $s$ and $\bar s$ are no longer forced to share the same $\Pi_z$, and similarly the topmost states at $K_M$ and $K'_M$ within one spin sector are no longer forced to carry opposite $\Pi_z$. Therefore the simple relation \cref{app:eqn:constraints from SBSO} in the \emph{single-orbital-single-branch} limit is lost once two orbitals/branches are included, and the topmost isolated band is no longer forced to be trivial.

Since the topological property of the isolated band is parameter dependent, we now derive the condition under which the isolated band set in layer group $p312$ can carry a nonzero $C_{3z}$ indicator for the spin Chern number. For a single isolated band in spin sector $s$, the $C_{3z}$ symmetry-indicator formula reads
\begin{equation}\label{app:eqn:P312-general-indicator}
    e^{-i\frac{2\pi}{3}C_s}
    =
    -\,\xi_{\Gamma_M,s}\,\xi_{K_M,s}\,\xi_{K'_M,s}.
\end{equation}
Because $\mathcal T$ and $C_{2x}$ map $\Gamma_M$ ($K_M$) to $\Gamma_M$ ($K'_M$) while flipping the spin, the isolated states satisfy
\begin{equation}\label{app:eqn:P312-KKp-opposite-spin}
    \xi_{\Gamma_M,\bar s}\xi_{\Gamma_M,s}
    =
    \xi_{K'_M,\bar s}\xi_{K_M,s}
    =
    1.
\end{equation}
Substituting \cref{app:eqn:P312-KKp-opposite-spin} into \cref{app:eqn:P312-general-indicator}, and using $(\xi_{\Gamma_M,s})^3=-1$ for a spinful $C_{3z}$ eigenvalue, we obtain
\begin{equation}\label{app:eqn:P312-general-indicator-ratio}
    e^{-i\frac{2\pi}{3}C_s}
    =
    \frac{\xi_{K_M,s}\,\xi_{\Gamma_M,\bar s}}
    {\xi_{K_M,\bar s}\,\xi_{\Gamma_M,s}}.
\end{equation}
For the \emph{two-orbital-single-branch} case, we assume the topmost isolated state at $K_M$ in spin sectors $s$ and $\bar s$ belongs to the subspaces labeled by $\Pi_{z,1}$ and $\Pi_{z,2}$, respectively. In each such two-dimensional subspace labeled by $\Pi_{z,j}$, the two basis states carry the same total $C_{3z}$ eigenvalue. Hence the topmost state in the $\Pi_{z,1}$ block has the same $C_{3z}$ phase as $|1,\Pi_{z,1},s\rangle$, and similarly for the $\Pi_{z,2}$ block in spin sector $\bar s$. The corresponding $C_{3z}$ eigenvalue is
\begin{equation}\label{app:eqn:P312-two-orbital-xi-K}
    \xi_{K_M,s}
    =
    e^{-i\frac{2\pi}{3}\left(\sigma m_{J,1}+\Pi_{z,1}\right)},\qquad \xi_{K_M,\bar s}
    =
    e^{-i\frac{2\pi}{3}\left(\bar \sigma m_{J,1}+\Pi_{z,2}\right)}
\end{equation}
For the topmost state, the corresponding $C_{3z}$ eigenvalue at $\Gamma_M$ is
\begin{equation}
    \xi_{\Gamma_M,s}
    =
    e^{-i\frac{2\pi}{3}\sigma m_{J,a_\Gamma}},\qquad a_\Gamma\in\{1,2\}.
\end{equation}
Therefore, \cref{app:eqn:P312-general-indicator-ratio} gives the spin Chern number as
\begin{equation}\label{app:eqn:P312-two-orbital-indicator-special}
    C_\uparrow
    \equiv
    \Pi_{z,1}-\Pi_{z,2}
    +2(m_{J,1}-m_{J,a_\Gamma})
    \pmod 3.
\end{equation}
For the present choice \(m_{J,1}=-\frac12\) and \(m_{J,2}=\frac32\), this gives
\begin{equation}\label{app:eqn:P312-two-orbital-indicator-special-cases}
\begin{aligned}
    &a_\Gamma=1:
    \qquad
    C_\uparrow
    \equiv
    \Pi_{z,1}-\Pi_{z,2}
    \pmod 3,\\
    &a_\Gamma=2:
    \qquad
    C_\uparrow
    \equiv
    \Pi_{z,1}-\Pi_{z,2}-1
    \pmod 3.
\end{aligned}
\end{equation}
For the \emph{single-orbital-two-branch} case, we assume the topmost isolated state at $K_M$ in spin sectors $s$ and $\bar s$ belongs to the subspaces labeled by $\Pi_{z,1}$ and $\Pi_{z,2}$, respectively. The corresponding $C_{3z}$ eigenvalue is
\begin{equation}
    \xi_{K_M,s}
    =
    e^{-i\frac{2\pi}{3}\left(\sigma m_{J,1}+\Pi_{z,1}\right)},\qquad \xi_{K_M,\bar s}
    =
    e^{-i\frac{2\pi}{3}\left(\bar \sigma m_{J,1}+\Pi_{z,2}\right)}
\end{equation}
For the topmost state, the corresponding $C_{3z}$ eigenvalue at $\Gamma_M$ is
\begin{equation}
    \xi_{\Gamma_M,s}
    =
    e^{-i\frac{2\pi}{3}\sigma m_{J,1}}.
\end{equation}
Therefore, \cref{app:eqn:P312-general-indicator-ratio} gives the spin Chern number as
\begin{equation}\label{app:eqn:P312-single-orbital-indicator-special}
    C_\uparrow
    \equiv
    \Pi_{z,1}-\Pi_{z,2}
    \pmod 3.
\end{equation}
For both cases, a practical route to change the $C_{3z}$ indicator is therefore to tune the ordering of $K_M$ states through a band inversion between two blocks with different \(\Pi_z\) in one spin sector. Provided that the topmost state at $\Gamma_M$ is unchanged during this process, once the gap reopens the inversion shifts $\Pi_{z,1}$ (if in spin $s$) or $\Pi_{z,2}$ (if in spin $\bar s$) to one of the other two allowed values, changing $\Pi_{z,1}-\Pi_{z,2}$ by \(\pm1\pmod 3\). Consequently, an isolated band that initially carries zero spin Chern number acquires \(C_\uparrow=\pm1\) after the inversion, as dictated by \cref{app:eqn:P312-two-orbital-indicator-special-cases,app:eqn:P312-single-orbital-indicator-special}.

Finally we illustrate this topological phase transition numerically. We first discuss the \emph{two-orbital-single-branch} case, where the two orbitals are characterized by total angular momentum $J_z=\pm m_J$ with $m_J=-\frac{1}{2}$ (\ie, $p_{-,\uparrow}$ and $p_{+,\downarrow}$) and $m_J=\frac{3}{2}$ (\ie, $p_{+,\uparrow}$ and $p_{-,\downarrow}$), respectively. With the basis 
\begin{equation}
    \{\cre{c}{\vk,\vQ,p_-,\uparrow},\cre{c}{\vk,\vQ,p_+,\downarrow},\cre{c}{\vk,\vQ,p_+,\uparrow},\cre{c}{\vk,\vQ,p_-,\downarrow}\},
\end{equation}
the exact symmetries of layer group $p321$ has the following representation matrices
\begin{equation}
\begin{aligned}
&D({C}_{3z})= \begin{pmatrix}
e^{-i\pi/3} & 0 & 0 & 0\\
0 & e^{i\pi/3} & 0 & 0\\
0 & 0 & -1 & 0\\
0 & 0 & 0 & -1
\end{pmatrix},\quad D({C}_{2x})= \begin{pmatrix}
0 & i & 0 & 0\\
i & 0 & 0 & 0\\
0 & 0 & 0 & i\\
0 & 0 & i & 0
\end{pmatrix},\quad D(\mathcal T)=\begin{pmatrix}
0 & -1 & 0 & 0\\
1 & 0 & 0 & 0\\
0 & 0 & 0 & -1\\
0 & 0 & 1 & 0
\end{pmatrix},
\end{aligned}
\end{equation}
Under the constraints from these exact symmetry and spin-\Uone, the $\vk$-dependent part up to $O(\vk^4)$ and moir\'e potential part up to first harmonic continuum model take the following form,
\begin{equation}\label{app:eqn:P312 SB limit}
\begin{aligned}
    \left[W(\vec{k})\right]^{\text{two-orbital}}_{\vQ,\vQ}=&a_1\Gamma_{z0}+\frac{(\vec k-\vQ)^2}{2m_1}\Gamma_{00}-\frac{\left[(k_+-Q_+)^2\Gamma_++(k_--Q_-)^2\Gamma_-\right]}{4m_2}+\frac{(\vec k-\vQ)^2}{2m_3}\Gamma_{z0}\\
    &+\frac{i}{2b_2}\left[(\vec{k}_+-\vec{Q}_+)^3-(\vec{k}_--\vec{Q}_-)^3\right]\Gamma_{0z}+\frac{i}{2b_4}\left[(\vec{k}_+-\vec{Q}_+)^3-(\vec{k}_--\vec{Q}_-)^3\right]\Gamma_{zz}\\
    &+\frac{1}{2b_6}(\vec k-\vQ)^2\left[(k_+-Q_+)\tilde{\Gamma}_+ +(k_--Q_-)\tilde{\Gamma}_- \right],\\
   T_{\vQ,\vQ'}^{\text{two-orbital}}=&\sum_{\alpha=\pm 1}\sum_{j=1}^3 \delta_{\vQ,\vQ'+\alpha \boldsymbol{g}_j} \left[ c_1 \Gamma_{00} - c_2 \left( z_j \Gamma_- + z_j^* \Gamma_+ \right)+c_3\Gamma_{z0} \right.\\
   &\left. +i\alpha c_5\Gamma_{00}+i\alpha c_6 \left( z_j \Gamma_- + z_j^* \Gamma_+ \right)+i\alpha c_7\Gamma_{z0}\right].
\end{aligned}
\end{equation}
The dispersion without moir\'e potential included is shown in \cref{app:fig:P312-SBSO:d} with $a_1=0.5$ and $m_1=-0.2$. The inclusion of the moir\'e potential ($c_1=c_2=-0.2$, $c_5=0.1$) gaps out the spectrum and yields isolated low-energy band sets in which, at each of $K_M$ and $K'_M$, the topmost two states come from opposite spin sectors. Tuning the relative energy of the two orbital families through $a_1$ drives a band inversion between the topmost and second-topmost states within the same spin sector, realizing the phase transition from the trivial isolated band in \cref{app:fig:P312-SBSO:e} to the topological band in \cref{app:fig:P312-SBSO:f}.

We next discuss \emph{single-orbital-two-branch} case where the orbital has total angular momentum $J_z=\pm\frac{1}{2}$ (\ie, $s$-like orbital).
With the basis 
\begin{equation}
    \{\cre{c}{\vk,\vQ,\text{antibonding}, \uparrow},\cre{c}{\vk,\vQ,\text{antibonding},\downarrow},\cre{c}{\vk,\vQ,\text{bonding}, \uparrow},\cre{c}{\vk,\vQ,\text{bonding},\downarrow}\},
\end{equation}
the exact symmetries of layer group $p312$ has the following representation matrices
\begin{equation}
\begin{aligned}
&D({C}_{3z})= \tau_0\otimes e^{i\frac{\pi}{3}s_z},\quad D({C}_{2x})= -\tau_z\otimes is_x,\quad D(\mathcal T)=\tau_0\otimes -is_y.
\end{aligned}
\end{equation}
Here, the Pauli matrices $\tau_i$ and $s_i$ act on the branch degrees of freedom (bonding/antibonding) and spin space, respectively.
In this case, the $\vk$-dependent part up to $O(\vk^4)$ and moir\'e potential part up to first harmonic continuum model take the following form,
\begin{equation}\label{app:eqn:P312 SO limit}
\begin{aligned}
    \left[W(\vec{k})\right]^{\text{two-branch}}_{\vQ,\vQ}=&a'_1\tau_z\otimes s_0+a'_2\tau_y\otimes s_z+\frac{(\vec k-\vQ)^2}{2m'_1}\tau_0\otimes s_0+\frac{(\vec k-\vQ)^2}{2m'_2}\tau_z\otimes s_0+\frac{(\vec k-\vQ)^2}{2m'_3}\tau_y\otimes s_z\\
    &-\left[(\vec{k}_+-\vec{Q}_+)^3-(\vec{k}_--\vec{Q}_-)^3\right]\left(\frac{1}{2b'_2}\tau_0\otimes s_z+\frac{1}{2b'_4}\tau_z\otimes s_z+\frac{1}{2b'_6}\tau_y\otimes s_0\right)\\
    &-\frac{1}{2b'_8}\left[(\vec{k}_+-\vec{Q}_+)^3+(\vec{k}_--\vec{Q}_-)^3\right]\tau_x\otimes s_z,\\
   T_{\vQ,\vQ'}^{\text{two-branch}}=&\sum_{\alpha=\pm 1}\sum_{j=1}^3 \delta_{\vQ,\vQ'+\alpha \boldsymbol{g}_j} \left( c'_1 \tau_0\otimes s_0 + c'_2 \tau_z\otimes s_0+c_3'\tau_y\otimes s_z-i\alpha c'_5 \tau_0\otimes s_0-i\alpha c'_6  \tau_z\otimes s_0-i\alpha c'_7  \tau_y\otimes s_z \right).\\
\end{aligned}
\end{equation}
The dispersion of the kinetic part with $a'_1=0.14$, $m'_1=-0.4$, $m'_3=-1$ and $b'_4=-0.4$ is shown in \cref{app:fig:P312-SBSO:g}.
After including the moir\'e potential with $c'_1=-0.01$, $c'_2=-0.2$ and $c'_5=-0.04$, the triplet in each branch at $K_M$ and $K'_M$ splits into three singlets as expected. By tuning the relative energy of the two branches through $a'_1$, we can induce the band exchange between the topmost and second-topmost bands within the same spin sector at $K_M$ and $K'_M$, which leads to the phase transition from the trivial isolated band in \cref{app:fig:P312-SBSO:h} to the topological band in \cref{app:fig:P312-SBSO:i}.

\subsection{Application to materials} 
Based on the analysis in \cref{app:sec: moire systems with different symmetry space groups}, we show that the emergence of an isolated topological band in twisted \ch{SnSe2} and \ch{ZrS2} originates from the inclusion of both bonding and antibonding branches and/or two effective orbitals.

\subsubsection{\ch{SnSe2} case} \label{app:sec: topology origin in SnSe2}
 We start with twisted AB-stacked \ch{SnSe2}, where we consider a single $\bar P$ orbital with $J_z=\pm \frac{3}{2}$ and both layers in the continuum model, corresponding to the \emph{single-orbital-two-branch} case discussed in \cref{app:sec:p3211' moire system}. To transform from the layer basis (top/bottom) to the bonding/antibonding basis, we apply the following unitary transformation to \cref{app:eqn:simplified model for AB-SnSe2}, 
\begin{equation}\label{app:eqn:antibond and bond transformation}
    U = \frac{1}{\sqrt{2}}\left(-i\tau_y+\tau_0\right)\otimes s_0.
\end{equation}
After the transformation, the Hamiltonian reads,
\begin{equation}
% \label{app:eqn:simplified model for AB-SnSe2}
\begin{aligned}
h^{\text{AB}} (\vec{r}) &=-\frac{ \hbar^2\nabla^2}{2m_1^*}\tau_0\otimes s_{0}+V_1\tau_0\otimes s_0\sum_{j=1}^3\sum_{\alpha=\pm 1} e^{i\alpha\boldsymbol{g}_j\cdot\vec{r}}\\
&+i\alpha V_2 \tau_x\otimes s_0\sum_{j=1}^3\sum_{\alpha=\pm 1} e^{i\alpha\boldsymbol{g}_j\cdot\vec{r}}+V_3\tau_0\otimes s_0\sum_{j=1}^3\sum_{\alpha=\pm 1} e^{i\alpha\boldsymbol{g}_{2j}\cdot\vec{r}}\\
&-\frac{ \hbar^2\nabla^2}{2m^*_2}\tau_z\otimes s_0+w_1\tau_z\otimes s_{0}+w_2\tau_z\otimes s_{0}\sum_{j=1}^3\sum_{\alpha=\pm 1} e^{i\alpha\boldsymbol{g}_j\cdot\vec{r}}\\
&+w_3\tau_z\otimes s_0\sum_{j=1}^3\sum_{\alpha=\pm 1} e^{i\alpha\boldsymbol{g}_{2j}\cdot\vec{r}}+i\alpha w_4\tau_y\otimes s_z\sum_{j=1}^3\sum_{\alpha=\pm 1} e^{i\alpha\boldsymbol{g}_{2j}\cdot\vec{r}}.
\end{aligned}
\end{equation}
Now we analyze the moir\'e terms, $(V_1-w_2)$ is the first-harmonic term for the branch closer to the Fermi energy. Because the sign of $(V_1-w_2)$ is negative as shown in \cref{app:tab:SnSe2_simplified_model_para_values}, the moir\'e potential maximizes at honeycomb sites and gives rise to the Dirac crossing in the \emph{single-orbital-single-branch} limit, as discussed in \cref{app:sec:p3211' moire system}. This is because the interlayer distance maximized at the honeycomb sites of the moir\'e unt cell as discussed in \cref{app:sec:p3211' moire system}. Furthermore, $V_2$ and $w_4$ are inter-branch terms that break the \emph{single-orbital-single-branch} limit. Specifically, the $w_4$ term breaks the effective $\bar{C}_{2y}$ symmetry represented by $D(C_{2y})=\tau_z\otimes s_0$. If we set $w_4=0$, both spin-$\mathrm{SU}(2)$ and the approximate $\bar{C}_{2y}$ emerge, forcing a four-fold degeneracy at $K_M$ and $K'_M$, as shown in \cref{app:fig:simplified-model-check:a}. 
The $C_{3z}$ eigenvalue of the four-fold degenerate state is $\xi_{K_M,\uparrow}=e^{i\frac{\pi}{3}}\oplus e^{-i\frac{\pi}{3}}$ and $\xi_{K_M,\downarrow}=e^{i\frac{\pi}{3}}\oplus e^{-i\frac{\pi}{3}}$, consistent with \cref{app:eqn:P321 doublet 1 in SBSO limit,app:eqn:P321 doublet 2 in SBSO limit} analyzed in \cref{app:sec:p3211' moire system} with $m_J=\frac{3}{2}$. For $w_4\neq 0$, the effective $C_{2y}$ is broken and the top band becomes isolated. As discussed in \cref{app:sec:p3211' moire system}, the isolated band split from the Dirac-like band represented by a 2D EBR. The isolated band cannot be represented by any EBR with $J_z=\pm\frac{3}{2}$ and carries a nonzero spin Chern number.

\subsubsection{\ch{ZrS2} case}\label{app:sec: topology origin in ZrS2}
\begin{figure}[!t]
    \centering
    \includegraphics[width=\textwidth]{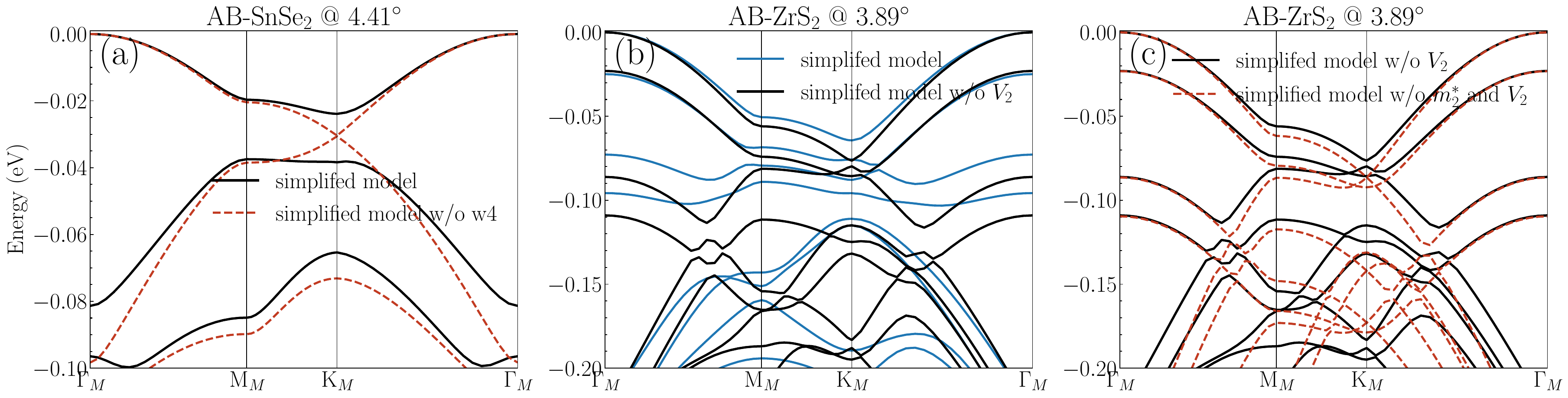}
     \subfloat{\label{app:fig:simplified-model-check:a}}
     \subfloat{\label{app:fig:simplified-model-check:b}}
     \subfloat{\label{app:fig:simplified-model-check:c}}
    \caption{Influence of inter-orbital/inter-branch couplings on the band structures of \ch{SnSe2} and \ch{ZrS2}. (a) Comparison of the band structures of the simplified model for \ch{SnSe2} with and without the $w_4$ term. (b) Comparison of the band structures of the simplified model for \ch{ZrS2} with and without the $V_2$ term. (c) Comparison of the band structures of the simplified model for \ch{ZrS2} without the $V_2$ term and without the $V_2$ and $m_2^*$ terms.}
\label{app:fig:SnSe2-ZrS2-simplified_model}
\end{figure}

We now turn to \ch{ZrS2}, which illustrates the four-fold Dirac-like crossing can be lifted by including two orbitals while keeping only the antibonding branch. For \ch{ZrS2}, we consider both layers and two orbitals ($J_z=\pm\frac{1}{2}$ and $J_z=\pm\frac{3}{2}$) in the continuum basis. Starting from the AB-stacked model in \cref{app:eqn:simplified model for AA/AB-ZrS2}, we transform to the bonding/antibonding basis. To isolate the multi-orbital effect, we set the bonding–antibonding coupling to zero by dropping the $V_2$ term. The resulting band structure is shown in \cref{app:fig:simplified-model-check:b}. We can rewrite the simplified model as,
\begin{equation}\label{app:eqn:antibond and bond AB-ZrS2}
\begin{aligned}
h^{\text{AB}}_{\eta} (\vec{r}) &=-\frac{ \hbar^2\nabla^2}{2m_1^*}\Gamma_{00}-\frac{ \hbar^2}{2m^*_2}\left[(\nabla_x^2-\nabla_y^2)\Gamma_{x0}+2\nabla_x\nabla_y\Gamma_{yz}\right]\\
&+V_1\Gamma_{z0} +\eta w_1\Gamma_{00}+ \eta w_2\Gamma_{00}\sum_{j=1}^3\sum_{\alpha=\pm 1} e^{i\alpha\boldsymbol{g}_j\cdot\vec{r}},
\end{aligned}
\end{equation}
with $\eta=+1$ for the bonding block and $\eta=-1$ for the antibonding block. We define $\Gamma_{ij}=\sigma_i\otimes s_j$, where $\sigma_i$ and $s_i$ ($i=0,x,y,z$) are the identity/Pauli matrices acting in the molecular-orbital and spin spaces. We focus on $\eta=+1$, which hosts the topmost band. The $m_2^*$ term is the only term in \cref{app:eqn:antibond and bond AB-ZrS2} that describes the inter-orbital coupling. To demonstrate the role of emergent $\bar{C}_{2y}$ with $D(\bar C_{2y})=\Gamma_{z0}$, we restore the symmetry by dropping the $m_2^*$ term. In this case, \cref{app:eqn:antibond and bond AB-ZrS2} split into two decoupled sectors described by the molecular orbital with $J_z=\pm\frac{1}{2}$ and $J_z=\pm\frac{3}{2}$, each satisfying the \emph{single-orbital-single-branch} limit discussed in \cref{app:subsec:triple model}. 
As shown in \cref{app:fig:simplified-model-check:c}, $\bar{C}_{2y}$ emerge when we drop the $m_2^*$ term and results in a four-fold degeneracy at $K_M$ and $K'_M$. The Dirac-like band can be represented by a 2D EBR, \textit{i.e.,} $\mathrm{\bar{E}}\mathrm{\bar{E}}@2c$ as discussed in \cref{app:sec:p3211' moire system}. When we restore the $m_2^*$ term, the Dirac-like bands split. The resulted isolated band set cannot be represented by any EBR with $J_z=\pm\frac{3}{2}$ and carries a nonzero spin Chern number $C_z=1$.

Because the continuum Hamiltonians of twisted AA- and AB-stacked \ch{ZrS2} are related by a fractional moir\'e translation [\cref{app:eqn:simplified model for AA/AB-ZrS2}],
\begin{equation}
    T_{\boldsymbol{\tau}_0}h^{AA}(\vec{r})T_{\boldsymbol{\tau}_0}^{-1}=h^{AB}(\vec{r}), \quad T_{\boldsymbol{\tau}_0}=\{E|\boldsymbol{\tau}_0=\tfrac{2}{3}\vec a_{M,1}+\tfrac{1}{3}\vec a_{M,2}\},
\end{equation}
an analogous analysis applies to twisted AA-stacked \ch{ZrS2}. In this case, inter-orbital coupling breaks the effective $\tilde C_{2y}\equiv T_{\boldsymbol{\tau}0} C_{2y} T_{\boldsymbol{\tau}_0}^{-1}$, splitting the 2D EBR so that each isolated branch acquires nontrivial topology. However, it is important to note that the $C_{3z}$ eigenvalue of the topmost band differs from that in the AA case by a phase arising from the fractional moiré translation. Under the symmetry operation with fractional translation $\tilde{g}=\{g|\tau\}$, the creation operator $\cre{c}{\vk,l_1,s_1}$ transforms as
\begin{equation}
\label{app:eqn:nonsym_action_moire_mom}
\begin{aligned}
\tilde{g} \cre{c}{\vk,l_1,s_1} \tilde{g} ^{-1} &= \sum_{l_2,s_2} \frac{1}{\sqrt{\Omega}} \int d^2 r \, \psi_{l_2,s_2}^\dagger(\tilde{g}\rr) e^{i\kk \cdot\rr} \left[ D(\tilde{g}) \right]_{ l_2s_2; l_1s_1}\\
&= \sum_{l_2,s_2} \frac{1}{\sqrt{\Omega}} \int d^2 r \, \psi_{l_2,s_2}^\dagger(\rr) e^{ig\kk \cdot \left(\rr-\tau\right)} \left[ D(\tilde{g}) \right]_{ l_2s_2; l_1s_1}\\
&= \sum_{l_2,s_2} \cre{c}{g\vk,l_1,s_1}  \left[ D(\tilde{g}) \right]_{ l_2s_2; l_1s_1}e^{-ig\kk \cdot \tau}.
\end{aligned}
\end{equation}
Consequently, we arrive at 
\begin{equation}
    \xi_{K_M}^{AB}=e^{-iK_M\cdot\left(\boldsymbol{\tau}_0-C_{3z}\boldsymbol{\tau}_0\right)}\xi_{K_M}^{AA}=e^{-i\frac{2\pi}{3}}\xi_{K_M}^{AA},\quad  \xi_{K'_M}^{AB}=e^{-iK'_M\cdot\left(\boldsymbol{\tau}_0-C_{3z}\boldsymbol{\tau}_0\right)}\xi_{K'_M}^{AA}=e^{i\frac{2\pi}{3}}\xi_{K_M}^{AA},
\end{equation}
which agrees with the results in \cref{app:tab:spin_chern_number_AA_ZrS2,app:tab:spin_chern_number_AB_ZrS2}.

\section{Hartree-Fock and Exact Diagonalization Calculations}\label{app:sec:HF and ED}

In this section, we investigate the interacting physics of AA- and AB-stacked twisted bilayer ZrS$_2$, and AB-stacked twisted bilayer SnSe$_2$. In all cases, we will take the approximation that the non-interacting continuum model preserves spin-\Uone symmetry 

\subsection{Interacting continuum model}

We begin by describing the interacting continuum model defined as
\begin{equation}
    H_\text{tot}=\mathcal{H}+H_\text{int},
\end{equation}
where $\mathcal{H}$ is the numerically exact single-particle continuum Hamiltonian obtained in \cref{app:sec:extracted continuum model}, and we apply an explicit \Uone symmetrization to enforce $S_z$ conservation. $H_\text{int}$ captures the electron interactions, which we describe with dual-gate screened density-density Coulomb interactions with potential $V(\bm{q})=\frac{e^2}{2\epsilon_0\epsilon_r q}\tanh(qd_\text{sc})$. $d_\text{sc}$ is the gate-to-sample distance, while the overall strength of the interaction is set by the relative dielectric constant $\epsilon_r$. In the following, we treat $\epsilon_r$ as a theoretically tunable parameter.

The interacting part of the Hamiltonian requires specifying an interaction scheme~\cite{mfci3}. In particular, we write
\begin{equation}
    H_{\text{int}}=:{H}_\text{int}:-{H}_{\text{HF,int}}[P^\text{ref}], 
\end{equation}
where $:{H}_\text{int}:$ is the interaction normal-ordered with respect to the Fock vacuum of the many-body Hilbert space, i.e.~no fermions in the system. The normal-ordered interaction reads
\begin{align}
    :{H}_\text{int}:&=\frac{1}{2\Omega_\text{tot}}\sum_{\bm{q}}V(\bm{q}):\hat{\rho}_{\bm{q}}\hat{\rho}_{-\bm{q}}:\\\
    &=\frac{1}{2\Omega_{\text{tot}}}\sum_{\bm{q}\bm{p}\bm{p}'}\sum_{ss'}\sum_{ll'\sigma\sigma'}V(\bm{q})c^\dagger_{\bm{p},l,s}c^\dagger_{\bm{p}'-\bm{q},l',s'}c_{\bm{p}',l',s'}c_{\bm{p},l,s}\\
    \hat{\rho}_{\bm{q}}&=\sum_{\bm{p}sl}c^\dagger_{\bm{p}+\bm{q},l,s}c_{\bm{p},l,s}.
\end{align}
$c^\dagger_{\mathbf p,l,s}$ denotes the electron creation operator in the continuum model plane wave basis. The summations over $\bm{p}$ and $\bm{p}'$ run over all momenta within the plane wave cutoff that is used to diagonalize $\mathcal{H}$, $\bm{q}$ runs over all momentum transfers within the cutoff, $\Omega_{\text{tot}}$ is the total system area, and we have introduced the density operator $\hat{\rho}_{\bm{q}}$.
${H}_{\text{HF,int}}[P^\text{ref}]$ is the Hartree-Fock potential corresponding to the `reference' one-body density matrix $P^\text{ref}$
\begin{equation}\label{app:eqn:HF potential of one-body density}
    {H}_{\text{HF,int}}[P^\text{ref}]=\frac{1}{\Omega_{\text{tot}}}\sum_{\bm{q}\bm{p}\bm{p}'}\sum_{ss'}\sum_{ll'\sigma\sigma'}V(\bm{q})\bigg[P^\text{ref}_{\bm{p}'-\bm{q},l',s';\bm{p}',l',s'}
    c^\dagger_{\bm{p},l,s}c_{\bm{p},l,s}
    -
    P^\text{ref}_{\bm{p}'-\bm{q},l',s';\bm{p},l,s}c^\dagger_{\bm{p},l,s}c_{\bm{p}',l',s'}\bigg],
\end{equation}
whose purpose is to set a reference point from which interactions are measured from. We note there has been no band projection performed so far.

The interaction scheme corresponds to the choice of $P^\text{ref}$. For the specific materials ZrS$_2$ and SnSe$_2$ treated here, the monolayer band structure possesses a large $\sim $eV scale gap at charge neutrality between the valence and conduction subspaces. Hence, a natural choice of reference density corresponds to fully occupying all valence bands. Since the moir\'e continuum models of ZrS$_2$ and SnSe$_2$ capture just the valence degrees of freedom, $P^\text{ref}$ therefore consists of fully occupying all the single-particle states within the plane-wave cutoff. This interaction scheme is analogous to that used in $K$ valley moir\'e TMDs such as twisted bilayer MoTe$_2$~\cite{YU24a}. 

\subsection{Hartree-Fock results}
\begin{figure}[!t]
    \centering
    \includegraphics[width=0.4\textwidth]{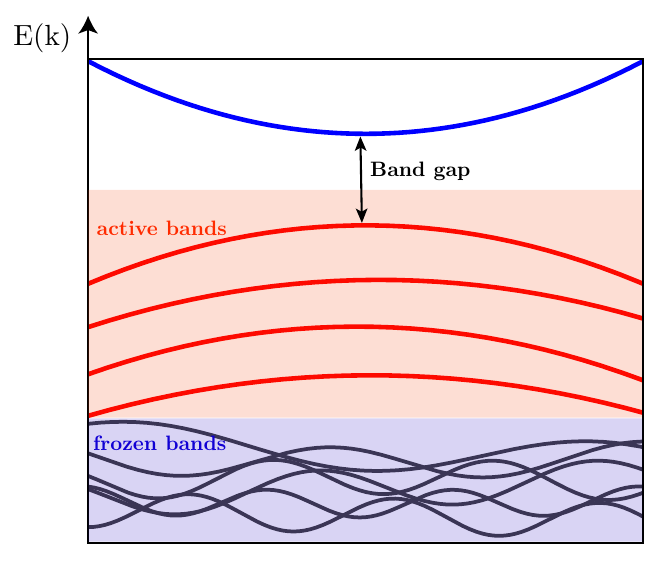}
    \caption{Schematic illustration of the band-projection scheme used in the Hartree–Fock calculations. The dashed line separates the valence bands from the conduction bands. In the valence-only continuum model, the conduction bands (blue) are not included in the modelling from the outset. Band-projection corresponds to retaining the highest $n_v$ valence bands as `active' degrees of freedom, while the remaining valence bands (violet region) are treated as frozen `remote' bands that are enforced to be fully occupied. The figure shows the case where $n_v=4$, as in the Hartree-Fock phase diagrams. The interaction calculation includes fluctuations only within the active bands. For the interaction scheme used in this work, the frozen remote bands do not contribute to the physics of the active bands.}
\label{app:fig:schematic_band_structure}
\end{figure}
To understand the interacting physics at integer filling factors $\nu$ of twisted bilayer ZrS$_2$ and SnSe$_2$, we perform self-consistent Hartree-Fock (HF) calculations. 

\subsubsection{Method}

For practical calculations of $H_\text{tot}$, it is useful to project into some subset of moir\'e bands to reduce the computational difficulty. Before doing so, it is useful to first rewrite $H_\text{tot}$ in the single-particle band basis. The latter is expressed in terms of the plane wave creation operators as
\begin{equation}
    \gamma^\dagger_{\bm{k},s,m}=\sum_{\bm{G},l}c^\dagger_{\bm{k}+\bm{G},l,s}U^s_{\bm{G}l,m}(\bm{k}),
\end{equation}
where $m$ labels the moir\'e band, and $U^s_{\bm{G},l,m}(\bm{k})$ is the Bloch eigenvector that diagonalizes the single-particle continuum model. We work in the periodic gauge $U^s_{\bm{G}-\bm{G}',l,m}(\bm{k}+\bm{G}')=U^s_{\bm{G},l,m}(\bm{k})$. The form factors are defined in terms of overlaps of Bloch functions
\begin{equation}
    M^s_{mn}(\bm{k},\bm{q}+\bm{G})=\sum_{\bm{G}',l}U^{s^*}_{\bm{G}+\bm{G}',l,m}(\bm{k}+\bm{q})U^s_{\bm{G}',l,n}(\bm{k}).
\end{equation}
The various contributions to the total Hamiltonian are then
\begin{gather}
    \mathcal{H}=\sum_{\bm{k},s,m}\epsilon_{s,m}(\bm{k})\gamma^\dagger_{\bm{k},s,m}\gamma_{\bm{k},s,m}\\
    :H_\text{int}:=\frac{1}{2\Omega_\text{tot}}\sum_{\bm{q},\bm{G}}\sum_{\bm{k},\bm{k}',m,m',n,n',s,s'}V(\bm{q}+\bm{G})M^s_{mn}(\bm{k},\bm{q}+\bm{G})M^{s'^*}_{n'm'}(\bm{k}'-\bm{q},\bm{q}+\bm{G})\gamma^\dagger_{\bm{k}+\bm{q},s,m}\gamma^\dagger_{\bm{k}'-\bm{q},s',m'}\gamma_{\bm{k}',s',n'}\gamma_{\bm{k},s,n}
\end{gather}
\begin{align}
    H_{\text{HF,int}}[P^\text{ref}]&=\sum_{\bm{G},\bm{k},m,n,s}\frac{V(\bm{G})}{\Omega_\text{tot}}M^s_{mn}(\bm{k},\bm{G})\left(\sum_{\bm{k}'m's'}M^{s'^*}_{n'm'}(\bm{k}',\bm{G})P^\text{ref}_{m',n'}(\bm{k},s')\right)\gamma^\dagger_{\bm{k},s,m}\gamma_{\bm{k},s,n}\\
    &-\sum_{\bm{G},\bm{q},\bm{k},m,m',n,n',s}\frac{V(\bm{q}+\bm{G})}{\Omega_\text{tot}}M^{s^*}_{n'm}(\bm{k},\bm{q}+\bm{G})M^s_{m'n}(\bm{k},\bm{q}+\bm{G})P^\text{ref}_{m',n'}(\bm{k}+\bm{q},s)\gamma^\dagger_{\bm{k},s,m}\gamma_{\bm{k},s,n},
\end{align}
where $\epsilon_{s,m}(\bm{k})$ is the non-interacting band dispersion, and the reference density matrix is now expressed in the band basis. In the choice of interaction scheme that we use, the reference density matrix is $P^\text{ref}_{mn}(\bm{k},s)=\delta_{mn}$.

The band projection is carried out by first specifying the number $n_v$ of highest valence bands of $\mathcal{H}$ that we choose to remain `active'. Then, we restrict the many-body Hilbert space to states of the form $\hat{\mathcal{O}} \prod_{\bm{k},s,n \in\text{rem. val.}}\gamma^\dagger_{\bm{k},s,n}\ket{\text{vac}}$, where `$\text{rem. val.}$' refers to all remote valence bands that do not belong to the active set of $n_v$ bands, and $\ket{\text{vac}}$ is the Fock vacuum of the electron operators~\cite{mfci3}. $\hat{\mathcal{O}}$ consists of an arbitrary combination of creation operators belonging to the active bands. Effectively, we are freezing the occupations of all non-active remote bands to be filled, but allowing arbitrary occupations within the active bands as shown in \cref{app:fig:schematic_band_structure}. For our choice of interaction scheme, the remote bands do not contribute to the physics of the active bands.

In the HF calculations, the one-body density matrix (projector) $P$ is constrained to take the following form
\begin{align}
    P_{\bm{k},s,m;\bm{k}',s',m'}&=\langle \gamma^\dagger_{\bm{k},s,m}\gamma_{\bm{k}',s',m'} \rangle\\
    &=f_{\uparrow\uparrow}(\bm{k})\delta_{s,\uparrow}\delta_{s',\uparrow}\delta_{\bm{k},\bm{k}'}+
    f_{\downarrow\downarrow}(\bm{k})\delta_{s,\downarrow}\delta_{s',\downarrow}\delta_{\bm{k},\bm{k}'}
    +f_{\uparrow\downarrow}(\bm{k})\delta_{s,\uparrow}\delta_{s',\downarrow}\delta_{\bm{k}+\bm{q},\bm{k}'}
    +f_{\downarrow\uparrow}(\bm{k})\delta_{s,\downarrow}\delta_{s',\uparrow}\delta_{\bm{k},\bm{k}'+\bm{q}},
\end{align}
where $m,m'$ are moir\'e band indices, and $\bm{q}$ is some moir\'e wavevector. In other words, we enforce moir\'e translation symmetry, with the possible exception of inter-spin coherence (ISC) at wavevector $\bm{q}$. Note that ISC breaks the spin-\Uone symmetry, which leads to non-zero $f_{\uparrow\downarrow},f_{\downarrow\uparrow}$. We constrain $\bm{q}$ to the $C_{3z}$-symmetric momenta $\Gamma_M,K_M,K'_M$.

We perform HF calculations with at least 25 initial seeds, and select the solution with the minimal HF energy. See Ref.~\cite{Kwan2025HFreview} for more details about solving the HF equations self-consistently.

\subsubsection{Phase diagrams: interaction strength vs twist angle}\label{subsubsec:phase_diagrams_int_strength_twist}

\begin{figure}[!t]
    \centering
    \includegraphics[width=0.7\textwidth]{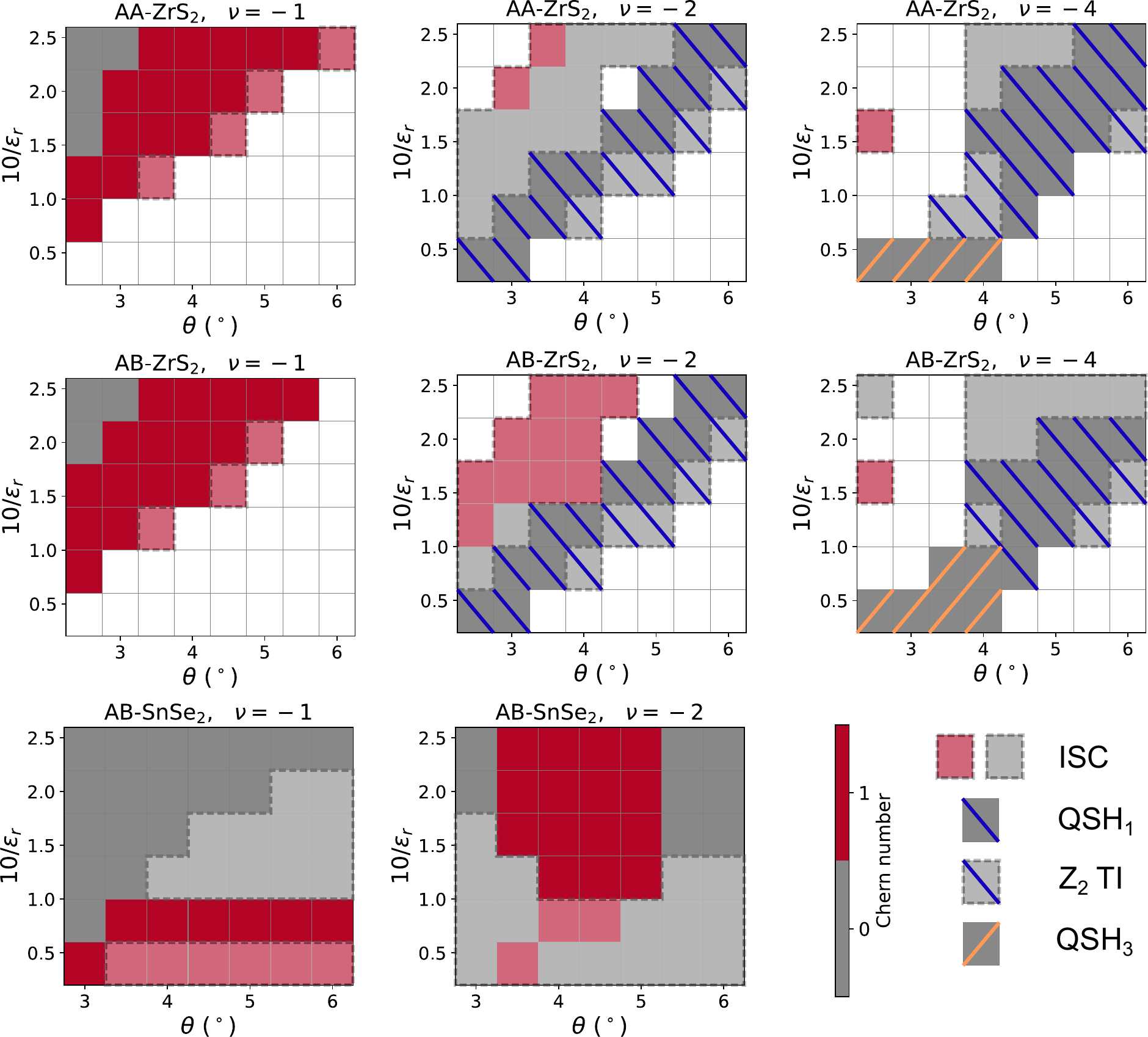}
    \caption{Hartree-Fock phase diagram as a function of interaction strength $10/\epsilon_r$ and twist angle $\theta$ for various integer filling factors $\nu$. For AA-ZrS$_2$ and AB-ZrS$_2$, we use the \Uone-symmetric continuum model parameters extracted at $\theta=3.89^\circ$, while for AB-SnSe$_2$, we use the \Uone-symmetric continuum model parameters extracted at $\theta=4.41^\circ$. White regions correspond to gapless phases. Light red and grey shading denote inter-spin coherence (ISC) with Chern number $C=1$ and $C=0$, respectively. `QSH$_n$' denotes a quantum spin 
    Hall phase with a spin Chern number of $n$. In the presence of finite ISC, `$Z_2$ TI' denotes a topological insulator with breaking spin-\Uone symmetry. System size is $12\times 12$, screening length is $d_\text{sc}=20\,$nm, and we project to four valence moir\'e bands per spin. 
    }\label{app:fig:interaction_theta_HF_phase_diagram}
\end{figure}

\begin{figure}[!t]
    \centering
    \includegraphics[width=0.5\textwidth]{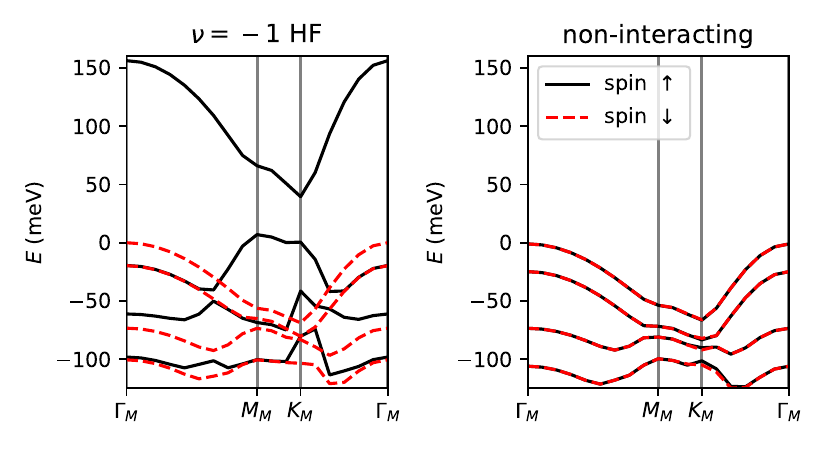}
    \caption{\textbf{Left:} $\nu=-1$ Hartree-Fock band structure of the $C=1$ insulator for AA-ZrS$_2$ using $10/\epsilon_r=1.6$ and $\theta=4.0^\circ$. The highest HF band (which is unoccupied) has $>99\%$ overlap with the highest non-interacting band. We have shifted the energies so that the highest state in the fully occupied spin $\downarrow$ sector has $E=0$. System size is $18\times 18$, screening length is $d_\text{sc}=20\,$nm, and we project to four valence moir\'e bands per spin. 
    \textbf{Right:} Corresponding non-interacting band structure.}\label{app:fig:HF_bandstruct_ZrS2_AA_nu-1_theta4.00_epsr6.25}
\end{figure}

In Fig.~\ref{app:fig:interaction_theta_HF_phase_diagram}, we present HF phase diagrams as a function of interaction strength $10/\epsilon_r$ and twist angle $\theta$. For a state with non-vanishing Chern number (which hence breaks TRS), the sign of $C$ can be flipped by applying TRS. We fix the sign of $C$ by imposing that the state has a net electron polarization in spin $\downarrow$ (i.e.~the doped holes for $\nu<0$ are primarily in the spin $\uparrow$ sector). We project to four moir\'e bands per spin for both \ch{ZrS2} and \ch{SnSe2}.

We first discuss the results for ZrS$_2$. We note that the phase diagrams for AA and AB stackings are similar. We discuss the fillings $\nu=-1,-2,-4$ in turn:
\begin{itemize}
    \item At $\nu=-1$ (left column of Fig.~\ref{app:fig:interaction_theta_HF_phase_diagram}), we find a gapped phase for sufficiently strong interactions for both AA-\ch{ZrS2} and AB-\ch{ZrS2}. The gapless region only has a small spin polarization, in contrast to the gapped region which is fully spin polarized, except near the phase boundary where a small amount of ISC develops. The threshold interaction strength required to stabilize a gapped state grows with twist angle, which is consistent with the increase in the non-interacting bandwidth of the top valence band. The gapped state is primarily a $C=1$ insulator, where the unoccupied HF band has a high overlap with the highest non-interacting valence band in one spin sector. For larger interaction strengths, we can observe a transition to a trivial $C=0$ spin-polarized insulator. The change in topology arises from a band inversion at the mBZ corner. 
    \item At $\nu=-2$ (middle column of Fig.~\ref{app:fig:interaction_theta_HF_phase_diagram}), we find that both stackings of ZrS$_2$ still require finite interactions to yield an indirect gap, but the threshold interaction strength is smaller than that of $\nu=-1$. Gapped symmetry-preserving states are allowed at $\nu=-2$, but exchange effects are still required to open an insulating gap due to the indirect energy overlap between the first and second non-interacting moir\'e valence bands within each spin. The first gapped phase we encounter from the metallic phase at large $\epsilon_r$ is a TRS-preserving topologically non-trivial state. This is either a quantum spin Hall insulator (QSH$_1$) with spin Chern number $1$, or a $Z_2$ topological insulator (TI) if there is a small amount of ISC which occurs near the phase boundary with the metal. For larger interaction strengths, a significant ISC develops, and the system is primarily a $C=0$ ($C=1$) insulator for AA (AB) stacking. 
    \item At $\nu=-4$ (right column of Fig.~\ref{app:fig:interaction_theta_HF_phase_diagram}), gapped states are possible for both stackings in the non-interacting limit at sufficiently small angles. This is because the second and third valence bands are already fully gapped from each other at the non-interacting level, leading to a QSH$_3$ with spin Chern number $C=3$. For larger interactions and twist angles, a band inversion between the second and third valence bands leads to a QSH$_1$ or Z$_2$ TI. 
\end{itemize}

In Fig.~\ref{app:fig:HF_bandstruct_ZrS2_AA_nu-1_theta4.00_epsr6.25}, we plot the HF band structure for a representative $C=1$ state in AA-\ch{ZrS2} at $\theta=4^\circ$, and compare it with the corresponding non-interacting dispersion. The HF bands in the spin $\downarrow$ sector, which is fully occupied, closely resembles the original non-interacting bands. This implies that the Hartree potential generated by the $C=1$ state is not spatially inhomogeneous enough to reshape the dispersion. On the other hand, the unoccupied HF band in spin $\uparrow$ has been substantially renormalized by interactions. Most notably, for the interaction strength $10/\epsilon_r=1.6$ considered here, bandwidth of the unoccupied HF band has been inflated to $\sim 110\,$meV compared to the original value of $\sim 60\,$meV. This is driven by the enhancement of the energy difference between $\Gamma_M$, which has a large Fock self-energy, and $K_M$, which has a smaller Fock self-energy. We note that the $\Gamma_M$ point has a smaller Berry curvature and quantum geometry than the $K_M$ point. This can be connected to the momentum-contrasting Fock self-energy according to the following argument. A significant Fock exchange between two HF orbitals requires that their Bloch wavefunctions have similar layer and plane-wave structure, and that their momenta are close to each other (since exchange is suppressed by the decay of the interaction potential). Since the region around $K_M$ has a large quantum geometry, corresponding to rapidly varying layer/plane-wave structure, the Fock self-energy there is suppressed.

Finally, we briefly comment on the phase diagrams for AB-SnSe$_2$. Since the first moir\'e valence band for twisted SnSe$_2$ is significantly narrower than that of twisted ZrS$_2$, we find that gapped states begin to appear in AB-SnSe$_2$ for significantly smaller interaction strengths. The existence of a full indirect gap below the first band for sufficiently small twist angles also means that symmetry-preserving gapped insulators are possible at $\nu=-2$ in the non-interacting limit, which would be a QSH$_1$. However, for the interaction strengths $10/\epsilon_r\geq0.4$ considered, we do not find any QSH$_1$ or Z$_2$ TI phases due to band inversions or strong ISC. We do not consider $\nu=-4$ as the continuum model does not correctly capture the second valence band in DFT.

\subsubsection{Phase diagrams: displacement field vs interaction strength}

\begin{figure}[!t]
    \centering
    \includegraphics[width=0.7\textwidth]{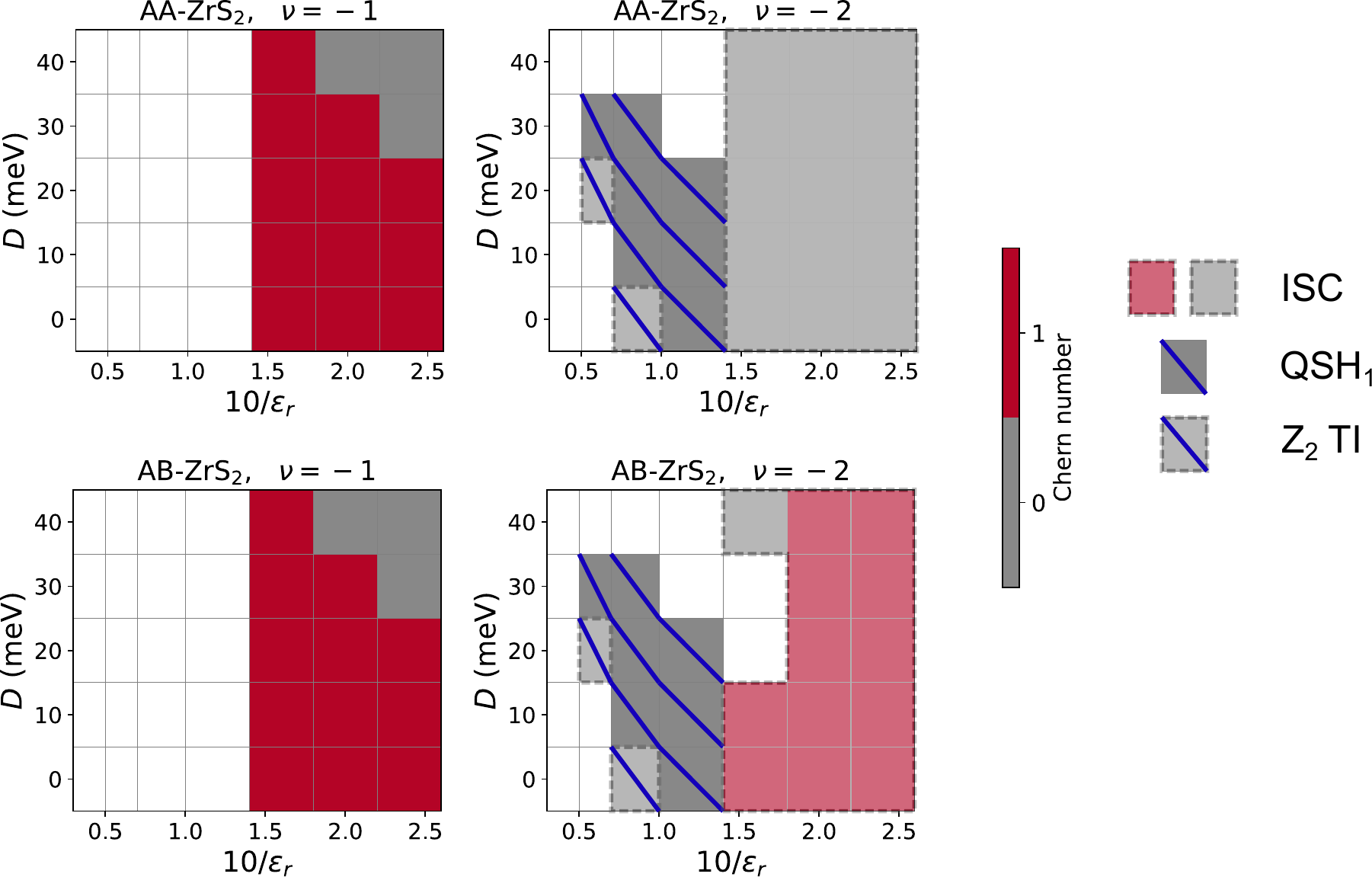}
    \caption{Hartree-Fock phase diagram as a function of displacement field $D$ and interaction strength $10/\epsilon_r$ at $\theta=3.89^\circ$ for various integer filling factors $\nu$. For AA-ZrS$_2$ and AB-ZrS$_2$, we use the \Uone-symmetric continuum model parameters extracted at $\theta=3.89^\circ$. White regions correspond to gapless phases. Light shading denotes inter-spin coherence (ISC). `QSH$_n$' denotes a quantum spin 
    Hall phase with a spin Chern number of $n$. In the presence of finite ISC, `$Z_2$ TI' denotes a topological insulator. System size is $12\times 12$, screening length is $d_\text{sc}=20\,$nm, and we project to four valence moir\'e bands per spin. 
    }\label{app:fig:D_interaction_HF_phase_diagram}
\end{figure}

In Fig.~\ref{app:fig:D_interaction_HF_phase_diagram}, we show the phase diagram of $\theta=3.89^\circ$ AA-ZrS$_2$ and AB-ZrS$_2$ as a function of displacement field $D$ and interaction strength. The displacement field is modelled as an energy difference $D$ between the two layers in the continuum model. At $\nu=-1$, we find that $D$ can drive a topological transition between the $C=1$ and $C=0$ insulators. At $\nu=-2$, a large enough displacement field can destroy the QSH$_1$/Z$_2$ TI phase and lead to a gapless state.

\subsection{Exact diagonalization results}

\begin{figure}[!t]
    \centering
    \includegraphics[width=1\textwidth]{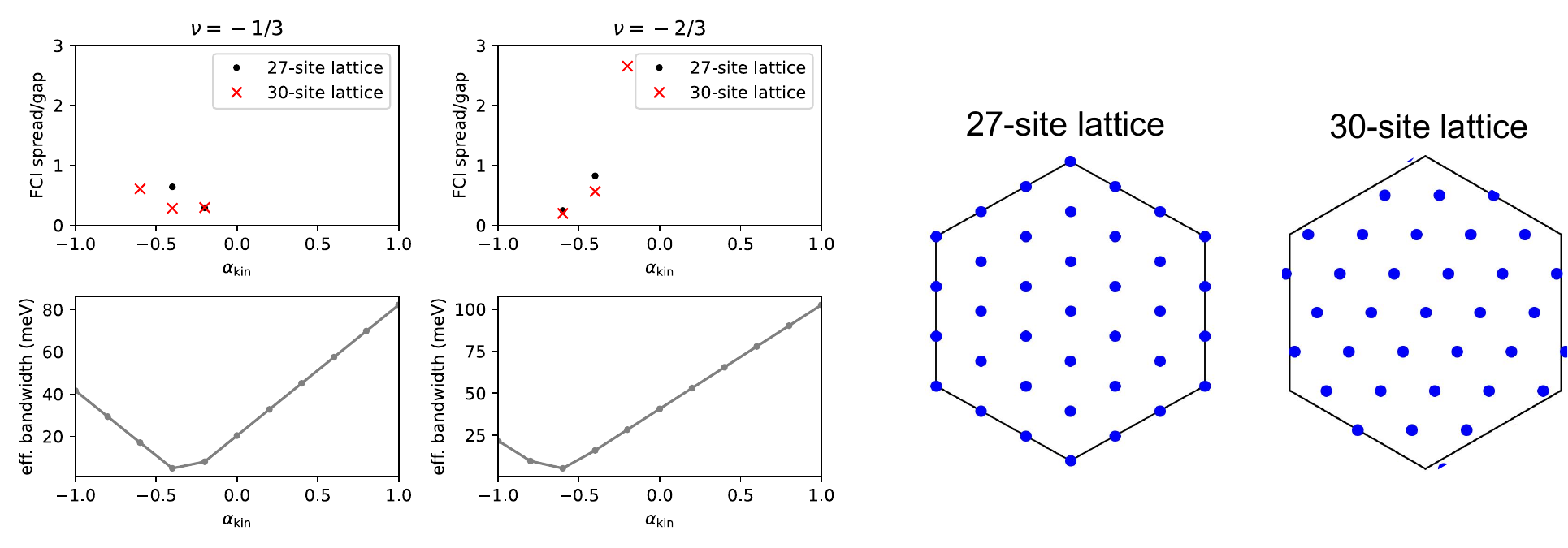}
    \caption{Calculations at fractional fillings in $\theta=3.89^\circ$ AA-ZrS$_2$ with $d_\text{sc}=20\,$nm and $\epsilon_r=5$.
    \textbf{Left:} Top row shows the FCI spread/gap ratio from exact diagonalization (ED) as a function of the scaling factor $\alpha_\text{kin}$ for $\nu=-1/3$ (left column) and  $\nu=-2/3$ (right column). We consider $\alpha_{\text{kin}}=-1.0,-0.8,\ldots,0.8,1.0$. Values of $\alpha_\text{kin}$ without any data markers indicate the absence of any FCI signatures. The ED calculation is restricted to the highest non-interacting valence band within a single spin sector.
    Bottom row shows the effective mean-field bandwidth $\nu=-1/3$ (left column) and  $\nu=-2/3$ (right column), computed on a $12\times12$ mesh (see text for details).
    \textbf{Right:} Momentum meshes in the mBZ for the 27- and 30-site lattices.}\label{app:fig:ED_eff_bandwidth}
\end{figure}

We also perform a preliminary investigation into the possibility of fractional Chern insulators (FCIs) using exact diagonalization (ED). In the following, we present results on $\theta=3.89^\circ$ AA-ZrS$_2$, but similar results are obtained for AB-ZrS$_2$ and AB-SnSe$_2$. We consider the simplest scenario of a $\nu=-1/3$ or $\nu=-2/3$ spin-polarized FCI restricted to the first non-interacting valence band. Hence, we do not include any potential band-mixing or spin depolarization effects.

We recall from \cref{subsubsec:phase_diagrams_int_strength_twist} that interactions tend to enhance the band width of the first valence band (see Fig.~\ref{app:fig:HF_bandstruct_ZrS2_AA_nu-1_theta4.00_epsr6.25}), which is primarily due to Fock exchange effects.
Motivated by this observation, we apply an artificial scaling factor $\alpha_\text{kin}$ to the non-interacting dispersion in the calculation
\begin{equation}
    E(\mathbf{k})\rightarrow \alpha_\text{kin}E(\mathbf{k}).
\end{equation}
Note that $\alpha_\text{kin}=1$ is the physical limit, while $\alpha_\text{kin}=0$ is the band-flattened limit. Negative $\alpha_\text{kin}$ corresponds to inverting the shape of the non-interacting dispersion. Our motivation is that inverting the kinetic dispersion may counteract the interaction-induced renormalization, and hence lead to a narrow `effective' band.

In the top left panels of Fig.~\ref{app:fig:ED_eff_bandwidth}, we show the results of ED calculations at $\nu=-1/3$ and $\nu=-2/3$ on 27- and 30-site lattices (see Fig.~\ref{app:fig:ED_eff_bandwidth} right) for $\epsilon_r=5$. An FCI on the torus exhibits a characteristic topological degeneracy whose momentum quantum numbers depend on the lattice choice. We diagnose the presence of an FCI by computing the spread/gap ratio. To do this, we first identify the putative topological ground state manifold, which consists of the lowest energy states with momenta consistent with the FCI momenta~\cite{regnault2011fractional,bernevig2012emergent}. The spread is defined as the energy difference between the lowest and highest energy states in this topological ground state manifold. The gap is defined as the energy difference between the lowest state not in this manifold, and the highest state in this manifold. A well-formed FCI in finite-size should exhibit a spread/gap ratio which is small and positive. A large or negative ratio indicates the absence of an FCI.

We find that an FCI only appears for a window of negative $\alpha_{\text{kin}}$ for both $\nu=-1/3$ and $\nu=-2/3$. To understand this, we plot an estimate of the `effective' bandwidth  in the bottom left panels of Fig.~\ref{app:fig:ED_eff_bandwidth}. This is computed by constructing the following effective dispersion
\begin{equation}
    E_\text{eff.}(\mathbf{k})=\alpha_\text{kin}E(\mathbf{k})+|\nu|E^\text{HF,int}(\mathbf{k}).
\end{equation}
Above, $E^\text{HF,int}(\mathbf{k})$ is defined as the interacting part of the Hartree-Fock dispersion generated by fully hole-occupying the valence band. Since this contribution is linear in the density matrix, we scale this by a factor $|\nu|$ to mimic the effective contribution at fractional filling. We find that the position (as a function of $\alpha_\text{kin}$) of the minimum effective bandwidth coincides well with the FCI region. We have checked other twist angles in the interval $\theta\in[2.0^\circ,6.0^\circ]$, and found that the FCI only appears for negative $\alpha_\text{kin}$.

\newpage
\section{Basis table for the continuum Hamiltonian}

\begin{longtable}{c c c c}
\caption{Symmetry-allowed terms and fitted coefficients of the first harmonic continuum Hamiltonian for twisted AA-stacked \ch{ZrS2} at \SI{3.89}{\degree}. The column ``Block'' indicates the matrix block in which each term appears, and ``Expression'' gives the corresponding operator structure.}
\label{app:tab:continuum_parameters for first harmonic AA ZrS2}\\
\toprule
$t_i$ & Value ($\mathrm{meV}{\cdot}\SI{}{\angstrom}^k$) & Block & Expression \\
\midrule
\endfirsthead
\toprule
$t_i$ & Value ($\mathrm{meV}{\cdot}\SI{}{\angstrom}^k$) & Block & Expression \\
\midrule
\endhead
\midrule
\multicolumn{4}{r}{\footnotesize\itshape continued on next page} \\
\endfoot
\bottomrule
\endlastfoot
$t_{1}$ & $4.1\times10^{1}$ & $\left[h_{\boldsymbol{Q},\boldsymbol{Q}}\right]_{l,l}$ & $\sigma_{z}\otimes s_{0}$ \\
\midrule
$t_{2}$ & $1.4\times10^{1}$ & $\left[h_{\boldsymbol{Q},\boldsymbol{Q}}\right]_{l,-l}$ & $\sigma_{0}\otimes s_{0}$ \\
\midrule
$t_{3}$ & $-2.5$ & $\left[h_{\boldsymbol{Q},\boldsymbol{Q}}\right]_{l,-l}$ & $\sigma_{z}\otimes s_{0}$ \\
\midrule
$t_{4}$ & $-1.7\times10^{-1}i$ & $\left[h_{\boldsymbol{Q},\boldsymbol{Q}}\right]_{l,-l}$ & $\sigma_{0}\otimes s_{z}$ \\
\midrule
$t_{6}$ & $4.1\times10^{-1}i$ & $\left[h_{\boldsymbol{Q},\boldsymbol{Q}}\right]_{l,-l}$ & $\sigma_{z}\otimes s_{z}$ \\
\midrule
$t_{7}$ & $-6.5\times10^{-1}$ & $\left[h_{\boldsymbol{Q},\boldsymbol{Q}}\right]_{l,l}$ & $k_x\sigma_{x}\otimes s_{x}+k_y\sigma_{x}\otimes s_{y}$ \\
\midrule
$t_{8}$ & $-4.8$ & $\left[h_{\boldsymbol{Q},\boldsymbol{Q}}\right]_{l,l}$ & $k_y\sigma_{x}\otimes s_{z}+k_x\sigma_{y}\otimes s_{0}$ \\
\midrule
$t_{9}$ & $4.6$ & $\left[h_{\boldsymbol{Q},\boldsymbol{Q}}\right]_{l,l}$ & $k_x\sigma_{0}\otimes s_{x}-k_x\sigma_{z}\otimes s_{x}-k_y\sigma_{0}\otimes s_{y}+k_y\sigma_{z}\otimes s_{y}$ \\
\end{longtable}

\begin{longtable}{c c c c}
\caption{Symmetry-allowed terms and fitted coefficients of the first harmonic continuum Hamiltonian for twisted AB-stacked \ch{ZrS2} at \SI{3.89}{\degree}. The column ``Block'' indicates the matrix block in which each term appears, and ``Expression'' gives the corresponding operator structure.}
\label{app:tab:continuum_parameters for first harmonic AB ZrS2}\\
\toprule
$t_i$ & Value ($\mathrm{meV}{\cdot}\SI{}{\angstrom}^k$) & Block & Expression \\
\midrule
\endfirsthead
\toprule
$t_i$ & Value ($\mathrm{meV}{\cdot}\SI{}{\angstrom}^k$) & Block & Expression \\
\midrule
\endhead
\midrule
\multicolumn{4}{r}{\footnotesize\itshape continued on next page} \\
\endfoot
\bottomrule
\endlastfoot
$t_{1}$ & $4.0\times10^{1}$ & $\left[h_{\boldsymbol{Q},\boldsymbol{Q}}\right]_{l,l}$ & $\sigma_{z}\otimes s_{0}$ \\
\midrule
$t_{2}$ & $1.3\times10^{1}$ & $\left[h_{\boldsymbol{Q},\boldsymbol{Q}}\right]_{l,-l}$ & $\sigma_{0}\otimes s_{0}$ \\
\midrule
$t_{3}$ & $-2.4$ & $\left[h_{\boldsymbol{Q},\boldsymbol{Q}}\right]_{l,-l}$ & $\sigma_{z}\otimes s_{0}$ \\
\midrule
$t_{4}$ & $-3.0\times10^{-1}i$ & $\left[h_{\boldsymbol{Q},\boldsymbol{Q}}\right]_{l,-l}$ & $\sigma_{0}\otimes s_{z}$ \\
\midrule
$t_{5}$ & $3.0\times10^{-2}i$ & $\left[h_{\boldsymbol{Q},\boldsymbol{Q}}\right]_{l,-l}$ & $\sigma_{0}\otimes s_{x}+\sigma_{z}\otimes s_{x}$ \\
\midrule
$t_{6}$ & $1.0i$ & $\left[h_{\boldsymbol{Q},\boldsymbol{Q}}\right]_{l,-l}$ & $\sigma_{z}\otimes s_{z}$ \\
\midrule
$t_{7}$ & $-1.4$ & $\left[h_{\boldsymbol{Q},\boldsymbol{Q}}\right]_{l,l}$ & $k_x\sigma_{x}\otimes s_{x}+k_y\sigma_{x}\otimes s_{y}$ \\
\midrule
$t_{8}$ & $3.0$ & $\left[h_{\boldsymbol{Q},\boldsymbol{Q}}\right]_{l,l}$ & $k_x\sigma_{x}\otimes s_{z}-k_y\sigma_{y}\otimes s_{0}$ \\
\end{longtable}

\begin{longtable}{c c c c}
\caption{Symmetry-allowed terms and fitted coefficients of the second harmonic continuum Hamiltonian for twisted AB-stacked \ch{SnSe2} at \SI{4.41}{\degree}. The column ``Block'' indicates the matrix block in which each term appears, and ``Expression'' gives the corresponding operator structure.}
\label{app:tab:continuum_parameters for second harmonic AB SnSe2}\\
\toprule
$t_i$ & Value ($\mathrm{meV}{\cdot}\SI{}{\angstrom}^k$) & Block & Expression \\
\midrule
\endfirsthead
\toprule
$t_i$ & Value ($\mathrm{meV}{\cdot}\SI{}{\angstrom}^k$) & Block & Expression \\
\midrule
\endhead
\midrule
\multicolumn{4}{r}{\footnotesize\itshape continued on next page} \\
\endfoot
\bottomrule
\endlastfoot
$t_{1}$ & $-6.9\times10^{1}$ & $\left[h_{\boldsymbol{Q},\boldsymbol{Q}}\right]_{l,-l}$ & $s_{0}$ \\
\midrule
$t_{2}$ & $8.6i$ & $\left[h_{\boldsymbol{Q},\boldsymbol{Q}}\right]_{l,-l}$ & $s_{x}$ \\
\midrule
$t_{3}$ & $-4.0\times10^{1}i$ & $\left[h_{\boldsymbol{Q},\boldsymbol{Q}}\right]_{l,-l}$ & $s_{z}$ \\
\midrule
$t_{4}$ & $-1.2\times10^{4}$ & $\left[h_{\boldsymbol{Q},\boldsymbol{Q}}\right]_{l,l}$ & $(k_x^2+k_y^2)s_{0}$ \\
\midrule
$t_{5}$ & $-5.9\times10^{3}$ & $\left[h_{\boldsymbol{Q},\boldsymbol{Q}}\right]_{l,-l}$ & $(k_x^2+k_y^2)s_{0}$ \\
\midrule
$t_{6}$ & $6.5\times10^{2}i$ & $\left[h_{\boldsymbol{Q},\boldsymbol{Q}}\right]_{l,-l}$ & $(k_x^2+k_y^2)s_{x}$ \\
\midrule
$t_{7}$ & $8.7\times10^{3}i$ & $\left[h_{\boldsymbol{Q},\boldsymbol{Q}}\right]_{l,-l}$ & $(k_x^2+k_y^2)s_{z}$ \\
\midrule
\multirow{6}{*}{$t_{8}$} & \multirow{6}{*}{$-4.0\times10^{-2}$} & $\left[h_{\boldsymbol{Q},\boldsymbol{Q}-\boldsymbol{b}_{M,1}}\right]_{l,l}$ & $s_{0}$ \\
\end{longtable}

\end{document}